\documentclass[a4paper,11pt]{article}
\pdfoutput=1 
\usepackage{jheppub} 
\usepackage{hyperref}
\usepackage{booktabs}
\usepackage[T1]{fontenc} 
\usepackage{siunitx} 
\usepackage{xcolor,graphicx}
\usepackage{dcolumn}
\usepackage{bm}
\usepackage[normalem]{ulem}
\usepackage{braket} 
\usepackage{amsmath}
\usepackage{cancel}
\usepackage{slashed}
\usepackage{multicol}
\usepackage[capitalise]{cleveref}

\allowdisplaybreaks

\newcommand{\nnl}{\nonumber \\}

\newcommand{\beq}{\begin{equation}}
\newcommand{\eeq}{\end{equation}}
\newcommand{\ba}{\begin{array}}
\newcommand{\ea}{\end{array}}
\newcommand{\bea}{\begin{eqnarray}}
\newcommand{\eea}{\end{eqnarray} }
\newcommand{\be}{\begin{eqnarray}}
\newcommand{\ee}{\end{eqnarray} }
\newcommand{\bal}{\begin{align}}
\newcommand{\eal}{\end{align}}
\newcommand{\bi}{\begin{itemize}}
\newcommand{\ei}{\end{itemize}}
\newcommand{\ben}{\begin{enumerate}}
\newcommand{\een}{\end{enumerate}}
\newcommand{\bc}{\begin{center}}
\newcommand{\ec}{\end{center}}
\newcommand{\bt}{\begin{table}}
\newcommand{\et}{\end{table}}
\newcommand{\btb}{\begin{tabular}}
\newcommand{\etb}{\end{tabular}}

\newcommand{\cL}{\mathcal{L}}
\newcommand{\cO}{\mathcal{O}}

\newcommand{\cM}{{\mathcal M}}

\newcommand{\re}{{\mathrm{Re}} \,}
\newcommand{\im}{{\mathrm{Im}} \,}
\newcommand{\tr}{\mathrm T \mathrm r}

\newcommand{\hc}{\rm h.c.}

\newcommand{\eS}{\epsilon_S}
\newcommand{\eT}{\epsilon_T}
\newcommand{\eP}{\epsilon_P}
\newcommand{\eL}{\epsilon_L}
\newcommand{\eR}{\epsilon_R}

\newcommand{\cvp}{C_V^+}

\newcommand{\csp}{C_S^+}

\newcommand{\capp}{C_A^+}

\newcommand{\ctp}{C_T^+}

\newcommand{\cspd}{\Bar{C}_S^+}

\newcommand{\cappd}{\Bar{C}_A^+}

\newcommand{\cppd}{\Bar{C}_P^+}

\newcommand{\ctpd}{\Bar{C}_T^+}

\newcommand{\cfvpd}{\Bar{C}_{FV}^+}
\newcommand{\cfapd}{\Bar{C}_{FA}^+}
\newcommand{\cftpd}{\Bar{C}_{FT}^+}

\newcommand{\cmmd}{\Bar{C}_M^+}

\newcommand{\ceed}{\Bar{C}_E^+}
\newcommand{\lzm}{\left(}
\newcommand{\dzm}{\right)}
\newcommand{\lzs}{\left[}
\newcommand{\dzs}{\right]}

\newcommand{\emax}{E_e^{\rm max}}

\usepackage{etoolbox,siunitx}
\robustify\bfseries

\begin{document}

\preprint{\parbox{10cm}{\flushright IFIC/21-54\\
FTUV-21-1213.6108\\}}

\title{Constraints on subleading interactions \\ in beta decay Lagrangian}

\author[a]{Adam Falkowski,}
\author[b]{Mart\'{i}n Gonz\'{a}lez-Alonso,}
\author[c]{Ajdin Palavri\'{c},}
\author[a]{Antonio Rodr\'iguez-S\'anchez}

\affiliation[a]{Universit\'{e} Paris-Saclay, CNRS/IN2P3, IJCLab, 91405 Orsay, France}
\affiliation[b]{Departament de F\'isica Te\`orica, IFIC, Universitat de Val\`encia - CSIC, Apt.  Correus 22085, E-46071 Val\`encia, Spain}
\affiliation[c]{Albert Einstein Center for Fundamental Physics, Institute for Theoretical Physics, University of Bern, CH-3012 Bern, Switzerland}

\emailAdd{adam.falkowski@ijclab.in2p3.fr,
          martin.gonzalez@ific.uv.es,
          ajdin.palavric@unibe.ch,
          arodriguez@ijclab.in2p3.fr
          }

\abstract{
We discuss the effective field theory (EFT) for nuclear beta decay. The general quark-level EFT describing charged-current interactions between quarks and leptons is matched to the nucleon-level non-relativistic EFT at the ${\cal O}$(MeV) momentum scale characteristic for beta transitions. The matching takes into account, for the first time, the effect of all possible beyond-the-Standard-Model interactions at the subleading order in the recoil momentum. We calculate the impact of all the Wilson coefficients of the leading and subleading EFT Lagrangian on the differential decay width in allowed beta transitions. As an example application, we show how the existing experimental data constrain the subleading Wilson coefficients corresponding to pseudoscalar, weak magnetism, and induced tensor interactions. The data display a 3.5 sigma evidence for nucleon weak magnetism, in agreement with the theory prediction based on isospin symmetry.
}
\maketitle

\section{Introduction} 

Nuclear beta transitions have been at the center of the particle physics research program since over a hundred years. Historically they have been essential for understanding various ingredients of the Standard Model (SM), such as the existence of neutrinos, non-conservation of parity, or the Lorentz structure of weak interactions~\cite{Pauli:1930pc,Fermi:1934hr,Cowan:1956rrn,Lee:1956qn,Wu:1957my,Weinberg:2009zz}. From the vantage point of a particle physicist today, their main role is twofold. On one hand they offer an opportunity for precision measurements of fundamental constants of the SM, notably of the $V_{ud}$ CKM matrix element~\cite{Abele:2008zz,Gonzalez-Alonso:2018omy,Hardy:2020qwl,Dubbers:2021wqv,Falkowski:2020pma}. They also provide insight into the complex non-perturbative dynamics emerging from the SM, for instance through phenomenological determinations of the axial nucleon charge $g_A$~\cite{Abele:2008zz,Falkowski:2020pma,Dubbers:2021wqv}. 
On the other hand, they yield important constraints on physics beyond the SM (BSM),  such as leptoquarks and other hypothetical particles contributing to scalar and tensor currents in weak interactions~\cite{Herczeg:2001vk,Gonzalez-Alonso:2018omy,Falkowski:2020pma}. On a more theoretical side, because of a wide range of scales and physical processes involved, beta transitions are a perfect laboratory to research and develop the concepts of Effective Field Theory (EFT).   

The main ingredients of the theory of beta transitions were worked out by the end of 1950s, see in particular Refs.~\cite{Lee:1956qn,Jackson:1957zz,Weinberg:1958ut}.
The flip side is that the habitual language in the literature may sometimes be unfamiliar to contemporary QFT practitioners. One of the goals of this paper is to reformulate beta transitions in the modern EFT language. The advantage, apart from the conceptual side, is that the theory can be smoothly incorporated  into the ladder of EFTs spanning various energy scales, from the TeV scale down to MeV. In particular, the EFT for beta transitions can be matched to the so-called WEFT (the general EFT of SM degrees of freedom {\em below} the electroweak scale), and via this intermediary to the SMEFT (the general EFT of SM degrees of freedom {\em above} the electroweak scale). This way, the general effects of heavy non-SM particles  can be naturally incorporated, along with the more studied SM effects,  into the low-energy effective theory of beta decay. 

An appropriate EFT framework to describe beta decay is the pionless EFT~\cite{vanKolck:1999mw}.  
The relevant degrees of freedom are the nucleons (protons and neutrons) and leptons (electrons and electron neutrinos).
Most of the details of the pionless EFT Lagrangian, such as the nucleon self-interactions describing the nuclear forces, are not relevant for our discussion. 
For the sake of this paper we will focus instead on the interactions mediating beta transitions, which are quartic terms connecting the proton, neutron, electron, and neutrino fields and their derivatives.  
Amplitudes for the neutron decay can be directly calculated starting from this Lagrangian. 
As for the beta decay of nuclei with the mass number $A>1$, the amplitudes involve matrix elements of the nucleon bilinears between the nuclear states. 
This is analogous to the usual treatment of hadrons in QCD, where the amplitudes involve matrix elements of quark operators between hadronic states.

The important difference between the present EFT approach and hadrons in QCD is in power counting. 
Since the 3-momentum transfer in beta transitions is much smaller than the nucleon mass $m_N$, the EFT Lagrangian can be organized in a non-relativistic expansion in powers of $\boldsymbol{\nabla}/m_N$.\footnote{In this paper we use the notation where 3-vectors are represented by {\bf bold-font} symbols.}
Focusing on the part of the Lagrangian  mediating beta transitions, the leading $\cO(\boldsymbol{\nabla}^0)$ term in this expansion encodes the usual Fermi and Gamow-Teller contributions to the allowed beta decays. This includes the SM-like contributions from the vector and axial currents, as well as the non-SM ones from the scalar and tensor currents.  
The resulting beta decay observables, such as the lifetime or angular correlations, are described exactly by the formulas obtained by Jackson, Treiman, and Wyld in the seminal Ref.~\cite{Jackson:1957zz}.\footnote{%
In the present EFT we assume the SM degrees of freedom, thus right-handed neutrinos are absent. More precisely, the leading order EFT interactions lead to the formulas of Ref.~\cite{Jackson:1957zz} in the limit where their couplings to right-handed neutrinos are set to zero. 
It is trivial to generalize our EFT to include right-handed neutrinos as well.}

The subleading corrections to these leading order expressions are the main focus of this paper. 
They vanish in the limit where 3-momenta of the parent and daughter nuclei are zero, hence in the literature they are referred to as the {\em recoil corrections}. 
We restrict to discussing the effects linear in 3-momenta of the nuclei. 
These originate from two sources. 
One is the $\cO(\boldsymbol{\nabla}^1)$ terms in the EFT Lagrangian, which gives a complete description of recoil effects in neutron decay.
For nuclei with $A > 1$ the other source is the  $\cO(\boldsymbol{\nabla}^0)$ Lagrangian with the nuclear matrix elements expanded to linear order in the 3-momenta. 
We give the full expressions for the amplitudes at the linear recoil level, as well as the corresponding differential decay width in a convenient parametrization. 
In the limit where non-standard currents are absent, our results can be matched to those in Ref.~\cite{Holstein:1974zf}.
The novelty of this paper is that we also present a complete treatment on non-standard corrections at the linear recoil level. 
We describe how the quark-level scalar and pseudoscalar interactions enter the beta decay observables.  
Moreover, we give a complete description of the effects of quark-level tensor interactions, which lead a number of distinct terms in the leading and subleading Lagrangian of our low-energy EFT. 

Parameters of the leading-order EFT interactions have been fit from data for more than 70 years. 
The existing experimental data on beta transitions~\cite{Abele:2008zz,Hardy:2020qwl,Gonzalez-Alonso:2018omy,Dubbers:2021wqv,Severijns:2021lsp} are nowadays precise enough to be sensitive to recoil corrections. Our formalism can be employed to place meaningful constraints on  Wilson coefficients of leading and subleading EFT operators.
We construct a global likelihood function for the Wilson coefficients based on the state-of-the art measurements of superallowed $0^+ \to 0^+$ transitions, neutron decay, mirror decay, and other allowed transitions.
This likelihood encodes confidence intervals for all the  Wilson coefficients. 
In addition to constraints on the leading Wilson coefficients, already obtained in Ref.~\cite{Falkowski:2020pma}, we derive constraints on certain Wilson coefficients of subleading EFT operators generated by BSM physics.  
In particular, we analyse the effects of pseudoscalar interactions on beta transitions, and obtain simultaneous constraints on non-standard pseudoscalar, scalar, tensor, and right-handed currents.   
Next, we discuss the Wilson coefficient of the  EFT operator describing the nucleon-level weak magnetism. 
Usually, its magnitude is determined by theory using isospin symmetry, which in this context is referred to as the conserved vector current (CVC) hypothesis.  We show that this Wilson coefficient is now efficiently constrained by the global data, which provides the first evidence for the nucleon-level weak magnetism.
Finally, we also discuss a subleading EFT operator describing the so-called induced tensor interactions (one of the second class currents in the classification of \cite{Weinberg:1958ut}).  
While isospin symmetry predicts that this Wilson coefficient should be negligibly small, the data show a 1.8 sigma preference for its non-zero value.   

This paper is organized as follows. 
In \cref{sec:EFT} we lay out the formalism connecting the general quark-level EFT below the electroweak scale to the low-energy EFT describing weak charged-current interactions of nucleons.  
Based on the latter EFT, in \cref{sec:recoil} we calculate the recoil corrections to the beta transition amplitudes and observables (the lifetime and correlation coefficients). 
Global fits to the Wilson coefficients are presented in \cref{sec:fits}. 
Our conclusions are contained in \cref{sec:conclusions}. 
\cref{app:clebsch} contains some useful mathematical details about spin representations,  while the contributions of all one-derivative EFT operators to the beta decay correlations coefficients are summarized in \cref{app:sub}.

\section{Effective Lagrangian for beta decay} 
\label{sec:EFT}

\subsection{WEFT}
The starting point is the so-called weak EFT (WEFT) Lagrangian, which is defined at the scale $\mu \simeq 2$~GeV and organized in an expansion in $\partial/m_W$, where  the $W$ boson mass $m_W$ plays the role of the cutoff scale.
The leading order term describing charged-current interactions between quarks and leptons is~\cite{Bhattacharya:2011qm}
\begin{align}
\cL_{\rm WEFT}   \supset   
- \frac{V_{ud}}{v^2} & \Big \{
\left( 1 +  \eL \right) \
\bar{e}  \gamma_\mu  \nu_L  \cdot \bar{u}   \gamma^\mu  (1 - \gamma_5)  d \Big.
+\eR   \,  \bar{e}  \gamma_\mu \nu_L
\cdot \bar{u}   \gamma^\mu  (1 + \gamma_5)  d
\nnl & 
+{1 \over 2} \eT    \   \bar{e}   \sigma_{\mu \nu} \nu_L    \cdot  \bar{u}   \sigma^{\mu \nu} d
+ \eS  \, \bar{e}  \nu_L  \cdot  \bar{u} d
- \eP  \,  \bar{e}  \nu_L  \cdot  \bar{u} \gamma_5 d
\Big \} 
+ {\rm h.c.}~ \qquad 
\label{eq:TH_Lweft}
\end{align}
where $u$, $d$, $e$, and $\nu_{L} \equiv (1-\gamma_5)\nu/2$  are the up quark, down quark, electron, and left-handed electron neutrino fields,
$\sigma_{\mu\nu} \equiv {i \over 2} [\gamma_\mu,\gamma_\nu]$, 
$V_{ud}$ is the CKM matrix element, 
and $v \equiv (\sqrt 2 G_F)^{-1/2} \approx 246.22$~GeV. 
The central assumption is that, below 2~GeV,  there are no other light degrees of freedom except for those of the SM. 
We treat the neutrino as massless, its tiny mass having no discernible effects on the observables studied in this paper. 
The Wilson coefficients $\epsilon_X$, 
$X = L,R,S,P,T$, parametrize possible effects of non-SM particles heavier than 2~GeV.
In the SM limit,  $\epsilon_X = 0$ for all $X$.

The Lagrangian in \cref{eq:TH_Lweft} is convenient to connect to new physics at {\em high} scales. 
For example, integrating out the so-called $S_1$ leptoquark~\cite{Dorsner:2016wpm,Angelescu:2021lln} with mass $M$ and Yukawa couplings $y$ can be approximated by the Wilson coefficients $\epsilon_S \approx -\epsilon_P \approx -\epsilon_T = {|y|^2 \over 4 V_{ud} M^2}$, up to loop and RG corrections. 
More generally, $\epsilon_X$ can be matched at the scale $\mu \simeq m_W$ to the  Wilson coefficients of the SMEFT (see e.g.~\cite{Falkowski:2019xoe}), which captures a broad range of new physics scenarios with heavy particles~\cite{deBlas:2017xtg}. 
In this paper, however, we are interested in {\em low-energy} physics of beta transitions. 
In these processes, the relevant degrees of freedom are not quarks but {\em nucleons} (protons and neutrons) or composite states thereof. In the following we connect the EFT in \cref{eq:TH_Lweft} to another EFT describing charged-current interactions of nucleons and leptons.   

\subsection{Nucleon matrix elements}

As a first step toward this end, we define the matrix elements of quark currents between nucleon states~\cite{Weinberg:1958ut,Holstein:1974zf,Bhattacharya:2011qm}:\footnote{%
For the vector and axial matrix elements our notation is close to that of  Ref.~\cite{Holstein:1974zf}, except that we trade $g_{II} \to g_{IT}$, and $g_{S,P} \to g_{IS,IP}$ (to remove the conflict with the $g_{S,P}$ form factors in the scalar and pseudoscalar currents). The notation for the scalar, pseudoscalar, and  tensor matrix elements follows that of  Ref.~\cite{Bhattacharya:2011qm}, 
up to $i$ and $m_N$ factors to make the $g_T^{(i)}$ form factors real and dimensionless. Compared to~\cite{Bhattacharya:2011qm} we also omit the $g_T^{(2)}$ form factor because 
its effect is equivalent to that of $g_T^{(1)}$ up to $\cO(\boldsymbol{q}^2/m_N^2)$.}
\begin{align}
\label{eq:TH_matrixelements}
\bra{p(k)}  \bar{u}   \gamma_\mu  d  \ket{n(p)}  =& 
  \bar  u_p \bigg [ g_V(q^2) \gamma_\mu + { g_{IS}(q^2)  \over 2 m_N } q_\mu 
- i {g_M(q^2) - g_V(q^2) \over 2 m_N} \sigma_{\mu \nu} q^\nu \bigg ] u_n , 
\nnl 
\bra{p(k)}  \bar{u}   \gamma_\mu  \gamma_5 d  \ket{n(p)}  =& 
 \bar  u_p \bigg [ g_A(q^2) \gamma_\mu \gamma_5  + { g_{IP}(q^2)  \over 2 m_N }\gamma_5 q_\mu 
- i {g_{IT}(q^2) \over 2 m_N} \sigma_{\mu \nu} \gamma_5 q^\nu \bigg ] u_n , 
\nnl 
\bra{p(k)}  \bar{u}   d  \ket{n(p)}  =& g_S(q^2) \bar u_p u_n   , 
\nnl 
\bra{p(k)} \bar u \gamma_5  d  \ket{n(p)}  =& 
g_P(q^2) \bar  u_p \gamma_5 u_n,
\nnl 
\bra{p(k)}  \bar{u}   \sigma_{\mu\nu}  d  \ket{n(p)}  =& 
\bar  u_p \bigg [ g_T(q^2) \sigma_{\mu\nu}  
+  {i g_T^{(1)}(q^2) \over m_N}  \big ( q_\mu \gamma_\nu -   q_\nu \gamma_\mu \big )  
\nnl  & 
+   {i g_T^{(3)}(q^2) \over m_N}  \big ( \gamma_\mu \gamma_\rho  \gamma_\nu -  \gamma_\nu  \gamma_\rho  \gamma_\mu  \big )   q^\rho
 \bigg ] u_n . 
\end{align}
Above $p$ and $k$ are the momenta of an incoming neutron and an outgoing proton, 
and $q = p- k$ is the momentum transfer. 
Next,  $m_N \equiv (m_n + m_p)/2$ is the nucleon mass, and $u_N$ is the Dirac spinor wave function of the neutron or proton, 
which implicitly depends on the respective momentum and polarization.
The matrix elements take the most general form allowed by the Lorentz symmetry and the discrete $P$ and $T$ symmetries of QCD. 
The dependence on the momentum transfer is encoded in the form factors  $g_X(q^2)$, which must be real by $T$ invariance.
We also define nucleon charges $g_X \equiv g_X(0)$, which are the relevant parameters for beta transitions, where $q^2 \ll m_N^2$. Symmetries impose important restrictions on the possible values of the charges coming from the different quark currents:
\begin{itemize}
\item \textit{Vector current}. The Ademollo-Gatto theorem implies that, up to second order in isospin breaking, $g_{V}=1$ \cite{Ademollo:1964sr,Donoghue:1990ti}. The induced scalar coupling, $g_{IS}$, vanishes in the isospin limit and then an $\cO(10^{-2}-10^{-3})$ value is expected for it. Finally, $g_{M}$ can be related, through an isospin rotation (CVC), to the response of the nucleons to an external magnetic field and can be fixed, up to relative $\mathcal{O}(10^{-2})$ isospin breaking corrections, from the experimentally known difference of the magnetic moment of the nucleons \cite{GellMann:1964tf,Holstein:1974zf,Cirigliano:2022hob},
\begin{equation}
g_M=\frac{\mu_{p}-\mu_{n}}{\mu_{N}}=4.706, 
\end{equation}
where $\mu_{N}$ is the nuclear magneton $\frac{e}{2 m_p}$.
\item \textit{Axial current}. The axial charge $g_A$ is not known from symmetry considerations alone and in practice is fixed from experimental data or by lattice calculations. 
As for the latter, the FLAG'21 average quotes 
$g_A = 1.246(28)$~\cite{Aoki:2021kgd,Gupta:2018qil,Chang:2018uxx,Walker-Loud:2019cif}. 
The induced pseudoscalar charge, $g_{IP}$, only enters into the observables at second order in the recoil expansion and then, in principle, it is beyond the scope of this analysis. Notice however that  $g_{IP}$ is enhanced by the pion pole. 
Indeed, using partial conservation of axial current one has~\cite{Gonzalez-Alonso:2013ura}
\begin{equation}
\label{eq:PCAC}
-\frac{q^2 g_{IP}(q^{2})}{4 m_{N}^2}=\frac{\overline{m}_{q}}{m_{N}}g_{P}(q^{2})-g_{A}(q^{2}) \, , 
\end{equation}
where $\overline{m}_{q}$ is the average of the light quark masses.
Using that $g_{A}(q^{2})$ is a very smooth function of $q^{2}$ for $q^{2}\lesssim m_{\pi}^2$  and that $g_{P}(q^{2})$ is dominated by the pion pole contribution (e.g. see \cite{Donoghue:1992dd,Chen:2021guo}), one obtains, up to a few per-cent level correction,
\begin{equation}
g_{IP}=-\frac{4 m_{N}^{2}}{m_{\pi}^{2}}g_{A} \approx -226(5)  \, , 
\end{equation}
where the displayed uncertainty is due to the error on the lattice determination of $g_A$.
Finally the induced tensor, $g_{IT}$, which by itself enters suppressed by one power in the recoil expansion, vanishes in the isospin limit, and then its expected size is $\mathcal{O}(10^{-2}-10^{-3})$.
\item \textit{Non-standard currents.} 
The  $q^{2} \to 0$ limit of  \cref{eq:PCAC} fixes the pseudoscalar charge in terms of the axial one~\cite{Gonzalez-Alonso:2013ura}: 
\begin{equation}
\label{eq:pseudocharge}
g_{P}=\frac{m_{N}}{\overline{m}_{q}}g_{A}= 346(9), 
\end{equation}
where we use $\overline{m}_q=3.381(40)\, \mathrm{MeV}$ from the lattice~\cite{Aoki:2021kgd,RBC:2014ntl,Durr:2010vn,Durr:2010aw,McNeile:2010ji,Bazavov:2010yq,FermilabLattice:2018est,EuropeanTwistedMass:2014osg}, and the error is again dominated by the lattice uncertainty of $g_A$. 
For the scalar and tensor charges we use the lattice values 
$g_S = 1.022(100)$ and $g_T = 0.989(34)$~\cite{Aoki:2021kgd,Gupta:2018qil} (see also Ref.~\cite{Gonzalez-Alonso:2013ura}).
Concerning the recoil level tensor charges, 
$g_{T}^{(1)}$ is expected to be $\mathcal{O}(1)$,
while $g_{T}^{(3)}$ vanishes in the isospin limit and then its expected size is $\cO(10^{-2}-10^{-3})$.
\end{itemize}

\subsection{Pionless EFT}

Beta transitions are characterized by a 3-momentum transfer much smaller than the nucleon mass, typically 
$|\boldsymbol{q}| \sim$1-10~MeV.
Therefore, the nucleon degrees of freedom are non-relativistic (in an appropriate reference frame), and can be described by non-relativistic quantum fields $\psi_N$, $N=p,n$, which are 2-component spinors.  
On the other hand, we continue describing leptons by the relativistic fields $\nu_L$ and $e$. 
The effective Lagrangian is constructed as the most general function of $\psi_N$, $e$, and $\nu_L$ and their derivatives,  respecting the rotational symmetry and Galilean boosts. 
It is organized in an expansion in $\boldsymbol{\nabla}/m_N$, where $\boldsymbol{\nabla}$ denotes spatial derivatives.\footnote{%
Notice that all hadronic degrees of freedom, including pions, have been integrated out and, as a consequence, certain Wilson Coefficients can get enhanced by the large $m_N/m_\pi$ ratio. 
In particular, the coefficient of pseudo-scalar interactions $C_P^+$ is enhanced by $m_N^2/m_\pi^2$, cf. \cref{eq:pseudocharge}.
This enhancement is however significantly smaller than the recoil suppression of $C_P^+$ effects in nuclear beta transitions, 
and therefore it is consistent to treat the pseudo-scalar interactions as subleading. }
This framework is known as the {\em pionless EFT}~\cite{vanKolck:1999mw}. 
In this paper we are interested in the subset of the pionless EFT Lagrangian relevant for beta transitions, that is in the quartic interactions between a proton, a neutron, an electron, and a neutrino.\footnote{%
Interactions with more than two nucleon fields also contribute to nuclear beta transitions (A>1), in particular the ones with four nucleons and two leptons are referred to as two-body currents in the literature. We comment on this issue in \cref{app:twobody}.  }
We organize these interactions as 
\begin{equation}
\label{eq:TH_NRleeyang}
\cL_{\pi \!\!\! / \rm EFT} \supset 
\cL^{(0)} + \cL^{(1)}  + \cO( \boldsymbol{\nabla}^2/m_N^2) + \hc \, ,
\end{equation} 
where $\cL^{(n)}$ refers to $\cO(\boldsymbol{\nabla}^n/m_N^n)$ terms.
At the zero-derivative level, the most general Lagrangian respecting the rules mentioned above is\footnote{For brevity, we do not display the overall $e^{\pm i (m_p-m_n) t}$ factors multiplying the Lagrangian terms away from the isospin limit.} 
\begin{equation}
\label{eq:TH_NRleeyang0}
 \cL^{(0)}   =  
 - (\psi_p^\dagger \psi_n)   \bigg  [    
C_V^+   \bar e_L \gamma^{0} \nu_L   +   C_S^+ \bar e_R  \nu_L  
  \bigg ] 
+  (\psi_p^\dagger \sigma^k \psi_n)   \bigg  [    
  C_A^+ \bar e_L \gamma^{k}  \nu_L  +   C_T^+ \bar e_R \gamma^0\gamma^{k}  \nu_L    \bigg ] ~,
\end{equation}
where $\sigma^k$ are the Pauli matrices, 
$\gamma^\mu = \begin{pmatrix} 0 & \sigma^\mu \\ \bar \sigma^\mu & 0 \end{pmatrix}$, 
$\sigma^\mu = (\sigma^0,\sigma^k)$,  
$\bar \sigma^\mu = (\sigma^0,-\sigma^k)$.   
At this level we have only two distinct nucleon bilinears,  $\psi_p^\dagger \psi_n$ and $\psi_p^\dagger \sigma^k \psi_n$, which, up to two-body current effects, mediate the so-called allowed Fermi and Gamow-Teller (GT) transitions, respectively.
The labels and normalization of the Wilson coefficients $C_X^+$ are chosen such that they simply relate to the familiar parameters of the Lee-Yang Lagrangian~\cite{Lee:1956qn}: 
$C_X^+ \equiv C_X + C_X'$ for $X=V,A,S,T$. 
In order to match $C_X^+$ to the parameters of the quark-level Lagrangian we calculate the $n \to p e^- \bar \nu$ amplitude in two ways:   using \cref{eq:TH_NRleeyang0}, and using \cref{eq:TH_Lweft} together with \cref{eq:TH_matrixelements}. 
We then demand that both calculations give the same result in the limit $\boldsymbol{q} \to 0$. 
This procedure leads to the matching equations 
\begin{align}
\label{eq:TH_matching0}
C_V^+  =&  {V_{ud} \over v^2} \bigg \{  
g_V \big ( 1 +  \epsilon_L + \epsilon_R \big )\sqrt{1 + \Delta_R^V}
- {m_e \over m_N} g_T^{(1)}   \epsilon_T  \bigg \}  , 
\nnl 
C_A^+  =&  - {V_{ud} \over v^2} \bigg \{    
g_A \big ( 1 +  \epsilon_L - \epsilon_R \big ) \sqrt{1 + \Delta_R^A} 
+  2{m_e \over m_N} g_T^{(3)} \epsilon_T  \bigg \} , 
\nnl 
C_S^+  =&  {V_{ud} \over v^2} \bigg \{  
g_S \epsilon_S + {m_e \over 2 m_N} g_{IS}  \big ( 1 +  \epsilon_L + \epsilon_R \big )  \bigg \}  ,
\nnl 
C_T^+  =&  {V_{ud} \over v^2} g_T \epsilon_T , 
\end{align}
up to $\cO(\boldsymbol{q}^2/m_N^2)$ effects which are consistently neglected in our analysis. 
On the other hand, we keep track of $\cO(\boldsymbol{q}/m_N)$ effects, which appear in this matching as the terms  suppressed by $m_e/m_N$.
One important thing to notice is that the  quark-level pseudoscalar interactions, parametrized by $\epsilon_P$ in \cref{eq:TH_Lweft}, do not affect the leading order Lagrangian of pionless EFT. 
The matching in \cref{eq:TH_matching0} is essentially tree-level, 
 but for the vector and axial Wilson coefficients we also included the short-distance (inner) radiative corrections, where we use the numerical values 
$\Delta_R^V = 0.02467(22)$~\cite{Gorchtein:2018fxl} and 
$\Delta_R^A - \Delta_R^V = 0.00026(26)$~\cite{Gorchtein:2020quo} (see also \cite{Cirigliano:2022hob}).
Other radiative corrections in this matching are not relevant from the phenomenological point of view and are omitted.

At the next-to-leading (one-derivative) order we consider the following interactions:
\begin{align}
\label{eq:TH_NRleeyang1}
\cL^{(1)}   = &   {1 \over 2 m_N} \bigg \{  
i C_P^+   ( \psi_p^\dagger \sigma^k\psi_n) \nabla_k \big (\bar e_R \nu_L \big )
- C_M^+ \epsilon^{ijk}  ( \psi_p^\dagger \sigma^j \psi_n)   
\nabla_i   \big  (\bar e_L \gamma^k  \nu_L \big ) 
 \nnl  &
- i C_E^+  ( \psi_p^\dagger \sigma^k\psi_n) \nabla_k \big (\bar e_L \gamma^{0} \nu_L \big )
- i C^+_{E'}  ( \psi_p^\dagger \sigma^k\psi_n) \partial_t \big (\bar e_L \gamma^k \nu_L \big ) 
 \nnl  &
- i  C^+_{T1}  ( \psi_p^\dagger \psi_n) 
\nabla_k \big  (\bar e_R \gamma^0 \gamma^k \nu_L  \big )  
+ i C^+_{T2} (\psi_p^\dagger \psi_n)  
(\bar e_R  \overleftrightarrow{\partial_t} \nu_L) 
+ 2 i C^+_{T3} (\psi_p^\dagger \sigma^k \psi_n)   
(\bar e_R  \overleftrightarrow{\nabla}_k \nu_L) \
\nnl & 
-  i C^+_{FV}  (\psi_p^\dagger  \overleftrightarrow{\nabla}_k \psi_n)   
 (\bar e_L \gamma^k \nu_L)    
+   i C^+_{FA}   (\psi_p^\dagger \sigma^k  \overleftrightarrow{\nabla}_k \psi_n)    (\bar e_L \gamma^{0} \nu_L)
+ C^+_{FT} \epsilon^{ijk} (\psi_p^\dagger \sigma^i  \overleftrightarrow{\nabla}_j \psi_n)   (\bar e_R \gamma^{0} \gamma^k \nu_L)  
\bigg \}  , 
\end{align}
where  $f_1 \Gamma \overleftrightarrow{\nabla} f_2    \equiv   f_1   \Gamma \nabla  f_2  - \nabla f_1 \Gamma f_2$. 
Again, we calculate the $n \to p e^- \bar \nu$ amplitude in two ways:   using \cref{eq:TH_NRleeyang1} and using \cref{eq:TH_Lweft} together with \cref{eq:TH_matrixelements}, but now we concentrate on linear terms in $\boldsymbol{q}$ in the limit $|\boldsymbol{q}|/m_N \ll 1$.   Matching these linear terms requires the following identification of the nucleon-level and quark-level Wilson coefficients:
\begin{align}
\label{eq:TH_matching1}
C_P^+  =&  {V_{ud} \over v^2} g_P \epsilon_P, 
\nnl 
C^+_{M}  =&  {V_{ud} \over v^2} g_M \big ( 1 +  \epsilon_L + \epsilon_R \big ) 
= {g_M \over g_V} C_V^+  ,
\nnl 
C^+_{E}  =& C^+_{E'} = {V_{ud} \over v^2} g_{IT} \big ( 1 +  \epsilon_L -\epsilon_R \big )  =  - {g_{IT} \over g_A} C_A^+  , 
\nnl 
C^+_{T1}  =&  {V_{ud} \over v^2} g_T \epsilon_T   =  C_T^+ ,  
\nnl 
C^+_{T2}  =  & 
 - {2 V_{ud} \over v^2}  g_T^{(1)}  \epsilon_T  = 
-2  {g_T^{(1)} \over g_T} C_T^+  ,  
\nnl 
C^+_{T3}  =  & - {2 V_{ud} \over v^2} g_T^{(3)} \epsilon_T 
= -2 {g_T^{(3)} \over g_T}  C_T^+  , 
\nnl
C^+_{FV}  =&  {V_{ud} \over v^2} g_V \big ( 1 +  \epsilon_L + \epsilon_R \big ) 
= C_V^+  , 
\nnl 
C^+_{FA}  =&  - {V_{ud} \over v^2}    g_A \big ( 1 +  \epsilon_L - \epsilon_R \big )  
= C_A^+  , 
\nnl 
C^+_{FT}  =&  {V_{ud} \over v^2} g_T \epsilon_T 
= C_T^+  , 
\end{align}
where all the equalities are true up to $\cO(\boldsymbol{q}/m_N)$ corrections, i.e  
up to terms suppressed by $m_e/m_N$ or $(m_n-m_p)/m_N$.
Note that  we should not keep such terms in the matching of the Wilson coefficients of the {\em subleading} Lagrangian, as they correspond to $\cO(\boldsymbol{q}^2/m_N^2)$ effects in our EFT counting.
We can see that quark-level pseudoscalar interactions,  which 
arise only beyond the SM, induce the interaction term proportional to $C_P^+$ in \cref{eq:TH_NRleeyang1}.\footnote{%
Within the SM one has the contribution to $C_P^+$ through the induced pseudoscalar coupling, 
$\Delta C_P^+  =  {V_{ud} \over v^2} {m_e \over 2 m_N} g_{IP}$. This is however suppressed by an additional factor of $\boldsymbol{q}/m_N$, and thus is neglected here as an $\cO(\boldsymbol{q}^2/m_N^2)$ effect.} 
The interaction term proportional to $C^+_M$ is known as the {\em weak magnetism}, 
while that proportional to $C^+_E$ is referred to as the induced tensor term.\footnote{%
This is a misnomer because it has nothing to do with the bona fide tensor interactions at the quark level, which contribute to completely different structures in the non-relativistic EFT Lagrangian. We will however use this name, to conform with the bulk of the beta decay literature.}
The Wilson coefficients $C^+_{T1}$-$C^+_{T3}$ are all proportional to $\epsilon_T$  parametrizing tensor interactions at the quark-level. 
They are less important phenomenologically, as we expect to first observe effects of $\epsilon_T$ through its contributions to the leading order Lagrangian in \cref{eq:TH_NRleeyang0}. 
The interactions in the first two lines of \cref{eq:TH_NRleeyang1} depend on the same  nucleon bilinears $\psi_p^\dagger \psi_n$ and $\psi_p^\dagger \sigma^k \psi_n$ as  the leading Lagrangian in \cref{eq:TH_NRleeyang0}. 
On the other hand, the last line contains bilinears where the derivative acts on the nucleon field. 
They appear in the subleading Lagrangian for the first time, and lead to three new nuclear matrix elements entering the decay amplitude at the linear level in recoil.

From the matching in \cref{eq:TH_matching1} it follows that the subleading Wilson coefficients $C_E^+$, $C_{E'}^+$, and $C_{T3}^+$ vanish in the isospin limit. Since the beta decay kinematics relates the maximum value of the 3-momentum transfer $\boldsymbol{q}$ to the isospin-breaking nuclear mass difference  
$\Delta \equiv m_{\cal N}- m_{\cal N'}$, 
in principle one could consider the corresponding terms as $\cO(\boldsymbol{q}^2/m_N^2)$ and relegate them to the   $\cO(\boldsymbol{\nabla}^2/m_N^2)$ Lagrangian. 
In this paper we keep them in \cref{eq:TH_NRleeyang1} for completeness, however, in  practice, recoil and isospin suppression together results in $\cO(10^{-5})$ effects at most, which are unobservable in current experiments.

\section{Recoil corrections to beta decay observables } 
\label{sec:recoil}

Starting from the EFT Lagrangian in \cref{eq:TH_NRleeyang} one can calculate the associated amplitude and differential distributions for  the nuclear beta transitions ${\cal N} \to {\cal N'} e \nu$.
Here, ${\cal N} ({\cal N'})$ denotes the parent (daughter) nucleus, $e$ denotes the beta particle ($e^\mp$ for $\beta^\mp$ transitions),  and $\nu$ denotes  the antineutrino (for $\beta^-$) or the neutrino (for $\beta^+$). 
In this paper we focus on the allowed beta transitions where ${\cal N}$ and ${\cal N'}$ have the same  spin,  $J' = J$;  the discussion of the allowed decays with   $J' = J \pm 1$ is analogous.
We first introduce the general structure of the amplitude including the recoil effects, and then move to discussing differential distributions. 

\subsection{Structure of the decay amplitude}

For concreteness, we present the formulas for the $\beta^-$ decay amplitude,  ${\cal M_{\beta^-}} \equiv {\cal M}( {\cal N} \to {\cal N}'e^- \bar \nu)$. 
The amplitude can be expanded in powers of recoil momenta: 
${\cal M_{\beta^-}} \equiv {\cal M}_{\beta^-}^{(0)} +{\cal M}_{\beta^-}^{(1)} + \dots$.
In this expansion, all 3-momenta involved 
($\boldsymbol{p}_{\cal N}$, $\boldsymbol{k}_{\cal N'}$, $\boldsymbol{k}_{e}$, $\boldsymbol{k}_{\nu}$), the lepton energies and electron mass ($E_e$, $m_e$, $E_\nu \equiv E_e^{\rm max} - E_e$)  as well as the nuclear mass difference $\Delta$ count as one order in recoil, and we denote these collectively as ${\cal O}(\boldsymbol{q}/m_N)$.

Only the leading order Lagrangian in \cref{eq:TH_NRleeyang0} contributes to the leading part of the amplitude  ${\cal M}_{\beta^-}^{(0)}$. 
One finds 
\beq
\label{eq:BETA_M1}
\cM_{\beta^-}^{(0)} =   -  \langle \psi_p^\dagger \psi_n \rangle  \big [  C_V^+ L^0 + C_S^+ L \big ]  
+  \langle \psi_p^\dagger \sigma^k \psi_n \rangle    \big [  C_A^+ L^k + C_T^+ L^{0k}  \big ] .  
\eeq 
Above, $C_X^+$ are the Wilson coefficients in the  Lagrangian of \cref{eq:TH_NRleeyang0}.
The leptonic currents are defined as 
$L^\mu \equiv \bar u_L(k_e) \gamma^\mu v_L(k_\nu)$,
$L \equiv \bar u_R(k_e) v_L(k_\nu)$, 
$L^{0k} \equiv \bar u_R(k_e) \gamma^0 \gamma^k v_L(k_\nu)$,
where $k=1\dots 3$,  and $u$ and $v$ are the spinor wave functions of the outgoing electron and antineutrino (the $L/R$ subindex denote the chirality projection). 
Finally, $\langle \cdot \rangle \equiv \bra{{\cal N'} (\boldsymbol{k}_{\cal N'})} \cdot \ket{{\cal N}  (\boldsymbol{p}_{\cal N}) } $ denotes matrix elements of the non-relativistic nucleon fields sandwiched between the daughter and parent nuclear states in relativistic normalization $\langle {\cal N}(\boldsymbol{p'})  \ket{{\cal N}(\boldsymbol{p})} = 2 E_{\cal N} (2 \pi)^3 \delta^3(\boldsymbol{p} - \boldsymbol{p'})$.
The matrix elements entering into \cref{eq:BETA_M1} are labeled as Fermi and GT, respectively.
We will write them down in the limit of unbroken isospin symmetry.\footnote{%
Isospin breaking should be included separately, whenever it is phenomenologically relevant.
Given the current precision of the beta decay experiments, isospin breaking effects must be taken into account in the case of the Fermi matrix element. 
In our analysis, these will be included in the $\delta_R$ correction in \cref{eq:BETA_dGammaTemplate}.
For the GT one, since the matrix element is proportional to the unknown parameter $r$ which has to be anyway fixed from experiment, the isospin breaking corrections do not play an important role. 
}
As we discuss in great detail in \cref{app:SHF}, rotational and Galilean symmetry, parity, time-reversal, and isospin invariance constrain the Fermi and GT matrix elements to take the most general form:
\begin{align}
\label{eq:BETA1_FGTmatrixelements} 
\langle \psi_p^\dagger  \psi_n  \rangle  = & \kappa \,  M_F  \delta_{J_z' J_z}  
 + \cO(\boldsymbol{q}^2/m_N^2) , 
\nnl 
\langle \psi_p^\dagger \sigma^k \psi_n  \rangle  = &    r \,  \kappa  \, M_F {[{\cal T}_{(J)}^{k}]_{J_z'}^{\, J_z} \over \sqrt{J(J+1)} }
+  \cO(\boldsymbol{q}^2/m_N^2) ,   
\end{align}
where $\kappa \equiv 2 \sqrt{E_{\cal N} E_{\cal N'}}$, and 
${\cal T}_{(J)}^k$ are the spin-J generators of the rotation group defined in~\cref{eq:SHF_Tk}.
For $J = 0$ one should take the limit $r\to 0$ while ignoring all apparent $1/J$ singularities.  
Terms linear in $\boldsymbol{p}$ or $\boldsymbol{p'}$ cannot appear in \cref{eq:BETA1_FGTmatrixelements} due to parity conservation. 
The common normalization factor $M_F$ can be calculated when parent and daughter nuclei are members of the same isospin multiplet. 
For $\beta^\mp$ transitions one has 
$M_F =   \delta_{j'j} \delta_{j_3',j_3\pm 1}  \sqrt{ j (j+1) - j_3( j_3 \pm 1)}$, 
where $(j,j_3)$ and $(j',j_3')$ are the isospin quantum numbers of the parent and daughter nuclei. 
The parameter $r$, which is real by time-reversal invariance,  is referred to as the ratio of GT and Fermi matrix elements in the literature. For the neutron decay $r=\sqrt{3}$.
For decays with $A>1$, it cannot currently be calculated from first principles with high accuracy, and instead has to be extracted from experiment or estimated in nuclear models.
In this notation the so-called mixing ratio is given by $\rho = r \,C_A^+/C_V^+$, up to radiative corrections.  
Finally, $\cO(\boldsymbol{q^2}/m_N^2)$ refers to corrections of the second order in recoil, which are consistently neglected in this paper. 

The subleading Lagrangian in \cref{eq:TH_NRleeyang1} contributes at the next-to-leading order in the recoil expansion.  
The amplitude at this order takes the form
\begin{align}
\label{eq:BETA_M2}
\cM_{\beta^-}^{(1)} = &  {1 \over 2 m_N} \bigg \{ 
\langle \psi_p^\dagger  \psi_n \rangle \bigg [
- C^+_{T1} q^k L^{0k} 
- C^+_{T2} (E_\nu - E_e) L  
\bigg ] 
\nnl + &  
\langle \psi_p^\dagger \sigma^k \psi_n \rangle \bigg [ 
C_P^+ q^k L  
- C^+_E  q^k L^0 
+ C^+_{E'}  (E_\nu + E_e) L^k 
-  i C^+_{M}  \epsilon^{ijk} q^i  L^j
+  2 C^+_{T3} (k_\nu^k -k_e^k) L  \bigg ] 
\nnl -&   
 i  \langle \psi_p^\dagger  \overleftrightarrow{\nabla}_k \psi_n \rangle   
C^+_{FV}  L^k 
+   i   \langle \psi_p^\dagger \sigma^k  \overleftrightarrow{\nabla}_k \psi_n \rangle C^+_{FA} L^0 
+ \epsilon^{ijk} \langle  \psi_p^\dagger \sigma^i  \overleftrightarrow{\nabla}_j \psi_n \rangle  C^+_{FT}  L^{0k}
\bigg \}. 
\end{align}
The matrix elements in the first two lines are the Fermi and GT ones, 
already discussed around \cref{eq:BETA1_FGTmatrixelements}. 
In the last line,  three new nuclear matrix elements enter at the next-to-leading order in recoil. 
Lorentz symmetry, parity, time-reversal, and isospin invariance constrain their form as
\begin{align}
\label{eq:BETA1_dFGTmatrixelements} 
i \langle \psi_p^\dagger  \overleftrightarrow{\nabla}_k   \psi_n  \rangle  = & \kappa \, M_F
\bigg \{ - {1 \over A} P^k  \delta_{J_z' J_z}  
- i  \beta_{FV} {r \over \sqrt{J(J+1)} }   \epsilon^{klm} q^l  [{\cal T}_{(J)}^m]_{J_z'}^{\, J_z}  
 \bigg \} 
 + \cO(\boldsymbol{q^2}/m_N^2) , 
\nnl 
i \langle \psi_p^\dagger \sigma^k  \overleftrightarrow{\nabla}_k  \psi_n  \rangle  = &   
r \, \kappa \, M_F \bigg \{ 
 -  {1 \over A \sqrt{J(J+1)} }  P^k  [{\cal T}_{(J)}^k]_{J_z'}^{\, J_z}   
\bigg \} 
+  \cO(\boldsymbol{q^2}/m_N^2) ,
\nnl 
\epsilon^{ijk} \langle \psi_p^\dagger \sigma^{i}  \overleftrightarrow{\nabla}_j \psi_n \rangle  = & 
 \kappa \, M_F 
\bigg \{ 
 i  {r   \over A \sqrt{J(J+1)}}  \epsilon^{klm}  P^l  [{\cal T}_{(J)}^m]_{J_z'}^{\, J_z}  
+    \alpha_{FT} q^k  \delta_{J_z'}^{\, J_z}  
  +  \gamma_{FT} { r \over J \sqrt{J(J+1)}}  q^l [{\cal T}_{(J)}^{kl} ]_{J_z'}^{\, J_z} 
 \bigg \} 
\nnl & + \cO(\boldsymbol{q^2}/m_N^2) , 
\end{align}
where $\boldsymbol{P} \equiv \boldsymbol{p}_{\cal N} +\boldsymbol{k}_{\cal N'}$, 
$A = {m_{\cal N} \over m_N}$ is approximately the mass number of the parent nucleus, 
and the matrix ${\cal T}_{(J)}^{jk}$ is defined in \cref{eq:SHF_Tkl}. 
For the subleading matrix elements, the isospin breaking corrections are not phenomenologically relevant, given the current experimental sensitivity.
The coefficients of the terms proportional to $\boldsymbol{P}$ are related by Lorentz invariance to those in \cref{eq:BETA1_FGTmatrixelements}~\cite{Holstein:1974zf}, as we also derive in \cref{app:SHF}. 
On the other hand, the form factors $\beta_{FV}$, $\alpha_{FT}$,  and $\gamma_{FT}$ are transition-dependent and are determined by strong dynamics.
Time-reversal invariance implies they are real, while isospin symmetry (CVC)  relates $\beta_{FV}$ to the magnetic moments of the parent and daughter nuclei~\cite{Holstein:1974zf}: 
\begin{equation}
\label{eq:BETA1_CVCrelation} 
{\mu_+ - \mu_- \over \mu_N}   =    \big (g_M + \beta_{FV} \big )   r  \sqrt{J \over J+1 },   
\end{equation}
where $\mu_+$ and $\mu_-$ denote the magnetic moments of the nuclei with larger and smaller $j_3$ quantum number, respectively. 
The form factors $\alpha_{FT}$, $\beta_{FV}$ and $\gamma_{FT}$ vanish for neutron decay. 
For nuclear transitions $\alpha_{FT}$ and $\gamma_{FT}$ can be non-zero, and they are not fixed by  isospin symmetry unlike $\beta_{FV}$. 
In order to analyze the effects of tensor interactions  at the recoil level, $\alpha_{FT}$ and $\gamma_{FT}$ have to be estimated for each transition using lattice or nuclear models.

Given the leading and subleading amplitudes in \cref{eq:BETA_M1} and \cref{eq:BETA_M2} it is straightforward if tedious to calculate the differential decay width.
In the next subsection we discuss the general parametrization of the differential width appropriate to incorporate the subleading corrections in the recoil expansion.

\subsection{Parametrization of the differential width}
\label{sec:template}

We focus on observables summed over the daughter and $\beta$ particle polarizations. 
We are interested in the differential decay width at the leading and subleading order in recoil momenta. 
In this observable, the effects of the Wilson coefficients in the EFT Lagrangian fit into the following template:\footnote{%
The contributions proportional to $\gamma_{FT}$ entering via the tensor matrix element in \cref{eq:BETA1_dFGTmatrixelements} generate additional correlations not included in the  template of \cref{eq:BETA_dGammaTemplate}. 
These are treated separately later, cf. \cref{eq:BETA_FTdeltaGammaTemplate}. 
} 
\begin{align}
\label{eq:BETA_dGammaTemplate}
{ d \Gamma \over  d E_e d \Omega_e   d \Omega_\nu  } = & 
 M_F^2 F(Z,E_e) (1 + \delta_R) {p_e E_e (E_e^{\rm max} - E_e)^2   \over 64 \pi^5  } 
 \hat \xi   \bigg \{  1  + b {m_e \over E_e} +  \xi_b(E_e)   
 \nnl +& 
 a(E_e) { \boldsymbol{k}_e \cdot \boldsymbol{k}_\nu \over E_e E_\nu} 
 + a'(E_e) \bigg[\bigg ( {\boldsymbol{k}_e \cdot \boldsymbol{k}_\nu \over E_e E_\nu}\bigg )^2 -  {p_e^2 \over 3 E_e^2} \bigg ]  
\nnl +& 
{\boldsymbol{J}  \cdot \boldsymbol{k}_e \over J E_e }  \bigg [ A(E_e) 
+ A'(E_e)  { \boldsymbol{k}_e \cdot \boldsymbol{k}_\nu   \over E_e E_\nu } 
- {B'(E_e) \over 3} \bigg ] 
+{\boldsymbol{J}  \cdot \boldsymbol{k}_\nu \over J E_\nu }  \bigg [ B(E_e) +
B'(E_e) {\boldsymbol{k}_e \cdot \boldsymbol{k}_\nu  \over E_e E_\nu } 
- {A'(E_e) \over 3} {p_e^2\over E_e^2} \bigg ]
 \nnl  +&  {\boldsymbol{J}  \cdot  (\boldsymbol{k}_e \times \boldsymbol{k}_\nu) \over J E_e E_\nu} \bigg [  D(E_e) 
 + D'(E_e) {\boldsymbol{k}_e \cdot\boldsymbol{k}_\nu \over E_e E_\nu } \bigg ]
\nnl +&  {J(J+1) - 3 (\boldsymbol{J}  \cdot \boldsymbol{j})^2 \over J (J+1)}
\bigg [
\left( \hat c(E_e) \, +\, c'(E_e) \, \frac{\boldsymbol{k}_e \cdot \boldsymbol{k}_\nu}{E_e E_\nu}\right) {(\boldsymbol{k}_e \cdot \boldsymbol{k}_\nu) -    3 (\boldsymbol{k}_e \cdot \boldsymbol{j}) (\boldsymbol{k}_\nu \cdot \boldsymbol{j}) \over 3 E_e E_\nu }
\nnl +&  
c_1(E_e) {\boldsymbol{k}_\nu^2 - 3 (\boldsymbol{k}_\nu \cdot  \boldsymbol{j})^2 \over E_\nu^2 }
+ c_2(E_e)  {\boldsymbol{k}_e^2 - 3 (\boldsymbol{k}_e \cdot  \boldsymbol{j})^2 \over E_e^2 }
\nnl +&   c_3(E_e) {( (\boldsymbol{k}_e \times \boldsymbol{k}_\nu)\cdot \boldsymbol{j}) (\boldsymbol{k}_\nu \cdot  \boldsymbol{j}) \over E_e E_\nu^2 }
+ c_4(E_e) {( (\boldsymbol{k}_e \times \boldsymbol{k}_\nu)\cdot \boldsymbol{j}) (\boldsymbol{k}_e\cdot  \boldsymbol{j}) \over E_e E_\nu^2 }
 \bigg ] 
\bigg \} , 
\end{align}
where $F$ is the Fermi function:
\begin{equation}
\label{eq:BETA_FermiFunction}
F(Z,E_e) \equiv 2(1+\gamma) (2 p_e R)^{-2 (1-\gamma)} {e^{\pi \eta} \Gamma (\gamma + i \eta)  \Gamma (\gamma - i \eta)  \over \Gamma(1+ 2 \gamma)^2 }, 
\quad \gamma \equiv \sqrt {1- \alpha^2 Z^2}, \quad \eta \equiv \pm {\alpha Z E_e \over p_e},  
\end{equation}
$R$ is the nuclear radius,
$\alpha$ is the fine structure constant, 
$\delta_R$ stands for radiative corrections\footnote{%
Customarily, one splits 
$1+\delta_R = (1+\delta_R')(1+\delta_{NS}^V - \delta_C^V)$, 
where $\delta_R'$ is the long-distance (outer) radiative correction, 
$\delta_{NS}$ is the nuclear-structure dependent correction, 
and $\delta_C$ is the isospin breaking correction.}, 
$Z$ is the daughter nucleus charge, 
$p_e \equiv \sqrt{E_e^2 - m_e^2}$,
$m_e$ is the electron mass, 
$E_e$ is the electron energy in the range $E_e \in [m_e ,E_e^{\rm max}]$, 
the endpoint electron energy is  $E_e^{\rm max} = \Delta + {m_e^2 - \Delta^2 \over 2 m_{\cal N}}$ with $\Delta= m_{\cal N} -m_{\cal N'}$. 
In \cref{eq:BETA_FermiFunction} and hereinafter, the upper (lower) sign applies to $\beta^-$($\beta^+$) transitions.
Finally, $\boldsymbol{k}_e$ and $\boldsymbol{k}_\nu$ denote the 3-momenta of the beta particle and neutrino,  $\boldsymbol{J}$ is the polarization vector of the parent nucleus, 
and $ \boldsymbol{j}$ is the unit vector in the polarization direction (if $J=0$ then  all terms proportional to $1/J$ should be set to zero). 
 
The bullets below offer some comments and rationale regarding our parametrization. 
\begin{itemize}
\item 
Only the first line in \cref{eq:BETA_dGammaTemplate} contributes to the total beta decay width (or lifetime/half-life) of the nucleus. 
The overall normalization $\hat \xi$ and the Fierz term $b$ are related to the parameters of the leading order Lagrangian in \cref{eq:TH_NRleeyang0} as~\cite{Jackson:1957zz}
\begin{align}
\hat \xi =  &   |C_V^+|^2   + |C_S^+|^2  +  r^2 \bigg [|C_A^+|^2  + |C_T^+|^2\bigg ] , 
\nnl  
b \hat \xi  =  & 
 \pm 2\,  \re  \bigg [  C_V^+ \bar C_S^+ +    r^2  C_A^+ \bar C_T^+ \bigg ] , 
\end{align}
and by definition do not receive any recoil-order corrections. 
The latter are encoded in $\xi_b(E_e)$.
At the linear order in recoil $\xi_b(E_e) =  b_{-1} {m_e \over E_e}   + b_0  + b_1 {E_e \over m_e}$, with $b_{-1,0,1}$ independent of $E_e$. 
Above and hereafter we use a bar to denote complex conjugation. 
The relation with the traditional notation of Ref.~\cite{Jackson:1957zz} is simply $\xi = M_F^2 \,\hat{\xi}/2$.
\item For unpolarized decays, the differential distribution should be averaged over the possible quantized values of $\boldsymbol{J}$ in the range  $\boldsymbol{J} \in [-J,J]$.
Only the first two lines of \cref{eq:BETA_dGammaTemplate} survive the averaging. 
In addition to the parameters affecting the total width, the relevant parameters for unpolarized decays include the $\beta$-$\nu$ correlation $a(E_e)$, and $a'(E_e)$ parametrizing recoil effects quadratic in $\cos(\boldsymbol{k}_\beta, \boldsymbol{k}_\nu)$.
The former can be represented as $a(E_e) = a_0 + \Delta a(E_e)$, 
where $a_0$ is generated by the leading order Lagrangian in \cref{eq:TH_NRleeyang0} and is independent of $E_e$~\cite{Jackson:1957zz}: 
\begin{equation}
a_0 \hat \xi = |C_V^+|^2  - |C_S^+|^2  - {r^2 \over 3 } \bigg  [  |C_A^+|^2 - |C_T^+|^2 \bigg ] ,
\end{equation}
while $\Delta a(E_e)$ arises at the recoil level and in general does depend on the energy of the  beta particle. 
As for $a'(E_e)$, at the linear order in recoil we have $a'(E_e) = b_a {E_e \over m_e}$ with $b_a$ independent of $E_e$. 
\item 
Effects linear in the polarization vector $\boldsymbol{J}$ are described by the  third and fourth lines of \cref{eq:BETA_dGammaTemplate}. 
Only $A(E_e)$, $B(E_e)$, $D(E_e)$ arise at the leading order in recoil. 
We again split 
$A(E_e) = A_0 + \Delta A(E_e)$, $B(E_e)=  B_0 +b_B \frac{m_e}{E_e}+ \Delta B(E_e)$, $D(E_e) =  D_0 + \Delta D(E_e)$ where $A_0$, $B_0$, $b_B$, $D_0$ generated by the leading order Lagrangian in \cref{eq:TH_NRleeyang0} and are independent of $E_e$~\cite{Jackson:1957zz}: 
\begin{align}
A_0 \hat \xi = & 
 - 2 { r \sqrt{J} \over \sqrt{J+1}}   \re \bigg  [   C_V^+ \bar C_A^+   - C_S^+ \bar C_T^+   \bigg ] 
\mp  {r^2 \over  J+1 }  \bigg [   |C_A^+|^2 -  |C_T^+|^2     \bigg ] , 
\nnl 
B_0  \hat \xi = & 
-2 { r \sqrt{J} \over \sqrt{J+1}}  \re  \bigg [ 
 C_V^+ \bar C_A^+     +  C_S^+ \bar C_T^+   \bigg ] 
 \pm {r^2\over  J+1 }  \bigg [  |C_A^+|^2    +  |C_T^+|^2  \bigg ] ,
\nnl 
b_B \hat \xi = & \mp 2 {r\sqrt{J} \over \sqrt{J+1}}  
 \re  \bigg [  C_V^+ \bar C_T^+   + C_S^+ \bar C_A^+     \bigg ]
+  2 {r^2\over  J+1 } \re \big [ C_A^+ \bar C_T^+ \big ] , 
\nnl 
D_0  \hat \xi = & 
 - 2 { r \sqrt{J} \over \sqrt{J+1}}   
 \im \bigg [ C_V^+ \bar C_A^+ - C_S^+ \bar C_T^+ \bigg ]  , 
\end{align}
while  $\Delta A(E_e)$,  $\Delta B(E_e)$,  $\Delta D(E_e)$, and also  
$A'(E_e)$,  $B'(E_e)$,  $D'(E_e)$ arise only at the recoil level. 
Note that the latter group of coefficients also depends on the beta particle energy in general,  but we do not stress this fact in our notation. 
\item 
Out of the terms linear in $\boldsymbol{J}$,  only $A(E_e)$ survives when the distribution is integrated over the neutrino direction $d \Omega_\nu$.  
Thus, the remaining coefficients will not affect experiments measuring the $\beta$-asymmetry,  where the neutrino momentum is not reconstructed. 
Similarly,  only $B(E_e)$ survives when these terms are integrated over the electron direction $d \Omega_e$, 
and the remaining  coefficients will not affect experiments measuring the $\beta$-asymmetry once the information about $e^\pm$ kinematics is integrated out.
These features make our parametrization  more convenient in practice than e.g. the (equivalent) one used in Refs.~\cite{Holstein:1974zf,Ando:2004rk,Bhattacharya:2011qm}. 
\item 
The last three lines in \cref{eq:BETA_dGammaTemplate} describe effects relevant only for polarized decays  with $J \geq 1$ (thus in particular  they are absent in neutron decay).   
We have $\hat c(E_e) = \hat c_0 + \Delta c(E_e)$  where $\hat c_0$ generated by the leading order Lagrangian~\cite{Jackson:1957zz}:\footnote{%
Compared to Ref.~\cite{Jackson:1957zz}, we have rescaled the $c$ coefficient by $(2J-1)/(J+1)$ and renamed $c \to \hat c$. 
}
\begin{equation}
\hat c_0 \hat \xi= r^2 \bigg [ |C_A^+|^2 -  |C_T^+|^2 \bigg ], 
\end{equation}
while  $\Delta c(E_e)$, $c'(E_e)$, and $c_{1,2,3,4}$ arise at the recoil level.
To our knowledge these correlations have never been experimentally measured, even the leading order $\hat c_0$, and thus their phenomenological role is currently null.  
We nevertheless quote them here for completeness. 
\item The coefficients $a'(E_e)$, $c'(E_e)$ and $D'(E_e)$ are actually not generated by the interactions in \cref{eq:TH_NRleeyang1}. They are however generated via kinematic recoil corrections to the 3-body decay phase space, which we treat on the same footing.
\item The leading electromagnetic corrections are included in \cref{eq:BETA_dGammaTemplate} via the Fermi function $F(Z,E_e)$ and via $\delta_R$. 
The correlation coefficients receive other $\cO(\alpha)$ corrections (``outer radiative corrections"), see e.g.~\cite{Ando:2004rk}.  In this paper we do not discuss these explicitly, but they are taken into account in the experimental analyses that we use as input of our phenomenological analysis in the next section,  whenever they are required for the theory predictions to match the experimental accuracy. 
\end{itemize}

The complete dependence of all the correlation coefficients in \cref{eq:BETA_dGammaTemplate} on the Wilson coefficients in the subleading EFT Lagrangian in \cref{eq:TH_NRleeyang1} is summarized in \cref{app:sub}.
Compared to earlier works~\cite{Holstein:1974zf,Wilkinson:1982hu,Ando:2004rk,Bhattacharya:2011qm} our results include complete BSM effects arising at the subleading order in the EFT, that is at the linear order in recoil.
Let us note that our results include terms that are quadratic in BSM couplings. 
These results will be employed in the next section to perform global fits constraining the EFT Wilson coefficients. 
We will focus on the effects of the pseudoscalar interactions ($C_P^+$), nucleon-level weak magnetism ($C_M^+$), and induced tensor interactions ($C^+_E$), 
but our results allow one to constrain any pattern of the Wilson coefficient entering the leading and subleading EFT Lagrangian in \cref{eq:TH_NRleeyang1}.

\subsection{Observables}

Before embarking on that analysis, let us first discuss how common experimental  observables are related to the correlation coefficients in \cref{eq:BETA_dGammaTemplate}. 
The total width is given by the expression\footnote{%
In the formulas presented here we omit the Coulomb corrections to non-standard terms, see Ref.~\cite{Jackson:1957auh}.}
\beq
\Gamma =  {M_F^2 (1+\delta_R) m_e^5 f  \over 4 \pi^3 }  \hat \xi \left( 1 + b \bigg \langle {m_e \over E_e} \bigg \rangle + \langle \xi_b(E_e) \rangle\right)~,
\eeq 
where
\begin{equation}
   f  \equiv \frac{1}{m_e^5} \int_{m_e}^{E_e^{\rm max}}  dE_e p_e E_e (E_e^{\rm max} - E_e)^2 F~,   
\end{equation}
and
\begin{equation}
\label{eq:BETA_Average}
\langle X \rangle  \equiv 
 {\int_{m_e}^{E_e^{\rm max}}  dE_e p_e E_e (E_e^{\rm max} - E_e)^2 F~X 
   \over \int_{m_e}^{E_e^{\rm max}}  dE_e p_e E_e (E_e^{\rm max} - E_e)^2 F } . 
\end{equation}
Concerning the $\beta$-asymmetry, 
experiments typically measure the number of events $N_{\uparrow}$ and $N_{\downarrow}$  with the beta particle emitted, respectively,  into the northern and southern hemisphere in reference to the parent polarization direction $\boldsymbol{j}$.  
The corresponding up-down asymmetry is given by 
\begin{equation}
{N_{\uparrow} - N_{\downarrow} \over N_{\uparrow} + N_{\downarrow} }  = 
P {\int_{m_e}^{E_e^{\rm max}}  d E_e p_e^2  (E_e^{\rm max} - E_e)^2  A(E_e) F
\over 
 2 \int_{m_e}^{E_e^{\rm max}}  dE_e p_e E_e (E_e^{\rm max} - E_e)^2  
 \big [  1 + b {m_e \over E_e}  +  \xi_b(E_e)   \big ]F },    
\end{equation} 
where $P$ the parent polarization fraction. 
We substitute $A(E_e) = A_0 + \Delta A(E_e)$ and expand the above to linear order in recoil: 
\begin{align} & 
{N_{\uparrow} - N_{\downarrow} \over N_{\uparrow} + N_{\downarrow} }   =  P 
{\int_{m_e}^{E_e^{\rm max}}  d E_e p_e^2  (E_e^{\rm max} - E_e)^2   F  
\over 
2 \int_{m_e}^{E_e^{\rm max}}  dE_e p_e E_e (E_e^{\rm max} - E_e)^2  F  \big [ 1 + b {m_e \over E_e} \big ]   } 
\bigg \{ 
\nnl  & 
A_0  + 
 {\int_{m_e}^{E_e^{\rm max}}  d E_e p_e^2  (E_e^{\rm max} - E_e)^2   F   \Delta A(E_e) \over 
\int_{m_e}^{E_e^{\rm max}}  d E_e p_e^2  (E_e^{\rm max} - E_e)^2  F   } 
 - A_0  {
\int_{m_e}^{E_e^{\rm max}}  dE_e p_e E_e (E_e^{\rm max} - E_e)^2  F  \xi_b(E_e)     \over 
\int_{m_e}^{E_e^{\rm max}}  dE_e p_e E_e (E_e^{\rm max} - E_e)^2 F  \big [ 1 + b {m_e \over E_e} \big ]   } 
\bigg \}.  
\end{align} 
Experimental collaborations often translate the observable asymmetry into a measurement of $A_0$, assuming vanishing Fierz term and the recoil corrections calculated in the absence of new physics beyond the SM. 
BSM physics can contribute as an $E_e$-independent shift of $A_0$ and $b$, or as an  $E_e$-dependent shift of  $\Delta A(E_e)$ and $\xi_b(E_e)$.  
We split the SM and BSM contributions at the recoil level as  
$\Delta A(E_e) = \Delta A^{\rm SM} +\Delta A^{\rm BSM}$,
$\xi_i = \xi_i^{\rm SM} + \xi_i^{\rm BSM}$.  
With this notation, we reinterpret
the experimental extraction of $A_0$ as a measurement of $\tilde A$, defined as\footnote{
We neglect the small SM contribution to $b$, which enters at the linear level in recoil and is in addition  suppressed by isospin breaking.} 
\begin{equation}
\label{eq:BETA_tildeArecoil}
\tilde A   =   {1 \over 1 + b \big \langle  {m_e \over E_e} \big \rangle  }
\bigg \{ 
A_0  +   \langle  \Delta A^{\rm BSM} \rangle_p  \,-\, b \big \langle  {m_e \over E_e} \big \rangle\langle \Delta A^{\rm SM}\rangle_p 
- A_0 {  \langle \xi_b^{\rm BSM} \rangle \,-\, 2 b \big \langle  {m_e \over E_e} \big \rangle\langle \xi_b^{\rm SM}\rangle  - b^2 \big \langle  {m_e \over E_e} \big \rangle^2 \langle \xi_b^{\rm SM}\rangle \over 1 + b \big \langle  {m_e \over E_e} \big \rangle }  
\bigg \} , 
\end{equation} 
where 
\begin{equation}
\label{eq:BETA_AverageP}
   \langle X \rangle_p  \equiv 
  {\int_{m_e}^{E_e^{\rm max}}  d E_e p_e^2  (E_e^{\rm max} - E_e)^2 F~X \over 
\int_{m_e}^{E_e^{\rm max}}  d E_e p_e^2  (E_e^{\rm max} - E_e)^2 F } .        
\end{equation}
At the leading order in recoil this reduces to the usual tilde prescription: $\tilde A = {A_0 \over 1 + b \langle m_e /E_e \rangle}$~\cite{Gonzalez-Alonso:2016jzm,Falkowski:2020pma}.  Eq.~\eqref{eq:BETA_tildeArecoil} generalizes this prescription to the subleading order in recoil.   

By the same logic one can define the tilde prescription for the neutrino asymmetry and for the $\beta$-$\nu$ asymmetry:
\begin{align}
\label{eq:BETA_tildeBrecoil}
\tilde B =  &   {1 \over 1 + b \big \langle  {m_e \over E_e} \big \rangle  }
\bigg \{ 
B_0  +   b_B   \bigg \langle  {m_e \over E_e} \bigg  \rangle 
+  \langle  \Delta B^{\rm BSM} \rangle
- b \big \langle  {m_e \over E_e} \big \rangle\langle \Delta B^{\rm SM}\rangle 
\nnl & 
- B_0 {  \langle \xi_b^{\rm BSM} \rangle \,-\, 2 b \big \langle  {m_e \over E_e} \big \rangle\langle \xi_b^{\rm SM}\rangle  - b^2 \big \langle  {m_e \over E_e} \big \rangle^2 \langle \xi_b^{\rm SM}\rangle \over 1 + b \big \langle  {m_e \over E_e} \big \rangle }  
-  b_B   \big \langle  {m_e \over E_e} \big  \rangle  { \langle \xi_b \rangle \over 1 + b \big \langle  {m_e \over E_e} \big \rangle } 
\bigg \} , 
\nnl 
\tilde a =  &   
{1 \over 1 + b \big \langle  {m_e \over E_e} \big \rangle  }
\bigg \{ 
a_0  +   \langle  \Delta a^{\rm BSM} \rangle_p 
 - b  \bigg \langle  {m_e \over E_e} \bigg  \rangle  \langle  \Delta a^{\rm SM} \rangle_p
- a_0 {  \langle \xi_b^{\rm BSM} \rangle \,-\, 2 b \big \langle  {m_e \over E_e} \big \rangle\langle \xi_b^{\rm SM}\rangle  - b^2 \big \langle  {m_e \over E_e} \big \rangle^2 \langle \xi_b^{\rm SM}\rangle \over 1 + b \big \langle  {m_e \over E_e} \big \rangle } 
\bigg \} . 
\end{align}
If the correlations are measured as a function of energy of the beta particle, we can likewise define $\tilde X(E_e)$ 
with the $dE_e$ integration restricted to a given energy bin.  
Note also that experimental conditions often imply that only a part of the phase space is effectively detected, and that should be taken into account when the $\langle \cdot \rangle$ averages are calculated~\cite{Gonzalez-Alonso:2016jzm}.

\section{Global fit to recoil effects} 
\label{sec:fits}

In our phenomenological analysis we use the experimental data summarized in~\cref{app:data}. The differences of this dataset with the one used in the recent analysis of leading-order effects~\cite{Falkowski:2020pma} are the following:
\begin{enumerate}
    \item An older measurement of the $\beta$-$\nu$ correlation of the neutron~\cite{Darius:2017arh} by the aCORN collaboration is superseded by the new result $\tilde a_n = -0.1078(18)$~\cite{Hassan:2020hrj}.
    \item The latest UCN$\tau$ measurement of the neutron lifetime~\cite{UCNt:2021pcg} leads to the improved combined result $\tau_n = 878.64(59)$, where we include both bottle and beam measurements and average the errors {\it \`a la} PDG with the scale factor $S=2.2$. 
    \item We update the lattice value of the axial coupling of the nucleon to the FLAG'21 average $g_A = 1.246(28)$~\cite{Aoki:2021kgd,Gupta:2018qil,Chang:2018uxx,Walker-Loud:2019cif}. 
    \item We do not use the pure GT transitions and beta polarization ratios $P_F/P_{GT}$, because the recoil corrections to these observables are not calculated in this paper.  Currently, these observables (calculated at leading-order) have a negligible impact on the global fit. 
\end{enumerate}
For the sake of these fits we assume all Wilson coefficients are real, since the sensitivity to the imaginary parts of  the considered observables is limited.\footnote{%
Constraints on the imaginary parts can be improved by including in the analysis experimental measurements of CP-violating beta decay observables, such as the $D$~\cite{Chupp:2012ta} or $R$~\cite{Kozela:2011mc} parameter of the neutron~\cite{Gonzalez-Alonso:2018omy}. This is left for future work.
In addition, in specific new physics scenarios the imaginary parts will be correlated with the strongly constrained  Wilson coefficients of CP-violating neutral current operators contributing to EDMs~\cite{Ng:2011ui,Vos:2015eba,Alioli:2017ces}.   }

\subsection{Pseudoscalar}

This subsection is focused on the pseudo-scalar interactions:
\begin{equation}
\cL^{(1)}  \supset    C_P^+   {i \over 2 m_N}
(\psi_p^\dagger \sigma^k\psi_n) \nabla_k \big (\bar e_R \nu_L \big ) + \hc  
\end{equation}
entering into the subleading EFT Lagrangian of \cref{eq:TH_NRleeyang1}. 
The complete effect of this interaction term on the correlation coefficients of \cref{eq:BETA_dGammaTemplate} is given in \cref{sec:subP}. 

\begin{table}[tb]
    \centering
    \begin{tabular}{|c|c|c|c|c|c|c|c|}
    \hline
 & $n$ & ${}^{17}$F & ${}^{19}$Ne & ${}^{21}$Na & ${}^{29}$P & ${}^{35}$Ar & ${}^{37}$K \\    \hline    
$\langle\xi_b(E_e)\rangle$ & -0.9 & -0.9&  -1.1 & -0.5 & -0.4 & -0.1 & -0.5  
\\   \hline
$\Delta  a(E_e)$ & -1.2 & -0.9 & -1.0 &  -0.5 & -0.3 & -0.1 & -0.4 
\\   \hline
$\Delta  A(E_e)$ & -0.9 & -1.8 &  1.1 &  -1.6 & -1.0 & -0.9 & 1.5  
\\    \hline
$ \langle \Delta B(E_e) \rangle$ &  -0.7 & 1.7 &  1.2 & -1.8 & -1.3 & -1.1 &  1.9  
\\    \hline
    \end{tabular}
\caption{%
Sensitivity of the correlation coefficients in neutron and mirror decays  to pseudoscalar interactions parametrized by the Wilson coefficient $C_P^+$. 
The entries are in units of $10^{-4}$ and correspond to the averaged function multiplying $C_P^+$ in \cref{eq:FIT_DeltaP} for different parent nuclei. 
For the sake of this table,  $C_V^+$, $C_A^+$, and $r$ are set to their minimum values in the SM fit (although they are kept as free parameters in the fits of this section).
For $\xi_b(E_e)$ and $\Delta B(E_e)$ the average is defined in \cref{eq:BETA_Average},
while in the case of $\Delta  a(E_e)$ and $\Delta  A(E_e)$ the pseudoscalar contribution is independent of the $\beta$ particle energy and thus no average is needed.  
\label{tab:pseudoscalar}
} 
\end{table}

It is instructive to first discuss a simpler setting where, with the exception of $C_P^+$,  only the leading order Wilson coefficients $C_V^+$ and $C_A^+$ are present.\footnote{We note that SM recoil corrections are included, since they have been taken into account in the experimental analyses that we use as inputs.} 
In particular, we assume $C_S^+ = C_T^+ = 0$ in the leading order Lagrangian in \cref{eq:TH_NRleeyang0}.  
In this limit the pseudoscalar contributions to the correlations simplify:
\begin{align}
\label{eq:FIT_DeltaP}
\xi_b(E_e)   = & {m_e \over 3  m_N}  \bigg [ {E_e^{\rm max}  \over E_e} -1 \bigg ]   
{r^2  C_A^+ C_P^+  \over (C_V^+)^2 + r^2 (C_A^+)^2 }      ,
\nnl
\Delta a(E_e)  = &   {m_e \over 3 m_N } 
{r^2 C_A^+ C_P^+ \over (C_V^+)^2 + r^2 (C_A^+)^2 }  , 
\nnl
\Delta A (E_e)    = &   - {m_e \over m_N }  \sqrt{J \over J+1}  
{ r C_V^+ C_P^+   \over (C_V^+)^2 + r^2 (C_A^+)^2 }   , 
\nnl 
\Delta B(E_e) =  &  -  {m_e \over m_N}    \bigg [ {E_e^{\rm max}  \over E_e} -1 \bigg ] \sqrt{J \over J+1} { r  C_V^+ C_P^+   \over (C_V^+)^2 + r^2 (C_A^+)^2 }   ~.
\end{align}
There are also contributions to $\Delta c(E_e)$ and $c_1(E_e)$, which are not relevant for our analysis. One can observe that, in each case, the pseudoscalar contribution is multiplied  by $m_e/m_N \sim 10^{-3}$, which is a typical suppression factor for recoil corrections in beta transitions. For this reason one expects much weaker constraints on $C_P^+$ compared to those on $C_V^+$ and $C_A^+$. 
Furthermore, the pseudoscalar corrections vanish in the limit $r\to 0$, that is they are zero for pure Fermi transitions.
In particular, they do not affect the superallowed $0^+ \to 0^+$ transitions,  currently the most accurate nuclear measurement, which further diminishes the experimental sensitivity to $C_P^+$. Let us note that the $\xi_b(E_e)$ expression given above agrees with the result of Ref.~\cite{Gonzalez-Alonso:2013ura} for the neutron decay case. Finally, let us stress that linear $C_P^+$ effects appear not only in the beta spectrum (as with scalar and tensor interactions) but also in the angular correlations.

\begin{figure}
    \centering
    \includegraphics[width=0.8\textwidth]{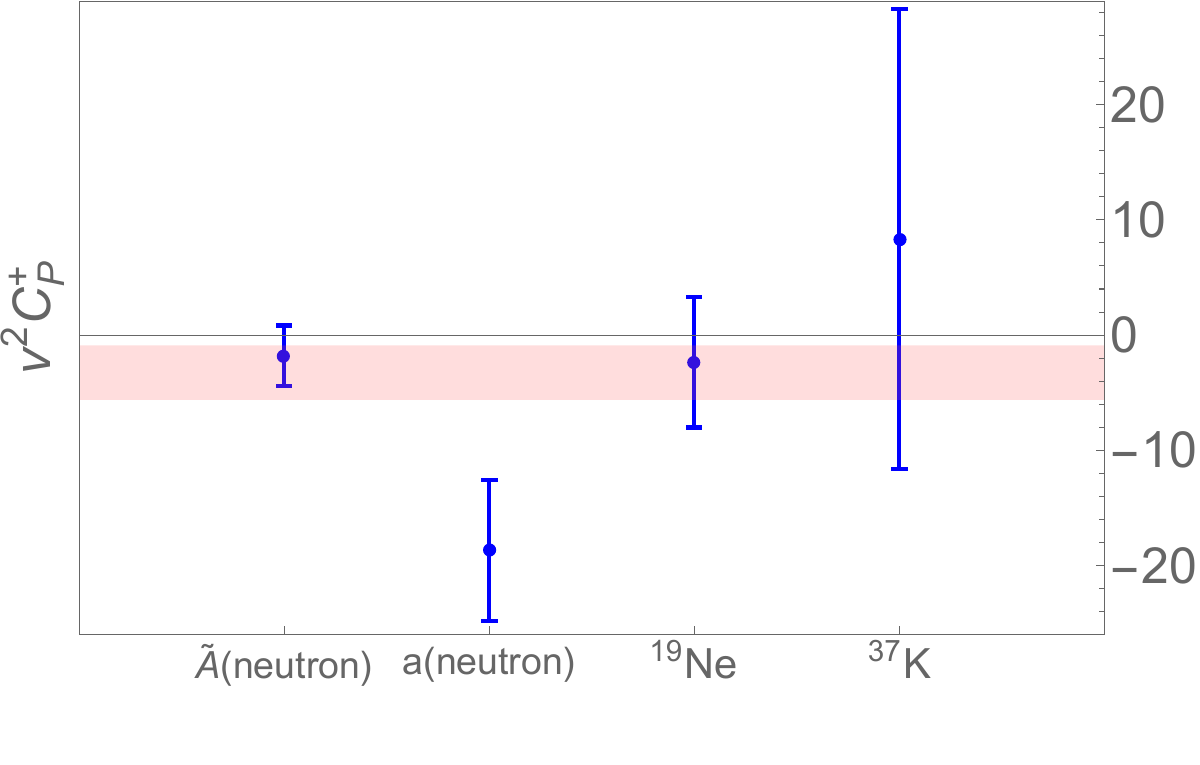}
    \caption{%
    Constraints on the pseudo-scalar coefficient $C_P^+$ from a subset of most sensitive observables. 
    We show the fit to $C_P^+$ marginalized over $C_V^+$ and $C_A^+$,  assuming other Wilson coefficients vanish. 
The error bars show the 68\% CL intervals obtained using 
${\cal F}t(0^+ \to 0^+)$,  neutron's lifetime, and one of the following: 
neutron's beta asymmetry, neutron's $\beta$-$\nu$ correlation, 
${\cal F} t$ and beta asymmetry of ${}^{19}$Ne, 
or ${\cal F} t$, beta asymmetry, and neutrino asymmetry of ${}^{37}$K. The pink band corresponds to the global fit in \cref{eq:FIT_CpseudoscalarSim}. 
    }
    \label{fig:CPp}
\end{figure}

Comparing \cref{eq:FIT_DeltaP} with the leading contributions to correlations, cf. \cref{sec:recoil}, one can see clearly that the pattern of the pseudoscalar contributions to the correlation is distinct than that of $C_V^+$ or $C_A^+$. 
Therefore,  a global fit with enough different observables can discriminate $C_P^+$ from other Wilson coefficient.  
Indeed  we obtain the following simultaneous constraints on all three Wilson coefficients:
\begin{equation}
\label{eq:FIT_CpseudoscalarSim}
    v^2\begin{pmatrix}
    C_V^+ \\ C_A^+ \\ C_P^+ 
    \end{pmatrix} = 
        \begin{pmatrix}
   \phantom{-}  0.98559(26)  \\ -1.25778(42) \\ -3.3(2.4)
    \end{pmatrix}, 
    \qquad      \rho =
\left(
\begin{array}{ccc}
 1 & -0.02 & 0.49 \\
 \text{} & 1 & 0.41 \\
 \text{} & \text{} & 1 \\
\end{array}
\right). 
\end{equation}
As expected, the uncertainty of $C_P^+$ is $\cO(10^4)$ larger than that of $C_V^+$ of $C_A^+$.
The relative sensitivity of various observables contributing to this result is visualized in \cref{fig:CPp}. 

We can relax our assumptions by allowing new physics to enter also via scalar and tensor interactions at the quark level,  which leads to  $C_S^+$ and  $C_T^+$ being non-zero in the leading order Lagrangian in \cref{eq:TH_NRleeyang0}.   
The existing experimental data  allows for a simultaneous fit of all five Wilson coefficients: 
\begin{equation}
\label{eq:FIT_Cpseudoscalar}
    v^2\begin{pmatrix}
    C_V^+ \\ C_A^+\\ C_S^+ \\ C_T^+  \\ C_P^+ 
    \end{pmatrix} = 
        \begin{pmatrix}
   \phantom{-} 0.98544(49) \\ -1.25818(82) 
   \\  -0.0005(12) \\  0.0008(16) \\ -5.9(4.4)
    \end{pmatrix}, 
\qquad \rho =     
\left(
\begin{array}{ccccc}
 1 & -0.07 & 0.85 & 0.22 & 0.45 \\
 \text{} & 1 & -0.06 & -0.84 & 0.73 \\
 \text{} & \text{} & 1 & 0.24 & 0.38 \\
 \text{} & \text{} & \text{} & 1 & -0.64 \\
 \text{} & \text{} & \text{} & \text{} & 1 \\
\end{array}
\right). 
\end{equation}
Introduction of $C_S^+$ and $C_T^+$ increases the degeneracy of the parameter space, leading to the constraints on $C_V^+$, $C_A^+$ and $C_P^+$ being relaxed by approximately a factor of two compared to the simplified scenario in \cref{eq:FIT_CpseudoscalarSim}. 
This is enough to disentangle the five independent WEFT couplings present in Eq.~(\ref{eq:TH_Lweft}). At linear order in new physics, they can be identified with the new physics parameters $\epsilon_{R,S,P,T}$ and the ``polluted'' CKM element corresponding to the combination $\hat V_{ud} \equiv V_{ud}(1+ \epsilon_L + \epsilon_R)$ \cite{Falkowski:2020pma}.\footnote{%
Let us stress that one cannot simultaneously extract
$V_{ud}$ and $\epsilon_L$ from any set of observables associated to the Lagrangian of \cref{eq:TH_Lweft}. Beyond the linear order, one may take $\{\hat{V}_{ud},\hat{\epsilon}_{R,S,P,T}\equiv\frac{\epsilon_{R,S,P,T}}{1+\epsilon_{L}+\epsilon_{R}}\}$ as a basis of WEFT couplings.} Indeed, using the matching in \cref{eq:TH_matching0,eq:TH_matching1}, we can translate the fit above into constraints on the parameters of the quark-level Lagrangian in \cref{eq:TH_Lweft}, in such a way that all the new physics parameters except for $\epsilon_L$ can be disentangled. We obtain
\begin{equation}
\label{eq:FIT_EpsilonP}
    \begin{pmatrix}
\hat V_{ud}  \\   \epsilon_S  \\  \epsilon_T \\  \epsilon_R \\  \epsilon_P   
    \end{pmatrix} = 
        \begin{pmatrix}
    0.97353(49)     \\
   -0.0004(12) \\ 0.0008(17) 
   \\  -0.010(11) \\  -0.017(13) 
    \end{pmatrix}, 
    \qquad 
    \rho =\left(
\begin{array}{ccccc}
 1 & 0.83 & 0.21 & 0.02 & 0.44 \\
 \text{} & 1 & 0.24 & 0.02 & 0.38 \\
 \text{} & \text{} & 1 & -0.02 & -0.64 \\
 \text{} & \text{} & \text{} & 1 & 0.03 \\
 \text{} & \text{} & \text{} & \text{} & 1 \\
\end{array}
\right). 
\end{equation}
This is the first time such general constraints, including those on the pseudoscalar parameter $\epsilon_P$,  are extracted from nuclear data.  
Inclusion of $\epsilon_P$ in the fit does not increase substantially the uncertainty on the remaining new physics parameters (see the fit without $\epsilon_P$ in Ref.~\cite{Falkowski:2020pma}), as also indicated by moderate off-diagonal entries in the last row/column of the correlation matrix.  
Curiously, the uncertainty on $\epsilon_P$ is only percent-level,
which is even slightly smaller than that on $\epsilon_R$. The lower sensitivity to the latter is in part a consequence of the relatively large uncertainty of the lattice input for $g_{A}$ when we match to \cref{eq:TH_matching0}, but it also shows that beta transitions have decent sensitivity to pseudoscalar interactions, even though their effects are suppressed by $\cO(m_e/m_N) \sim 10^{-3}$,  thanks to the large value of the pseudoscalar charge $g_P$~\cite{Gonzalez-Alonso:2013ura}. 
All in all, the magnitudes of $\epsilon_X$ effectively probed by beta  transitions is well within the validity range of the quark-level EFT. 
However, the sensitivity of beta transitions to $\epsilon_P$ is well inferior to that of pion decays. 
Very recently,  Ref.~\cite{Cirigliano:2021yto} obtained the constraint $\epsilon_P = 3.9(4.3)\times 10^{-6}$ in the combined fit to beta and pion decay data.
This is more than three orders of magnitude better than the result in \cref{eq:FIT_EpsilonP} based nuclear beta transitions alone.\footnote{If one allows for cancellations between the linear and quadratic pseudoscalar contributions to pion decays, then the bound is relaxed to $|\epsilon_P|\sim 4\times 10^{-4}$~\cite{Gonzalez-Alonso:2018omy}.}

\subsection{Weak magnetism}

Recoil corrections generated by scalar and tensor interactions have a negligible impact in the previous fit, because we are not yet sensitive to much larger leading BSM contributions. 
Recoil effects in the SM vector and axial currents are, however, relevant and automatically incorporated in the experimental data used for the previous fit. Here we study the sensitivity to the dominant recoil effect, the weak magnetism.
The name {\em weak magnetism} refers to a sum of two distinct effects entering at the subleading order in recoil. 
One is the contribution to the decay amplitude due to the operator 
\begin{equation}
\label{eq:WM_CM}
\cL^{(1)}  \supset  - C_M^+ { 1 \over 2 m_N}
\epsilon^{ijk}  ( \psi_p^\dagger \sigma^j \psi_n)  \nabla_i   \big  (\bar e_L \gamma^k  \nu_L \big )         
\end{equation}
in the subleading EFT Lagrangian in \cref{eq:TH_NRleeyang1}.
We will refer to this effect as the {\em universal}  weak magnetism, because it is common for all nuclear transitions. 
It turns out that a part of the contribution of another operator in \cref{eq:TH_NRleeyang1}, 
\begin{equation}
\label{eq:WM_FV}
    \cL^{(1)}  \supset
    -  C^+_{FV}   {i \over 2 m_N}   (\psi_p^\dagger  \overleftrightarrow{\nabla}_k \psi_n)   
 (\bar e_L \gamma^k \nu_L)  , 
\end{equation}
has the same tensor structure as that originating from \cref{eq:WM_CM}. 
Namely, using the parametrization of the 
$\langle \psi_p^\dagger  \overleftrightarrow{\nabla}_k \psi_n \rangle$ matrix element in \cref{eq:BETA1_dFGTmatrixelements} and retaining only the part proportional to $\beta_{FV}$, the subleading decay amplitude is affected by the operators in \cref{eq:WM_CM} and \cref{eq:WM_FV} as 
\begin{equation}
\cM_{\beta^-}^{(1)} \supset 
 i   \big (C_M^+ + \beta_{FV} C^+_{FV} \big ) r M_F  A 
 \epsilon^{ijk}  \boldsymbol{q}^i  {[{\cal T}_{(J)}^j]_{J_z'J_z}\over \sqrt{J(J+1)} }   L^k . 
\end{equation}
We will refer to the contribution entering via the operator in \cref{eq:WM_FV} as the nucleus-dependent contribution to weak-magnetism (because the form factor $\beta_{FV}$ depends on the nuclei participating in the transition) or, in short, {\em nuclear} weak magnetism.
Once again, it is instructive to first display the correlations in  a simplified setting where we only take into account the interference between weak magnetism and the leading SM effects proportional to $C_V^+$ and $C_A^+$. 
Defining $C^+_{WM} \equiv C_M^+ + \beta_{FV} C^+_{FV}$, at the linear order in recoil the total (universal+nuclear) weak magnetism  enters the relevant correlations as:
\begin{align}
\label{eq:WM_correlations}
\xi_b(E_e)  =  &   
\pm r^2   {2 ( E_e^{\rm max }   -  2 E_e  + m_e^2/E_e)  \over 3 m_N } { C_A^+ C^+_{WM}  \over (C_V^+)^2 + r^2 (C_A^+)^2 }  , 
\nnl 
 \Delta a(E_e) = &    
\pm r^2  {  2(2 E_e -  E_e^{\rm max}) \over 3 m_N }  { C_A^+ C^+_{WM}  \over (C_V^+)^2 + r^2 (C_A^+)^2 } , 
\nnl 
 \Delta A (E_e)  = &   
 \mp  r \sqrt{J \over J+1}    {2 (E_e^{\rm max} - E_e) \over 3 m_N } {C_V^+  C^+_{WM}  \over (C_V^+)^2 + r^2 (C_A^+)^2 }   
+ {r^2 \over J+1}  {5 E_e  - 2 E_e^{\rm max} \over 3 m_N} {C_A^+  C^+_{WM}  \over (C_V^+)^2 + r^2 (C_A^+)^2 }   , 
 \nnl 
 \Delta B(E_e)  = &   
 \pm  r \sqrt{J \over J+1}    {2 p_e^2  \over 3  m_N E_e  }   { C_V^+ C^+_{WM}  \over (C_V^+)^2 + r^2 (C_A^+)^2 }  
+ {r^2 \over J+1}  {3 E_e E_e^{\rm max}   - 5 E_e^2 + 2 m_e^2 \over 3 m_N E_e } { C_A^+ C^+_{WM}  \over (C_V^+)^2 + r^2 (C_A^+)^2 }  . 
\end{align}
See \cref{app:correlationsWM} for the complete expressions including the interference with scalar and tensor currents and for the remaining correlations. Much as the pseudoscalar interactions in the previous subsections, weak magnetism is zero for pure Fermi transitions, in particular it does not affect the superallowed $0^+ \to 0^+$ transitions.  Unlike pseudoscalar, weak magnetism leads to all correlation coefficients picking up a dependence on the beta particle energy $E_e$. 

\begin{table}[tb]
    \centering
    \begin{tabular}{|c|c|c|c|c|c|c|c|}
    \hline
 & $n$ & ${}^{17}$F & ${}^{19}$Ne & ${}^{21}$Na & ${}^{29}$P  & ${}^{35}$Ar & ${}^{37}$K \\    \hline    
$\langle\xi_b(E_e)\rangle$ & -0.1 &  -0.1 &   -0.1 & -0.1 & 0.0 &  0.0 & -0.1  
\\   \hline
$\langle \Delta  a(E_e) \rangle_p$ & -1.8 & 1.1 &  1.2 &  0.5 & 0.3 & 0.1 & 0.3 
\\   \hline
$\langle \Delta  A(E_e)\rangle_p$ & -3.1 & 1.3 &   -4.6 &  2.0 & 1.7 & 2.7 & -6.1 
\\    \hline
$ \langle \Delta B(E_e) \rangle$ &  -0.3 & -2.8 &  0.2 & -3.5 & -3.8 & -3.3 &  4.5  
\\    \hline
    \end{tabular}
\caption{%
Sensitivity of the correlation coefficients in neutron and mirror decays to the nucleon-level weak magnetism parametrized by the  Wilson coefficient $C_M^+$. 
The entries are in units of $10^{-4}$ and correspond to the averaged function multiplying $C_M^+$ in \cref{eq:WM_correlations}  for different parent nuclei. 
For the sake of this table,  $C_V^+$, $C_A^+$, and $r$ are set to their minimum values in the SM fit (although they are kept as free parameters in the fits of this section).
The averages $\langle \cdot \rangle_{(p)}$ over the $\beta$ particle energy $E_e$ are defined in \cref{eq:BETA_Average,eq:BETA_AverageP}. 
\label{tab:weakmagnetism}
}
\end{table}

As discussed in \cref{sec:EFT}, the numerical value of $C_{WM}^+$ for each transition is fixed by isospin symmetry (CVC), 
assuming the absence of large contributions to $C_M^+$ from higher-dimensional operators in the quark-level Lagrangian.
More precisely, $C_{WM}^{+,\rm CVC} \approx  (g_M^{\rm CVC} + \beta_{FV}^{\rm CVC}) C_V^+$ up to small isospin breaking effects, where 
$g_M^{\rm CVC} = (\mu_p - \mu_n)/\mu_N \approx 4.7$, 
and $\beta_{FV}^{\rm CVC}$ is transition-dependent and can be related to nuclear magnetic moments via  \cref{eq:BETA1_CVCrelation}. 
The goal of this subsection is to show that the existing experimental data is powerful enough to pinpoint universal weak magnetism parametrized by $C_M^+$, {\em without} a  theoretical input from isospin symmetry. 

Notice first that  $\beta_{FV} = 0$ for the neutron decay. 
Therefore, restricting to the subset of data that includes only the $0^+ \to 0^+$ transitions and the neutron decay, we can directly determine $C_M^+$ (treated as a free parameter) along with other Wilson coefficients in the EFT Lagrangian. 
In the restricted framework where only $C_V^+$, $C_A^+$, and $C_M^+$ are non-zero, we obtain the simultaneous constraint on the remaining three Wilson coefficients: 
\begin{equation}
\label{eq:FIT_CMsim}
    v^2\begin{pmatrix}
    C_V^+ \\ C_A^+ \\ C_M^+ 
    \end{pmatrix} = 
        \begin{pmatrix}
  0.98563(26)  \\ -1.25785(52) \\ 3.6(1.1)
    \end{pmatrix}, 
    \qquad \rho = 
    \left(
\begin{array}{ccc}
 1 & 0.13 & 0.47 \\
 \text{} & 1 & 0.66 \\
 \text{} & \text{} & 1 \\
\end{array}
\right). 
\end{equation}
The fit is dominated by the measurements of ${\cal F} t(0^+ \to 0^+)$ (which fixes $C_V^+$), neutron's lifetime (which then fixes $C_A^+$), and the neutron's beta asymmetry (which then fixes $C_M^+$). 
We observe that the result is consistent with the CVC prediction $C_M^{+,\rm CVC} \approx 4.6/v^2$, and that  
$C_M^+ = 0$ is excluded at more than 3 sigma. 
To our knowledge this is the first experimental evidence for the universal (nucleon-level) weak magnetism.

\begin{figure}
    \centering
    \includegraphics[width=0.8\textwidth]{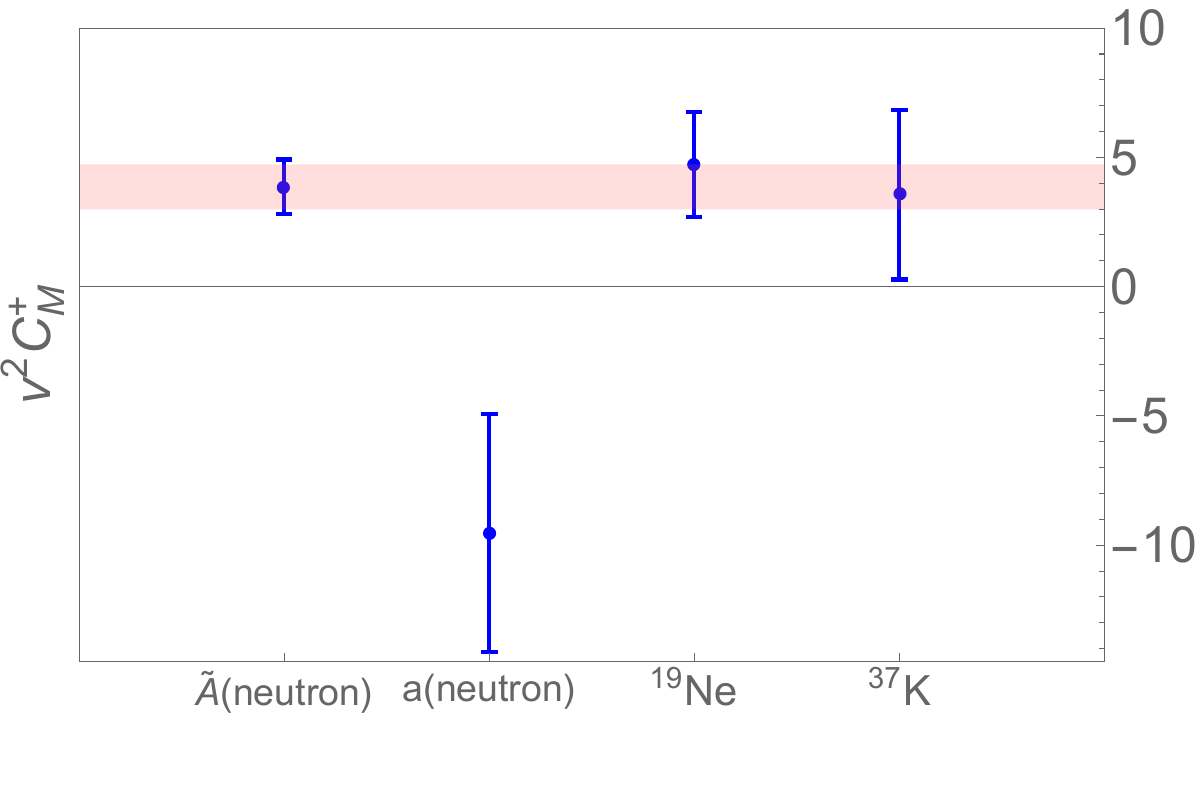}
    \caption{%
    Constraints on the universal weak-magnetism $C_M^+$ from a subset of most sensitive observables. 
    We show the fit to $C_M^+$ marginalized over $C_V^+$ and $C_A^+$,  assuming other Wilson coefficients vanish. 
The error bars show the 68\% CL intervals obtained  using 
${\cal F}t(0^+ \to 0^+)$,  neutron's lifetime, and one of the following: 
neutron's beta asymmetry, neutron's $\beta$-$\nu$ correlation, 
${\cal F} t$ and beta asymmetry of ${}^{19}$Ne, 
or ${\cal F} t$, beta asymmetry, and neutrino asymmetry of ${}^{37}$K. The pink band corresponds to the global fit in \cref{eq:FIT_CWMsim} and the dashed line marks the prediction of isospin symmetry. 
    }
    \label{fig:CMp}
\end{figure}

We can sharpen this evidence somewhat by including data from mirror transitions studied in Ref.~\cite{Falkowski:2020pma}. In those cases $\beta_{FV}$ is non-zero, therefore those transitions do not give us an unobstructed access to the $C_M^+$ Wilson coefficient. Ideally for this argument, $\beta_{FV}$ would be determined by first principle calculations, for example by ab initio nuclear computations.
 In absence of such, for the sake of the fits below we fix $\beta_{FV}$ from the CVC relation in \cref{eq:BETA1_CVCrelation}.  
Following this procedure we obtain 
\begin{equation}
\label{eq:FIT_CWMsim}
    v^2\begin{pmatrix}
    C_V^+ \\ C_A^+ \\ C_M^+ 
    \end{pmatrix} = 
        \begin{pmatrix}
   \phantom{-}  0.98570(24)  \\ -1.25778(48) \\ 3.86(87)
    \end{pmatrix}, 
    \qquad 
    \rho = \left(
\begin{array}{ccc}
 1 & 0.02 & 0.36 \\
 \text{} & 1 & 0.60 \\
 \text{} & \text{} & 1 \\
\end{array}
\right) , 
\end{equation}
where the uncertainty on $C_M^+$ is improved by $\sim 20\%$ and the experimental evidence for universal weak magnetism is strengthened above $4\sigma$. 
The relative sensitivity of various observables contributing to this result is visualized in \cref{fig:CMp}. 
Admittedly, our assumption that $\beta_{FV} = \beta_{FV}^{\rm CVC}$ makes the above argument a bit circular. 
Note however that the constraining power of the mirror data is currently dominated by the ${}^{19}$Ne decay~\cite{Rebeiro:2018lwo,Combs:2020ttz}. 
Indeed, discarding all other mirror data except for ${}^{19}$Ne leads to 
$C_M^+ = 3.86(90)$, a negligible difference compared to \cref{eq:FIT_CWMsim}. 
Furthermore, this result is very weakly sensitive to the precise value of 
$\beta_{FV}({}^{19}{\rm Ne})$ used in the fit. 
Using a much weaker assumption, 
$\beta_{FV}({}^{19}{\rm Ne}) = (1 \pm 1)\beta_{FV}^{\rm CVC}({}^{19}{\rm Ne})$, where $\beta_{FV}^{\rm CVC}({}^{19}{\rm Ne}) = 1.5$, leads to    
$C_M^+ = 3.80(94)$, which is still $4 \sigma$ away from zero. 
We conclude that the further evidence for fundamental weak magnetism provided by the mirror data is robust. 

Finally, we can also make a more general fit by allowing for new physics entering as scalar and tensor current at the leading order: 
\begin{equation}
\label{eq:FIT_CWMfull}
    v^2\begin{pmatrix}
    C_V^+ \\ C_A^+ \\ C_S^+ \\ C_T^+ \\ C_M^+ 
    \end{pmatrix} = 
        \begin{pmatrix}
 0.98576(44)  \\ -1.2578(13) 
   \\ 0.0002(11) \\ 0.0000(23) \\ 4.0(1.9)
    \end{pmatrix}, 
    \qquad \rho = 
\left(
\begin{array}{ccccc}
 1 & -0.41 & 0.83 & 0.53 & -0.16 \\
 \text{} & 1 & -0.36 & -0.93 & 0.91 \\
 \text{} & \text{} & 1 & 0.5 & -0.16 \\
 \text{} & \text{} & \text{} & 1 & -0.84 \\
 \text{} & \text{} & \text{} & \text{} & 1 \\
\end{array}
\right) . 
\end{equation}
Even in this more general framework the preference for non-zero $C_M^+$ is still at the $\sim 2\sigma$ level. 

We remark that all our fits in this section use only observables integrated over the energy spectrum of the beta particle, because only that information is provided by experimental collaborations in the readily usable form. 
On the other hand, as is clear from \cref{eq:WM_correlations}, weak magnetism predicts specific dependence of the correlations on $E_e$. 
Exploring this energy dependence will allow one to further tighten the theory-independent bounds on $C_M^+$, possibly pushing the evidence for weak magnetism beyond the $5 \sigma$ threshold.

\subsection{Induced tensor}

As the final example of application of our formalism, we present the fit to the Wilson coefficient parametrizing the so-called induced tensor interactions in the subleading EFT Lagrangian in \cref{eq:TH_NRleeyang1}: 
\begin{equation}
\label{eq:CE_L}
\cL^{(1)}   \supset -  C_E^+ { i  \over 2 m_N} 
 ( \psi_p^\dagger \sigma^k\psi_n) \nabla_k \big (\bar e_L \gamma^0 \nu_L \big )
 -  C_{E'}^+ { i  \over 2 m_N} 
 ( \psi_p^\dagger \sigma^k\psi_n) \partial_t \big (\bar e_L \gamma^k \nu_L \big ). 
\end{equation}
As discussed in \cref{sec:EFT}, the UV matching relations for these Wilson coefficients imply  $C^+_E = C_{E'}^+ = - C_A^+ g_{IT}/g_A$,
thus in the following we set $C_{E'}^+ = C_E^+$ and assume $C_E^+$ is real. 
We expect the nucleon-level parameter $g_{IT}$ to be suppressed by small isospin breaking effects, hence $v^2 |C_E^+|$ is $\cO(10^{-2}-10^{-3})$.
But, as in the preceding subsection, we can ask the question what does experiment tell about $C_E^+$ {\em without} any theoretical input from isospin symmetry. 
At the level of interference with $C_V^+$ and $C_A^+$, the operators in \cref{eq:CE_L} affect the correlations as 
\begin{align}
\label{eq:IT_correlations}
\xi_b(E_e) = & 
\pm r^2   {2 E_e^{\rm max}   + m_e^2/E_e \over 3 m_N}  { C_A^+  C_E^+   \over (C_V^+)^2 + r^2 (C_A^+)^2  }  , 
\nnl 
 \Delta a(E_e)  =  & 
\mp r^2   { 2 E_e^{\rm max}   \over 3 m_N}  { C_A^+  C_E^+   \over (C_V^+)^2 + r^2 (C_A^+)^2 }  , 
\nnl 
 \Delta A(E_e)  = &  
\mp r \sqrt{J \over J+1} {2 (E_e^{\rm max} - E_e) \over 3 m_N}  {C_V^+  C_E^+   \over (C_V^+)^2 + r^2 (C_A^+)^2 } 
- {r^2 \over J+1}  { 2 E_e^{\rm max}  + E_e  \over 3 m_N} { C_A^+ C_E^+  \over (C_V^+)^2 + r^2 (C_A^+)^2 } , 
\nnl 
  \Delta B(E_e)  = & 
\mp r \sqrt{J \over J+1} {2 E_e  + m_e^2/E_e \over 3 m_N} {C_V^+  C_E^+   \over (C_V^+)^2 + r^2 (C_A^+)^2 } 
\nnl  +& 
  {r^2 \over J+1}  {3 E_e^{\rm max} - E_e + m_e^2/E_e  \over 3 m_N}    {C_A^+  C_E^+  \over (C_V^+)^2 + r^2 (C_A^+)^2 }  .
\end{align}
See \cref{app:correlationsIT,app:correlationsITprime} for the complete expressions including the interference with scalar and tensor currents and for the remaining correlations. 
Note that, in principle, isospin breaking contributions to the 
$\langle \psi_p^\dagger \sigma^k  \overleftrightarrow{\nabla}_k  \psi_n  \rangle $ matrix element in \cref{eq:BETA1_dFGTmatrixelements} would lead to additional transition-dependent contributions to the correlations with the same functional form as \cref{eq:IT_correlations}. 
In this analysis we assume that such contributions are absent. 

\begin{table}[tb]
    \centering
    \begin{tabular}{|c|c|c|c|c|c|c|c|}
    \hline
 & $n$ & ${}^{17}$F & ${}^{19}$Ne & ${}^{21}$Na & ${}^{29}$P & ${}^{35}$Ar & ${}^{37}$K \\    \hline    
$\langle\xi_b(E_e)\rangle$ & -6.9 & 8.4 &  11.5 & 5.9 & 5.7 & 2.3 & 8.1  
\\   \hline
$\Delta  a(E_e)$ & 5.2 & -7.5 &   -10.7 &  -5.6 & -5.5 & -2.3 & -7.9 
\\   \hline
$\langle \Delta  A(E_e)\rangle_p$ & 4.8 & 5.1  &  7.6 &  5.8 & 7.6 &  4.2 & -1.1 
\\    \hline
$ \langle \Delta B(E_e) \rangle$ &  -6.6 & 0.3 & -11.7 & 0.6 & -1.5 &  2.1 & -9.6
\\    \hline
    \end{tabular}
\caption{Sensitivity of the correlation coefficients in neutron and mirror decays to the induced tensor interactions parametrized by the  Wilson coefficients $C_{E'}^+ = C_E^+$. 
The entries are in units of $10^{-4}$ and correspond to the averaged function multiplying $C_E^+$ in \cref{eq:IT_correlations} for different parent nuclei. 
For the sake of this table,  $C_V^+$, $C_A^+$, and $r$ are set to their minimum values in the SM fit (although they are kept as free parameters in the fits of this section).
The averages $\langle \cdot \rangle_{(p)}$ over the $\beta$ particle energy $E_e$ are defined in \cref{eq:BETA_Average,eq:BETA_AverageP}; for $\Delta  a(E_e)$  the induced tensor contribution is independent of $E_e$.  }
    \label{tab:inducedtensor}
\end{table}

Assuming that $C_E^+$ is the only free parameter in the fit in addition to the SM Wilson coefficients $C_V^+$ and $C_A^+$, 
we obtain 
\begin{equation}
\label{eq:FIT_CEsim}
   v^2\begin{pmatrix}
    C_V^+ \\ C_A^+ \\ C_E^+ 
    \end{pmatrix} = 
        \begin{pmatrix}
  0.98566(23)  \\ -1.25881(90) \\ 1.7(1.1)
    \end{pmatrix}, 
    \qquad \rho =
\left(
\begin{array}{ccc}
 1 & 0.16 & -0.30 \\
 \text{} & 1 & -0.90 \\
 \text{} & \text{} & 1 \\
\end{array}
\right). 
\end{equation}
Clearly, the existing data are sensitive only to $|C_E^+| \sim 1/v^2$ (similarly as for the other recoil-level Wilson coefficients, e.g. $C_P^+$ or $C_M^+$). 
This is 3 orders of magnitude larger than the theoretically expected magnitude
and hence experimental detection of $C_E^+$ is unlikely in the envisagable future. 
Furthermore, the current data show a mild $~1.6\sigma$ preference for a non-zero $C_E^+$, driven largely by the measurement of neutron's $\beta$-$\nu$ correlation in the aSPECT experiment~\cite{Beck:2019xye}. 
It will be interesting to see if this preference goes away, as experiments acquire more precise data.

\section{Conclusions}
\label{sec:conclusions} 

In this paper we discussed the EFT for beta transitions. Working in the framework of the pionless EFT, with nucleons as degrees of freedom, the effective interactions between nucleons and leptons were organized in a non-relativistic expansion in powers of $\boldsymbol{\nabla}/m_N$. The novelty of this paper is that the low-energy EFT was matched to the general quark-level EFT at higher energy. The latter, which we refer to as the WEFT, describes the effects of the SM weak interactions, as well as possible effects of new heavy particles from beyond the SM. We worked out the matching between the WEFT and the low-energy EFT up to the subleading order in $\boldsymbol{\nabla}/m_N$, that is including the linear recoil effects. The results in \cref{eq:TH_matching0} and \cref{eq:TH_matching1} describe the matching conditions for the Wilson coefficients of the leading and subleading Lagrangian in \cref{eq:TH_NRleeyang0} and  \cref{eq:TH_NRleeyang1}.  In particular, the matching of the non-standard tensor interactions in the WEFT to the recoil-level non-relativistic interactions at low energy was worked out for the first time.               

The EFT framework allows us to systematically describe how the standard and non-standard weak interactions affect beta decay observables, such as the lifetime, beta energy spectrum, and various angular correlations. 
We calculated the impact of all the terms in the leading and subleading Lagrangian on the differential decay width in allowed beta transitions summed over the beta particle and daughter nucleus polarizations.
In \cref{eq:BETA_dGammaTemplate,eq:BETA_FTdeltaGammaTemplate} we list all correlation coefficients that appear at the leading and subleading orders in recoil; the ones in \cref{eq:BETA_FTdeltaGammaTemplate} receive recoil-level contributions only in the presence of tensor interactions.
We express the correlation coefficients in terms of the Wilson coefficients of the effective Lagrangian and matrix elements of non-relativistic nucleon currents.
Partial results, most relevant for our numerical analysis,  were displayed in \cref{sec:fits}, while the complete results are collected in \cref{app:sub}. 

With the expressions for the observables at hand, we can use the existing experimental data on beta transitions to determine confidence intervals for the Wilson coefficients in the EFT Lagrangian. 
For the leading Lagrangian this exercise was already completed in Ref.~\cite{Falkowski:2020pma}, where $\cO(10^{-4})$ relative precision was found for the standard Wilson coefficients corresponding to the vector and axial currents, and stringent constraints were established on the non-standard scalar and tensor interactions. 
In this paper we extended this analysis to recoil-level Wilson coefficients. 
In particular, we performed the first ever comprehensive analysis of the pseudoscalar interactions in allowed beta decay. 
We find that nuclear decays set a robust constraint on the Wilson coefficient descending from the pseudoscalar interactions in the WEFT, even though it enters the observables only at the linear order in recoil. This translates into a percent-level constraint on the pseudoscalar WEFT parameter ($\epsilon_P$ in \cref{eq:TH_Lweft}), which is comparable to the sensitivity to the right-handed WEFT current ($\epsilon_R$), and one order of magnitude weaker than the sensitivity to the scalar and tensor currents ($\epsilon_S$ and $\epsilon_T$). 
One should note, however, that within the WEFT framework, constraints on $\epsilon_P$ from pion decays are 4 orders of magnitude stronger. 
It would be interesting to extend this analysis to the recoil-level effects of tensor interactions. In particular, at this order, tensor interactions contribute to the very precisely measured $0^+ \to 0^+$ transitions. However, a quantitative analysis of this kind would require (at least approximate) knowledge of the subleading tensor charges $g_T^{(1,3)}$, cf. \cref{eq:TH_matrixelements}, as well as of the nuclear form factors $\alpha_{FT}$ and $\gamma_{FT}$, cf. \cref{eq:BETA1_dFGTmatrixelements}.

Weak magnetism is another recoil-level effect to which experiment is sensitive. 
Our global analysis of the allowed beta decay data showed a 3 sigma evidence for a non-zero value of the EFT Wilson coefficient corresponding to the universal (nucleon-level)  weak magnetism. 
The evidence is dominated by the neutron decay measurements (lifetime and beta asymmetry), and is further strengthened by mirror decay data.  We also discussed the recoil-level EFT operator describing the so-called induced tensor interactions. 
The isospin symmetry of QCD predicts that this Wilson coefficient should be suppressed so as to give negligible contributions to observables. Instead,  our global analysis showed a small $1.8~\sigma$ preference for non-zero induced tensor interactions. 
Future measurements and better theoretical calculations will improve the understanding of the effects of these Wilson coefficients.

\section*{Acknowledgements}
AF and ARS are supported by the Agence Nationale de la Recherche (ANR) under grant ANR-19-CE31-0012 (project MORA).
AF is supported by the European Union’s Horizon 2020 research and innovation programme under the Marie Sklodowska-Curie grant agreement No. 860881 (HIDDe$\nu$ network). 
MGA is supported by the Generalitat Valenciana through the plan GenT program (CIDEGENT/2018/014), and by the Spanish Ministerio de Ciencia, Innovación y Universidades through grants PID2020-114473GBI00 and CNS2022-135595.
The work of AP has received funding from the Swiss National Science Foundation (SNF) through the Eccellenza Professorial Fellowship “Flavor Physics at the High Energy Frontier” project number 186866.

\appendix

\newpage
\section{Symmetry constraints on nuclear matrix elements} 
\label{app:SHF}
In this appendix we discuss the nuclear matrix elements of the form
\begin{equation}
\label{eq:SHF_matrixElement}
\bra{{\cal N'}} \bar p \Gamma n  \ket{{\cal N}} .
\end{equation}
Here, $\ket{\cal N}$ and $\ket{\cal N'}$ are nuclear states with spin $J$,
$J_z$ and $J_z'$ projection of the spin on the z-axis,
and momenta $p$ and $p'$.
They both belong to the same isospin multiplet with the isospin quantum number $j$ and the isospin projections related by 
$j_3' -  j_3 = 1$.
The operator sandwiched between the states is made of  relativistic neutron and proton fields $n$ and $p$ evaluated at $x=0$. 
Below we will consider the vector, axial, and tensor matrix elements, that is $\Gamma = \gamma^\mu,\gamma^\mu \gamma^5, \sigma^{\mu\nu}$. We do not consider the (pseudo)scalar cases because, as we will see, they are not needed to extract the non-relativistic matrix elements that we are interested in.

The aim of this appendix is to write down the most general expression for the matrix element in \cref{eq:SHF_matrixElement} consistent with Lorentz invariance and the discrete symmetries of the strong interactions: parity and time reversal invariance. We will work in the isospin limit where ${\cal N}$ and ${\cal N'}$ (and also $p$ and $n$) have the same mass, thus 
$p^2 = p'{}^2= m_{\cal N}^2$. 
In this limit, the matrix element is also invariant under another discrete symmetry called G-parity, which can be defined as a product of $P$, $T$, and a 180 degrees isospin rotation. 
Given the relativistic matrix elements, it will be straightforward to take the non-relativistic limit where $\boldsymbol{p}$ and $\boldsymbol{p}'$ are much smaller than the nucleon mass, 
and by this means to determine the most general form of the non-relativistic matrix elements used in our analysis. 

\subsection{Spin-J representation matrices}
\label{app:clebsch}

We first introduce the spin-J representation matrices ${\cal T}_{(J)}^k$, which appear in nuclear matrix elements at the leading and subleading order of recoil expansion.   
It is convenient to define them in terms of the Clebsch-Gordan coefficients. 
The latter are denoted by $C(J_1,m_1;J_2,m_2;J,m)$ and defined by 
$\ket{J,m} = \sum_{m_i = - J_i}^{J_i}  C(J_1,m_1;J_2,m_2;J,m) \ket{J_1,m_1}\otimes \ket{J_2,m_2}$ with the Condon–Shortley phase conventions. 
We can define the  $(2 J+1)\times (2 J+1)$-dimensional matrices  ${\cal T}_{(J)}^k$ as follows 
\begin{align}
\label{eq:SHF_Tk}
\, [{\cal T}_{(J)}^1]^{\, J_z}_{J_z'}   \equiv &  - \sqrt{J(J+1) \over 2} \bigg [   C(J,J_z;1,1;J,J_z') -   C(J,J_z;1,-1;J,J_z') \bigg ], 
\nnl 
\, [{\cal T}_{(J)}^2]^{\, J_z}_{J_z'}   \equiv &  \,i\, \sqrt{J(J+1) \over 2  } \bigg [   C(J,J_z;1,1;J,J_z') +   C(J,J_z;1,-1;J,J_z') \bigg ]  ,  
\nnl 
\, [{\cal T}_{(J)}^3]^{\, J_z}_{J_z'}   \equiv &  \,\sqrt{J(J+1)} \,C(J,J_z;1,0;J,J_z') . 
\end{align}
These are the familiar spin-J generators of the $SO(3)$ Lie algebra, normalized such that ${\cal T}_{(J)}^3 = {\rm diag}(J, J-1, \dots, -J)$.\footnote{%
In our conventions the rows (columns) of  ${\cal T}_{(J)}$ correspond to $J_z'$ ($J_z$) going from $+J$ to $-J$.   }
In particular  ${\cal T}_{(0)}^k = 0$, ${\cal T}_{(1/2)}^k = \sigma^k/2$, 
and 
\begin{equation}
{\cal T}_{(1)}^1 = \left(
\begin{array}{ccc}
 0 & \frac{1}{\sqrt{2}} & 0 \\
 \frac{1}{\sqrt{2}} & 0 & \frac{1}{\sqrt{2}} \\
 0 & \frac{1}{\sqrt{2}} & 0 \\
\end{array}
\right), \quad 
{\cal T}_{(1)}^2 =\left(
\begin{array}{ccc}
 0 & -\frac{i}{\sqrt{2}} & 0 \\
 \frac{i}{\sqrt{2}} & 0 & -\frac{i}{\sqrt{2}} \\
 0 & \frac{i}{\sqrt{2}} & 0 \\
\end{array}
\right), \quad 
{\cal T}_{(1)}^3 =\left(
\begin{array}{ccc}
 1 & 0 & 0 \\
 0 & 0 & 0 \\
 0 & 0 & -1 \\
\end{array}
\right). 
\end{equation}
One property of the ${\cal T}_{(J)}^{1,2}$ matrices is that, for any $J$, the only non-zero entries are those with $|J_z - J_z' |= 1$. 
For $J \geq 1$ we can also define the 2-index Hermitian and traceless matrices:
\beq
\label{eq:SHF_Tkl}
{\cal T}_{(J)}^{kl} \equiv {\cal T}_{(J)}^k {\cal T}_{(J)}^l +  {\cal T}_{(J)}^l {\cal T}_{(J)}^k - {2 \over 2 J+ 1} \tr [ {\cal T}_{(J)}^k {\cal T}_{(J)}^l]  \boldsymbol{1}_{2J+1} . 
\eeq 
These can appear in nuclear matrix elements at the subleading order in recoil expansion, and have non-zero entries for $|J_z - J_z' | \leq 2$.
The matrices ${\cal T}_{(J)}^{k}$ and  ${\cal T}_{(J)}^{kl}$ satisfy the useful sum rules (for any value of $J_z$): 
\begin{align}
\label{eq:SHF_SSdaughter}
\sum_{J_z' = -J}^J  [{\cal T}_{(J)}^k]^{\, J_z}_{J_z'} \delta_{J_z' J_z}   = & 
J^k, 
\nnl  
\sum_{J_z'=-J}^{J} [{\cal T}_{(J)}^k]^{\, J_z}_{J_z'}  [{\cal T}_{(J)}^{l\,*}]^{\, J_z}_{J_z'}
 = &    {J (J + 1) \over 3}  \delta^{kl}
 +   {J(J+1) - 3 (\boldsymbol{J} \boldsymbol{j})^2 \over 6} 
 \big ( \delta^{kl} - 3 j^k j^l \big  ) 
-   {i \over 2 }   \epsilon^{klr} J^r , 
\nnl 
\sum_{J_z' = -J}^J  [{\cal T}_{(J)}^{kl}]^{\, J_z}_{J_z'} \delta_{J_z' J_z}   = & 
{J(J+1) - 3 (\boldsymbol{J} \boldsymbol{j})^2 \over 3} 
 \big ( \delta^{kl} - 3 j^k j^l \big  ) , 
 \nnl 
 \sum_{J_z' = -J}^J  [{\cal T}_{(J)}^{kl}]^{\, J_z}_{J_z'} [{\cal T}_{(J)}^{j\,*}]^{\, J_z}_{J_z'}   = &  
 {(2J+3)(2J-1) \over 6} \big  ( \delta^{jk} J^l + \delta^{jl} J^k \big )
 - J^j\big  (4 J^k J^l - j^k j^l \big ) 
 \nnl + & 
 {J(J+1) - 3 (\boldsymbol{J} \boldsymbol{j})^2 \over 3}  \bigg [ 
  \big ( \delta^{kl} - 3 j^k j^l \big  ) J^j 
  +    \big ( \delta^{kj} - 3 j^k j^j \big  ) J^l 
+    \big ( \delta^{lj} - 3 j^j j^l \big  ) J^k 
 \bigg]
 \nnl + & 
  i {3 \over 2 }  \bigg [ \epsilon^{jkn}  J^l +  \epsilon^{jln} J^k\bigg ] J^n  
 -  i { J(J+1) \over 2 }  \bigg [ \epsilon^{jkn}  j^l +  \epsilon^{jln} j^k\bigg 
]j^n .  
\end{align}

\subsection{Spinor conventions}

We will express relativistic nuclear matrix elements in terms of commuting spinor variables using the formalism introduced in Ref.~\cite{Arkani-Hamed:2017jhn}. 
Here we provide a lightning review of this formalism.  

Let us first define our conventions for the 2-component spinor algebra. 
We work in four dimensions with the mostly-minus metric $\eta_{\mu \nu} = {\rm diag}(1,-1,-1,-1)$. 
The Lorentz algebra can be decomposed into $SL(2,\mathbb{C}) \times SL(2,\mathbb{C})$. 
Holomorphic and anti-holomorphic spinors $\chi_\alpha$ and $\tilde \chi_{\dot \alpha}$, $\alpha=1,2$,   transform under the respective $SL(2,\mathbb{C})$ factors with indices being raised and lowered by the antisymmetric epsilon tensor:
\beq
\label{eq:SHF_spinordef}
\chi^\alpha = \epsilon^{\alpha \beta} \chi_\beta ~,
\quad 
\chi_\alpha = \epsilon_{\alpha \beta} \chi^\beta ~,
\qquad
\tilde \chi^{\dot \alpha} = \epsilon^{\dot \alpha \dot \beta} \tilde \chi_{\dot \beta} ~,
\quad 
\tilde \chi_{\dot \alpha} = \epsilon_{\dot \alpha \dot \beta} \tilde \chi^{\dot \beta}~, 
\eeq 
where summing over repeated spinor indices is implicit, and we use the convention $\epsilon^{12} = 1$, $\epsilon_{12} = -1$. 
One can construct Lorentz invariants from two holomorphic or two  anti-holomorphic spinors: $\chi \psi \equiv \chi^\alpha \psi_\alpha$, $\tilde \chi\tilde\psi \equiv \tilde \chi_{\dot \alpha} \tilde \psi^{\dot \alpha}$. 
Furthermore, spinor indices can be traded for the vector ones with the  help of  the sigma matrices:  
$[\sigma^\mu]_{\alpha \dot \beta}  = (\mathbb{I}, \boldsymbol{\sigma} )$ and 
$[\bar \sigma^\mu]^{\dot \alpha \beta}  = (\mathbb{I}, -\boldsymbol{\sigma})$, where $\boldsymbol{\sigma}$ are the usual Pauli matrices. 
For example, 
$\chi \sigma^\mu \tilde \psi \equiv \chi^\alpha [\sigma^\mu]_{\alpha \dot \beta}  \tilde \psi^{\dot \beta}$ and 
$\tilde \chi \bar \sigma^\mu \psi \equiv \tilde \chi_{\dot \alpha} [\bar \sigma^\mu]^{\dot \alpha \beta}\psi_{\beta}$ both transform as Lorentz vectors. 
In the 2-component context we also define 
$\sigma^{\mu \nu} \equiv {i \over 2}(\sigma^\mu \bar \sigma^\nu-\sigma^\nu \bar \sigma^\mu)$ and 
$\bar \sigma^{\mu \nu} \equiv {i \over 2}(\bar \sigma^\mu \sigma^\nu-\bar \sigma^\nu \sigma^\mu)$, 
using which one can construct Lorentz tensors 
$\chi \sigma^{\mu \nu} \psi$ and $\tilde \chi \bar \sigma^{\mu \nu} \tilde \psi$. 

Consider now  a massive particle whose (four-)momentum $p$ satisfies the on-shell condition $p^2 = m^2$. 
We treat $p$ as incoming momentum, thus $p^0 > 0$ for initial state particles, 
and $p^0 < 0$ for final state particles. 
The momentum can be equivalently represented by four {\em commuting} two-component spinors  $\chi^K$ and $\tilde \chi_K$, where $K=1,2$: 
\begin{equation}
\label{eq:SHF_chidef}
[p_\mu \sigma^\mu]_{\alpha \dot \beta} 
=   \sum_{K=1}^2 \chi_{\alpha}^{\, K} \tilde \chi_{\dot \beta \, K}~  ,
\qquad (p  \bar \sigma)^{\alpha \dot \beta} =   \sum_{K=1}^2   \tilde \chi^{\dot \alpha}_{\, K} \chi^{\beta \, K} \, . 
\end{equation} 
In our conventions the spinors are normalized as 
\begin{equation}
\label{eq:SHF_chinorm}
(\chi^K \chi_L) = \delta^K_L  m ~,  \qquad    
(\tilde \chi_K \tilde \chi^L) =   \delta_K^L  m ~.
\end{equation} 
It follows that they satisfy the Dirac equation:
\begin{equation}
(p_\mu \sigma^\mu) \tilde \chi^K = m  \chi^K~,
\quad 
(p_\mu  \bar \sigma^\mu) \chi^K  =   m  \tilde \chi^K  ~, 
\quad 
\chi^K (p_\mu \sigma^\mu)  = -  m \tilde \chi^K~, 
\quad 
 \tilde \chi^K (p_\mu  \bar \sigma^\mu) = - m  \chi^K ~. 
\end{equation}
\cref{eq:SHF_chidef,eq:SHF_chinorm} are invariant under the  $SU(2)$ little group rotation of the spinors:  
$\chi^K \to  [U]^K_{\, L} \chi^L$, 
$\chi_K \to   \chi_L  [U^\dagger]^L_{\, K}$, 
$\chi_K \to  [U^*]_K^{\, L}  \chi_L$. 
where from now on summation over repeated little group indices is implicit. 
In complete analogy to spinor indices, the little group indices can  be raised and lowered by the epsilon tensor: $\chi_K = \epsilon_{KL}\chi^L$, etc.  
 For a real momentum $p$ the spinors $\chi$ and $\tilde \chi$ are related by complex conjugation. For an initial-state particle (positive $p^0$)  we have 
 \begin{equation}
 \label{eq:SHF_complexStar}
  (\chi_{\alpha}^{\, K})^* = \tilde \chi_{\dot \alpha \, K}~, 
  \quad 
 (\tilde \chi_{\dot \alpha \, K})^* =    \chi_{\alpha}^{\, K} ~, 
  \quad 
    (\tilde \chi_{\dot{\alpha}}^{\, K})^* = -  \chi_{\alpha \, K}~, 
      \quad 
      (\chi_{\alpha \, K})^* =    - \tilde \chi_{\dot \alpha}^{\, K} ~,  
 \end{equation}
 while for a final-state particle (negative $p^0$) the signs on the right-hand sides of \cref{eq:SHF_complexStar}  are reversed. 

Let us parametrize particle's 3-momentum as $\boldsymbol{p} = |p| \boldsymbol{n}$, where $\boldsymbol{n} \equiv (\sin \theta \cos \phi,\sin\theta \sin \phi,\cos \theta)$ is the unit 3-vector in the direction of  motion. 
A convenient representation of the corresponding spinors is
\begin{align}
 \label{eq:SHF_spinorExplicit}
\chi^K   = &  { \sqrt{E_p + |p|} + \sqrt{E_p- |p|} \over 2} \eta^K 
+  { \sqrt{E_p + |p|} - \sqrt{E_p- |p|} \over 2}  \zeta^K ~, 
\nnl
\tilde \chi^K   = &  \pm { \sqrt{E_p + |p|} + \sqrt{E_p-|p|} \over 2} \tilde \eta^K 
\pm  { \sqrt{E_p + |p|} - \sqrt{E_p- |p|} \over 2}  \tilde \zeta^K  ~, 
\end{align}
where $E_p = \sqrt{|p|^2+m^2}$, the upper (lower) sign refers to an initial- (final-) state particle, and we introduced
\begin{align}
\eta_\alpha^K  \equiv &  \delta_\alpha^K ~, 
\qquad 
\zeta_\alpha^{K=1} \equiv  \begin{pmatrix} - \cos \theta \\ - e^{i \phi} \sin \theta  \end{pmatrix}~, 
\quad 
\zeta_\alpha^{K=2} \equiv  \begin{pmatrix}  - e^{- i \phi} \sin \theta \\  \cos \theta  \end{pmatrix} ~. 
\end{align}
The representation of \cref{eq:SHF_spinorExplicit} is particularly useful in the non-relativistic limit, because the small-velocity expansion is the same as the expansion in powers of $\zeta$'s. 

Lorentz invariance implies that an amplitude describing scattering of $n$ massive particles with spins $J_i$, $i = 1 \dots n$ must be a scalar function of the relevant momenta $p_i^\mu$ and spinors $\chi_i$, $\tilde \chi_i$. Moreover, little group covariance requires that the amplitude contains exactly $2 J_i$ of the spinors $\chi_i$ or $\tilde \chi_i$ with uncontracted little group indices (spinors with contracted little group indices can be traded for momenta using \cref{eq:SHF_chidef}). 
Initial-state particles will be represented by spinors with raised little group indices, $\chi_{i}^{K_j}$ or $\tilde \chi_{i}^{K_j}$, $j=1 \dots 2 J_i$, while final-state particles will be represented by spinors with lowered little group indices, $\chi_{i \, L_j}$,  $\tilde \chi_{i \, L_j}$. 
The little group indices  corresponding to the same particle are always implicitly symmetrized. 
To reduce clutter, the little group indices are often omitted whenever it does not lead to ambiguities. 

In the representation of \cref{eq:SHF_spinorExplicit}, the little group index of a spinor or a twiddle spinor corresponds to the projection of particle's polarization $J_z$ on the z-axis. 
For example, an initial-state spin 1/2 particle with polarization $J_z=+1/2$ is represented by $\chi^{K=1}$, and  $J_z=-1/2$ is represented by $\chi^{K=2}$.
Likewise, for a final-state particle we use $\chi_{L=1}$ for $J_z = +1/2$, and $\chi_{L=2}$ for $J_z = -1/2$.
For an initial-state spin-1, a pair ($\chi^{K=1}$, $\chi^{K=1}$) represents $J_z=+1$, a pair ($\chi^{K=1}$, $\chi^{K=2}$) represents $J_z=0$ (after symmetrizing the little group index), and  ($\chi^{K=2}$, $\chi^{K=2}$) represents $J_z=-1$. 
And so on. 
Other spin quantization axes can be obtained from \cref{eq:SHF_spinorExplicit} by appropriate little group rotations, e.g. the helicity representation is obtained using the SU(2) rotation matrix $U = \begin{pmatrix}
\cos {\theta\over 2}   &  e^{i \phi }\sin {\theta\over 2}  \\
-e^{-i \phi }\sin {\theta\over 2} & \cos {\theta\over 2}   
\end{pmatrix}$. 

\subsection{Discrete symmetries}

The spinor formalism allows for a transparent description of the discrete symmetries corresponding to the parity ($\boldsymbol{x} \to - \boldsymbol{x}$) and time reversal ($t\to -t$) invariance of the strong interactions.  
The matrix element in \cref{eq:SHF_matrixElement} can be expressed as a functional $F$ of the momenta  $p^\mu$, $p'{}^\mu$ and of the spinor variables $\chi_1$, $\chi_{2}$ associated with these momenta. 
Parity and time reversal invariance can be re-formulated as constraints on the functional $F$.
In the discussion below, we will always assume that ${\cal N}$ and ${\cal N}'$ transform in the same way under parity and time reversal (in particular, they are both parity-even, or both parity-odd). 

The relevant nucleon bi-linears transform under the unitary Hilbert space parity operator 
${\cal P}$ as ${\cal P}[ \bar p \Gamma n(t,\boldsymbol{x})] {\cal P}^{-1} = \eta_\Gamma^P \bar p \Gamma n(t,-\boldsymbol{x})$, 
where the values of $\eta_\Gamma^P$ are collected in \cref{tab:SHF_discrete}. 
Then, parity conservation implies that $F \stackrel{P}{\to}  \eta_\Gamma^P F$ under the following transformation $P$ of the spinors and momenta:  
\begin{align}
\label{eq:SHF_paritySpinor}
\chi_{1 \, \alpha} \stackrel{P}{\leftrightarrow} &  \tilde \chi_1^{\dot{\alpha}}, 
 \qquad 
 \chi_1^{\alpha} \stackrel{P}{\leftrightarrow}  - \tilde \chi_{1 \, \dot{\alpha}}, 
 \qquad
 \chi_{2 \, \alpha} \stackrel{P}{\leftrightarrow}   - \tilde \chi^{\dot{\alpha}}_{2}, 
 \qquad 
 \chi_{2}^{\alpha} \stackrel{P}{\leftrightarrow}   \tilde \chi_{2 \, \dot{\alpha}} , 
 \nnl P^\mu \stackrel{P}{\to} &  (-)^\mu  P^\mu, 
 \qquad 
  q^\mu \stackrel{P}{\to}  (-)^\mu  q^\mu, 
\end{align}
where $P^\mu = p^\mu + p'{}^\mu$, $q^\mu = p^\mu - p'{}^\mu$, 
$(-)^{\mu =0} \equiv 1$, $(-)^{\mu =1,2,3} \equiv -1$,  and the sign difference between $\chi_1$ and $\chi_2$ transformations is due to the former (latter) corresponding to an initial-(final-) state particle.
Little group indices are not displayed in \cref{eq:SHF_paritySpinor}, but implicitly they always match, that is $\chi^K \stackrel{P}{\leftrightarrow}  \tilde \chi^K$, and $\chi_K \stackrel{P}{\leftrightarrow}  \tilde \chi_K$.
It follows for example that $\chi_1 \chi_2 \pm \tilde \chi_1 \tilde \chi_2$ transforms under $P$ into $\pm$ itself, while 
$\chi_1 \sigma^\mu \tilde \chi_2 \stackrel{P}{\to} (-)^\mu \chi_2 \sigma^\mu \tilde \chi_1$.

\begin{table}[]
    \centering
    \begin{tabular}{|c|c|c|c|}
    \hline 
&$\bar p \gamma^\mu n$ & $\bar p \gamma^\mu \gamma_5 n$ & $\bar p \sigma^{\mu\nu}n$  
\\    \hline 
$\eta_\Gamma^P$ & $(-)^\mu$ & $- (-)^\mu$ & $ (-)^\mu(-1)^\nu$ 
\\ \hline 
$\eta_\Gamma^T$ & $(-)^\mu $&  $(-)^\mu$ &   $-(-)^\mu(-)^\nu$ 
\\ \hline 
$\eta_\Gamma^{PT}$ & +1 &  -1 &   -1 
\\ \hline 
    \end{tabular}
    \caption{Transformation properties of the nucleon currents under parity and time reversal. We define $(-)^{\mu =0} \equiv 1$ and $(-)^{\mu =1,2,3} \equiv -1$. }
    \label{tab:SHF_discrete}
\end{table}

For the anti-unitary Hilbert space time-reversal operator ${\cal T}$, 
the action on the nucleon bi-linears is 
${\cal T}[ \bar p \Gamma n(t,\boldsymbol{x}) ]{\cal T}^{-1} = \eta_\Gamma^T \bar p \Gamma n(-t,\boldsymbol{x})$, see \cref{tab:SHF_discrete}. 
Time reversal implies that $F \stackrel{T}{\to}  \eta_\Gamma^T F^*$ under the following transformation $T$ of the spinors and momenta:
\begin{align}
\label{eq:SHF_timeSpinor}
\chi_{i \, \alpha} \stackrel{ T}{\to } &  [\chi_i^{\alpha}]^* , 
\quad 
\chi_i^\alpha \stackrel{T}{\to }   -[\chi_{i \, \alpha}]^* , 
\qquad 
\tilde \chi_{i \, \dot \alpha} \stackrel{ T}{\to }   [\tilde \chi_i^{\dot \alpha}]^* , \quad \tilde \chi_i^{\dot \alpha} \stackrel{ T}{\to }   -[\tilde \chi_{i \, \dot \alpha}]^* , 
\nnl 
P^\mu   \stackrel{T}{\to } &  (-)^\mu P^\mu, \qquad q^\mu   \stackrel{ T}{\to }   (-)^\mu q^\mu, 
\end{align}
for $i=1,2$, and again the matching little group indices are implicit. It follows for example that both $[\chi_1 \chi_2]^*$ and $[\tilde \chi_1 \tilde \chi_2]^*$ transform into their complex conjugates under $T$, while 
$[\chi_1 \sigma^\mu \tilde \chi_2]^* \stackrel{T}{\to} (-)^\mu \chi_1 \sigma^\mu \tilde \chi_2$.  

Finally, we define $G$-parity as ${\cal G} = e^{i \pi \tau^2} {\cal T} {\cal P}$, 
where $ e^{i \pi \tau^2}$ is the 180 degrees isospin rotation transforming 
$(p,n) \to (-n,p)$ and $({\cal N},{\cal N'}) \to (\mp {\cal N'},\pm {\cal N})$. 
Isospin invariance implies that $F \stackrel{G}{\to}  \eta_\Gamma^{PT} F$ under the following transformation $G$ of the spinors and momenta: 
\begin{align}
\label{eq:SHF_tildeG}
\begin{pmatrix} \chi_1^K \\ \chi_{2 \, L} \\    \tilde \chi_1^K \\  \tilde \chi_{2 \, L} \end{pmatrix}
\stackrel{G}{\to}
 \begin{pmatrix} - \chi_{2 \, L} \\  \chi_{1}^K \\  \tilde \chi_{2 \, L} \\  -\tilde \chi_1^K  \end{pmatrix}, 
 \qquad 
 P^\mu \stackrel{G}{\to }  P^\mu, 
 \quad 
q^\mu \stackrel{G}{\to } - q^\mu . 
\end{align}
It follows for example that both $\chi_1 \chi_2$ and $\tilde \chi_1 \tilde \chi_2$ transform into itself under $G$, 
while $\chi_1 \sigma^\mu \tilde \chi_2   \stackrel{G}{\to}    \chi_2 \sigma^\mu \tilde \chi_1$.

\subsection{Relativistic matrix elements}
\label{sec:SHF_relativistic}

Using the spinor formalism,  the most general form of the relativistic nuclear matrix elements in \cref{eq:SHF_matrixElement} consistent with Lorentz, parity, time reversal, and G-parity invariance is
\begin{align}
\hspace{-1cm}\bra{{\cal N'}} \bar p \gamma^\mu n  \ket{{\cal N}}  = &   {\alpha_{V} \over (2 m_{\cal N})^{2J} }  \big ( \chi_1 \chi_2  + \tilde \chi_1 \tilde \chi_2  \big )^{2 J}   P^\mu 
+ {\beta_{V}  \over(2 m_{\cal N})^{2J-1} } \big ( \chi_1 \chi_2  + \tilde \chi_1 \tilde \chi_2  \big )^{2 J-1}  \big ( \chi_1 \sigma^\mu \tilde \chi_2 +  \chi_2 \sigma^\mu \tilde \chi_1 \big )
  + \dots   
\nnl 
\hspace{-1cm}\bra{{\cal N'}} \bar p \gamma^\mu \gamma_5 n  \ket{{\cal N}}  = &
{\beta_{A} \over (2 m_{\cal N})^{2J-1}  }  \big ( \chi_1 \chi_2  + \tilde \chi_1 \tilde \chi_2  \big )^{2J-1}    \big ( \chi_1 \sigma^\mu \tilde \chi_2 -  \chi_2 \sigma^\mu \tilde \chi_1 \big )
+ \dots  
\nnl 
\hspace{-1cm}\bra{{\cal N'}} \bar p \sigma^{\mu\nu} n  \ket{{\cal N}} = & 
 i {\alpha_{T} \over (2 m_{\cal N})^{2J+1}}  \big ( \chi_1 \chi_2  + \tilde \chi_1 \tilde \chi_2  \big )^{2J}   (P^\mu q^\nu - P^\nu q^\mu )
\nnl  + &   {\beta_{T} \over (2 m_{\cal N})^{2J-1} }  \big ( \chi_1 \chi_2  + \tilde \chi_1 \tilde \chi_2  \big )^{2J-1}    \big ( \chi_1 \sigma^{\mu \nu} \chi_2  + \tilde \chi_1\bar \sigma^{\mu \nu}\tilde \chi_2  \big )  
 \nnl  + & 
    i {\gamma_T \over (2 m_{\cal N})^{2J} }  \big ( \chi_1 \chi_2  + \tilde \chi_1 \tilde \chi_2  \big )^{2J-2}   \big ( \chi_1 \chi_2  - \tilde \chi_1 \tilde \chi_2  \big )   \bigg  [  
\big ( \chi_1 \sigma^\mu \tilde \chi_2 - \chi_2 \sigma^\mu \tilde \chi_1 \big ) P^\nu   - (\mu \leftrightarrow \nu)  \bigg ]
\nnl & + \dots   
\label{eq:SHF_relativisticMatrixElements}
\end{align}
Above, the dots stand for terms leading to effects that are  quadratic or higher order in recoil; 
e.g. in the axial matrix element we have 
$\big ( \chi_1 \chi_2  + \tilde \chi_1 \tilde \chi_2  \big )^{2J-3}  ( \chi_1 \chi_2  - \tilde \chi_1 \tilde \chi_2  \big )^2   \big ( \chi_1 \sigma^\mu \tilde \chi_2 -  \chi_2 \sigma^\mu \tilde \chi_1 \big )$, 
$\big ( \chi_1 \chi_2  + \tilde \chi_1 \tilde \chi_2  \big )^{2J-5}  ( \chi_1 \chi_2  - \tilde \chi_1 \tilde \chi_2  \big )^4   \big ( \chi_1 \sigma^\mu \tilde \chi_2 -  \chi_2 \sigma^\mu \tilde \chi_1 \big )$, and so on. 
Other similar terms in the axial matrix element, e.g.  
$\big ( \chi_1 \chi_2  + \tilde \chi_1 \tilde \chi_2  \big )^{2J-2}  ( \chi_1 \chi_2  - \tilde \chi_1 \tilde \chi_2  \big )   \big ( \chi_1 \sigma^\mu \tilde \chi_2 -  \chi_2 \sigma^\mu \tilde \chi_1 \big )$ or 
$\big ( \chi_1 \chi_2  + \tilde \chi_1 \tilde \chi_2  \big )^{2J-1}   \big ( \chi_1 \sigma^\mu \tilde \chi_2 + \chi_2 \sigma^\mu \tilde \chi_1 \big )$
are forbidden by parity invariance. 
The form factors $\alpha_{V,T}$, $\beta_{V,A,T}$, $\gamma_T$ are in principle functions of $q^2$, but for our purpose they can be treated as constants. 
Time reversal invariance dictates that these form factors are all real. 
Eq.~\eqref{eq:SHF_relativisticMatrixElements} is valid for any integer $2 J \geq 0$ with the convention that whenever a naive evaluation of some term leads to spinors in a denominator then this term should be set to zero. 
In particular, for $J=0$ only the $\alpha_{V,T}$ structures survive, while for $J=1/2$ the $\gamma_T$ term should be set to zero. 

As we discussed, \cref{eq:SHF_relativisticMatrixElements} is obtained in the isospin limit, and in particular it does not contain G-parity-odd terms. 
These ``second-class currents" in the nomenclature of Weinberg~\cite{Weinberg:1958ut} are often considered in the literature, and searched for in many experiments.
For completeness, we list here the G-parity-odd contributions to the relativistic vector, axial, and tensor matrix elements that lead to effects of linear order in recoil: 
\begin{align}
\Delta \bra{{\cal N'}} \bar p \gamma^\mu n  \ket{{\cal N}}  = &   {\alpha_{V}' \over (2 m_{\cal N})^{2J} }  \big ( \chi_1 \chi_2  + \tilde \chi_1 \tilde \chi_2  \big )^{2 J}   q^\mu
\nnl + & 
{\beta_V' \over (2 m_{\cal N})^{2J-1} }  \big ( \chi_1 \chi_2  + \tilde \chi_1 \tilde \chi_2  \big )^{2J-2}   \big ( \chi_1 \chi_2  - \tilde \chi_1 \tilde \chi_2  \big )   
\big ( \chi_1 \sigma^\mu \tilde \chi_2 - \chi_2 \sigma^\mu \tilde \chi_1 \big ) ,
\nnl 
\Delta \bra{{\cal N'}} \bar p \gamma^\mu \gamma_5 n  \ket{{\cal N}}  = & 
 {\alpha_{A}' \over (2 m_{\cal N})^{2J} }   \big ( \chi_1 \chi_2  + \tilde \chi_1 \tilde \chi_2  \big )^{2J-1}     \big ( \chi_1 \chi_2  - \tilde \chi_1 \tilde \chi_2  \big ) P^\mu ,
 \nnl \Delta \bra{{\cal N'}} \bar p \sigma^{\mu\nu} n  \ket{{\cal N}}  = & 
    i {\alpha_T'   \over (2 m_{\cal N})^{2J} }  \big ( \chi_1 \chi_2  + \tilde \chi_1 \tilde \chi_2  \big )^{2J-1}  \bigg  [  
\big ( \chi_1 \sigma^\mu \tilde \chi_2 +  \chi_2 \sigma^\mu \tilde \chi_1 \big ) P^\nu  - (\mu \leftrightarrow \nu)     \bigg ]. 
\end{align}

\subsection{Non-relativistic limit}

We now take the non-relativistic limit of Eq.~\eqref{eq:SHF_relativisticMatrixElements}. 
On the right-hand side,  using the $|p| \to 0$ limit of the representation in \cref{eq:SHF_spinorExplicit}, we find the following approximations:
\begin{align}
 \big ( \chi_1 \chi_2  + \tilde \chi_1 \tilde \chi_2  \big )^{2J}   \approx &  (2 m_{\cal N})^{2J}  \delta_{J_z'}^{\, J_z} , 
 \nnl 
  \big ( \chi_1 \chi_2  + \tilde \chi_1 \tilde \chi_2  \big )^{2J-1} 
  \big ( \chi_1 \chi_2  - \tilde \chi_1 \tilde \chi_2  \big ) \approx &  
  -{(2 m_{\cal N})^{2J-1}  \over J} q^m  [{\cal T}_{(J)}^m]_{J_z'}^{\, J_z} , 
\end{align}
\begin{align}
   \big ( \chi_1 \chi_2  + \tilde \chi_1 \tilde \chi_2  \big )^{2J-1} \big ( \chi_1 \sigma^0 \tilde \chi_2 +  \chi_2 \sigma^0 \tilde \chi_1 \big ) \approx & 
    - (2 m_{\cal N})^{2J}  \delta_{J_z'}^{\, J_z} , 
 \nnl 
  \big ( \chi_1 \chi_2  + \tilde \chi_1 \tilde \chi_2  \big )^{2J-1}  \big ( \chi_1 \sigma^k \tilde \chi_2 +  \chi_2 \sigma^k \tilde \chi_1 \big ) \approx &
 - (2 m_{\cal N})^{2J-1}  \big ( P^k  \delta_{J_z'}^{\, J_z}   +   {i \over J}  \epsilon^{klm} q^l  [{\cal T}_{(J)}^m]_{J_z'}^{\, J_z}  \big) , 
\end{align}
\begin{align}
 \big ( \chi_1 \chi_2  + \tilde \chi_1 \tilde \chi_2  \big )^{2J-1}  \big ( \chi_1 \sigma^0 \tilde \chi_2 -   \chi_2 \sigma^0 \tilde \chi_1 \big )  
 \approx &  {(2 m_{\cal N})^{2J-1} \over J}   P^l  [{\cal T}_{(J)}^l]_{J_z'}^{\, J_z}  , 
 \nnl 
 \big ( \chi_1 \chi_2  + \tilde \chi_1 \tilde \chi_2  \big )^{2J-1}   \big (  \chi_1 \sigma^k \tilde \chi_2  -  \chi_2 \sigma^k \tilde \chi_1 \big )  
  \approx &  {(2 m_{\cal N})^{2J} \over J}  [{\cal T}_{(J)}^k]_{J_z'}^{\, J_z}  , 
\end{align}
\begin{align}
 \big ( \chi_1 \chi_2  + \tilde \chi_1 \tilde \chi_2  \big )^{2J-1}  \big (  \chi_1 \sigma^{0k} \chi_2  + \tilde \chi_1\bar \sigma^{0k} \tilde \chi_2 \big  )  \approx & 
 (2 m_{\cal N})^{2J-1} \big [ -  i  q^k  \delta_{J_z'}^{\, J_z}  + {1 \over J} \epsilon^{klm} P^l  [{\cal T}_{(J)}^m]_{J_z'}^{\, J_z} \big ] , 
\nnl 
\big ( \chi_1 \chi_2  + \tilde \chi_1 \tilde \chi_2  \big )^{2J-1}   \big ( \chi_1 \sigma^{ij} \chi_2  + \tilde \chi_1\bar \sigma^{ij}\tilde \chi_2 \big )    \approx &
-{(2 m_{\cal N})^{2J} \over J}  \epsilon^{ijm}  [{\cal T}_{(J)}^m]_{J_z'}^{\, J_z},
\end{align}
\begin{align}
 \big ( \chi_1 \chi_2  + \tilde \chi_1 \tilde \chi_2  \big )^{2J-2}   \big ( \chi_1 \chi_2  - \tilde \chi_1 \tilde \chi_2  \big )   
\big ( \chi_1 \sigma^0 \tilde \chi_2 - \chi_2 \sigma^0 \tilde \chi_1 \big )
   \approx & 0 , 
   \nnl 
    \big ( \chi_1 \chi_2  + \tilde \chi_1 \tilde \chi_2  \big )^{2J-2}   \big ( \chi_1 \chi_2  - \tilde \chi_1 \tilde \chi_2  \big )   
\big ( \chi_1 \sigma^k \tilde \chi_2 - \chi_2 \sigma^k \tilde \chi_1 \big )
  \approx & - (2  m_{\cal N})^{2J-1}  \bigg [    { 1\over 3 }  q^k \delta_{J_z'}^{\, J_z}  +  {1 \over J(2J-1)}  q^l [{\cal T}_{(J)}^{kl} ]_{J_z'}^{\, J_z} \bigg ] , 
\end{align}
up to quadratic terms in recoil. 
On the left-hand side of Eq.~\eqref{eq:SHF_relativisticMatrixElements} we trade the relativistic nucleon fields $N$ for their non-relativistic counterparts $\psi_N$. 
To this end, working in the isospin limit, we make the following replacement of  the left- and right-handed components of $N$:
\begin{align}
\label{eq:SHF_NRsubstitution}
N  = \begin{pmatrix}
N_{L\, \alpha}  \\ N_R^{\dot \alpha}  
\end{pmatrix}
\to 
{1 \over \sqrt 2} 
\begin{pmatrix}
\big [ 1 + i { \boldsymbol{\sigma}\cdot \boldsymbol{\nabla} \over 2 m_N} \big ]_{\alpha \beta }
\psi_{N \, \beta} 
\\
\big [ 1 -   i { \boldsymbol{\sigma}\cdot  \boldsymbol{\nabla} \over 2 m_N} \big ]_{\alpha \beta } \psi_{N \, \beta} 
\end{pmatrix}, 
\end{align}
where $\alpha = 1,2$ is the spinor index. 
Here,  $\psi_{N}$ are  2-component anti-commuting spinor fields containing only particle (and no anti-particle) modes.  
The rationale for this substitution is that, up to quadratic terms in $\boldsymbol{\nabla}/m_N$,  the equation of motion for $\psi_{N}$ is the Schr\"{o}dinger equation, and the kinetic terms of  particle and anti-particle modes are decoupled. 
Up to $\cO(\boldsymbol{\nabla}^2/m_N^2)$ corrections, one can derive the  following non-relativistic approximations for the relativistic nucleon bi-linear currents: 
\begin{align}
\label{eq:SHF_nrVector}
 \bar p  \gamma^0 n  =   &
   \psi_p^\dagger \psi_n    , 
\nnl 
 \bar p  \gamma^k n  =   & 
  -  { i  \over 2 m_N}   \bigg [    \psi_p^\dagger \overleftrightarrow{\nabla}_k  \psi_n 
  + i \epsilon^{klm} {\nabla}_l  \big [ \psi_p^\dagger\sigma^m\psi_n  \big ] \bigg]    ,  
\end{align}
\begin{align}
 \bar p  \gamma^0 \gamma_5 n  =   & 
  -  { i  \over 2 m_N}   \big [  \psi_p^\dagger \boldsymbol{\sigma}\cdot  \overleftrightarrow{\boldsymbol{\nabla}} \psi_n \big]  ,
 \nnl 
  \bar p  \gamma^k \gamma_5 n   =   & \psi_p^\dagger \sigma^k \psi_n    ,   
\end{align}
\begin{align}
\label{eq:EFT_nrTensor}
\bar p \sigma^{0k} n   =   & 
 {1 \over 2 m_N}   \bigg [
  \nabla_k \big [ \psi_p^\dagger \psi_n  \big ]  
  + i \epsilon^{klm}  \psi_p^\dagger\sigma^m   \overleftrightarrow{\nabla}_l   \psi_n  \bigg]    ,   
\nnl 
\bar p \sigma^{ij} n  =   &     \epsilon^{ijk} \psi_p^\dagger \sigma^k \psi_n    ,   
\end{align}
where  $\psi^\dagger   \Gamma \overleftrightarrow{\nabla}  \psi    \equiv   \psi^\dagger   \Gamma \nabla   \psi  - \nabla \psi^\dagger \Gamma \psi$. 
Plugging this on the left-hand side of Eq.~\eqref{eq:SHF_relativisticMatrixElements} and disentagling the non-relativistic matrix elements, we obtain 
\begin{align}
\label{eq:SHF_nrMatixElements}
\langle \psi_p^\dagger \psi_n \rangle = &  2 m_{\cal N}   \delta_{J_z'}^{\, J_z}  M_F  , 
\nnl 
\langle  \psi_p^\dagger \sigma^k \psi_n  \rangle = &  {r \over \sqrt{J(J+1)} }    2 m_{\cal N}    [{\cal T}_{(J)}^k]_{J_z'}^{\, J_z}  M_F  , 
\nnl 
{i \over  2 m_N} \langle \psi_p^\dagger \overleftrightarrow{\nabla}_k \psi_n \rangle 
= & \bigg \{ 
-  P^k  \delta_{J_z'}^{\, J_z}  - i  \beta_{FV} {r A \over \sqrt{J(J+1)} }   \epsilon^{klm} q^l  [{\cal T}_{(J)}^m]_{J_z'}^{\, J_z}  
 \bigg \} M_F   , 
\nnl \hspace{-2cm}
{i \over 2 m_N} \langle  \psi_p^\dagger \boldsymbol{\sigma}\cdot  \overleftrightarrow{\boldsymbol{\nabla}} \psi_n  \rangle = &  
  -  {r \over \sqrt{J(J+1)} }  P^k  [{\cal T}_{(J)}^k]_{J_z'}^{\, J_z}  M_F   , 
\nnl
 {1 \over 2 m_N} \epsilon^{k l m} \langle \psi_p^\dagger \sigma^m \overleftrightarrow{\nabla}_l \psi_n \rangle 
= & 
     \bigg \{  
   i  {r   \over \sqrt{J(J+1)}}  \epsilon^{klm}  P^l  [{\cal T}_{(J)}^m]_{J_z'}^{\, J_z}  
+    A \alpha_{FT} q^k  \delta_{J_z'}^{\, J_z}  
  +  \gamma_{FT}   {A r \over J \sqrt{J(J+1)} } q^l [{\cal T}_{(J)}^{kl} ]_{J_z'}^{\, J_z} 
        \bigg \} M_F  , 
\end{align}
where the parameters $M_F$, $\beta_{FV}$, $\alpha_{FT}$, $\gamma_{FT}$ are related to the form factors in Eq.~\eqref{eq:SHF_relativisticMatrixElements} as 
\begin{align} 
\alpha_{V} - \beta_{V}   = & M_F, \quad 
  \beta_{V}  = -   r A  \sqrt {J \over J+1}   \bigg [ \beta_{FV} + 1 \bigg ] M_F , 
\nnl  
 \beta_{A} = &   r \sqrt {J \over J+1}  M_F,  
 \nnl   
\alpha_T - \beta_T + {\gamma_T \over 3} = & A \big ( \alpha_{FT} + 1 \big ) M_F, 
   \quad
    \beta_{T} =   - r \sqrt {J \over J+1} M_F,  
\quad \gamma_T =      r {2 J - 1 \over \sqrt{J(J+1)} }  A \gamma_{FT} M_F .
\end{align}

\newpage
\section{Subleading corrections to correlations}
\label{app:sub}

In this appendix we present the dependence of the correlations in the mixed Fermi-GT beta decay ($J'=J$) on the Wilson coefficients of the subleading effective Lagrangian in \cref{eq:TH_NRleeyang1}.   
To organize the presentation, each subsection below deals with the contribution of a single Wilson coefficient.  
The left-hand-sides refer to the correlation coefficients defined by \cref{eq:BETA_dGammaTemplate}, and we recall the definition 
$\hat \xi \equiv   |C_V^+|^2   + |C_S^+|^2  +  r^2 (|C_A^+|^2  + |C_T^+|^2)$. 
We are interested in  $\cO(\boldsymbol{q}/m_N)$ effects, that is linear order in recoil, which are suppressed by one power of the nucleon mass $m_N$ or the nuclear mass $m_{\cal N}$.  
We neglect $\cO(\boldsymbol{q}^2/m_N^2)$ and higher order effects, in particular we neglect the contributions quadratic in the Wilson coefficients of  \cref{eq:TH_NRleeyang1}.
In expression containing $\pm$ or $\mp$,   the upper (lower) sign refers to $\beta^-$ ($\beta^+$) transitions. 
These results are new because they include the effects of subleading non-SM Wilson coefficients ($C_P^+$, $C^+_{T1}$, $C^+_{T2}$, $C^+_{T3}$, $C^+_{FT}$) at the linear order in recoil, as well as interference of the subleading Wilson coefficients with the non-SM leading order Wilson coefficients ($C_S^+$, $C_T^+$).   
If the SM is the UV completion of our  EFT, all these Wilson coefficients are zero (ignoring the tiny induced $C_S^+$), and moreover 
$C^+_{FA} = C_A^+$, $C^+_{FV} = C_V^+$, $C^+_{E'} = C^+_E$. 

Moreover, in \cref{sec:phasespace} we also quote the results for another class of recoil corrections to the correlations.
They appear because the recoil corrections to the phase space and the relativistic normalization of the nuclear states result in the overall factor $1 + {  3 E_e  - E_e^{\rm max}    \over  m_{\cal N}}  -   3 {\boldsymbol{k}_e \cdot \boldsymbol{k}_\nu \over E_\nu m_{\cal N}} + \cO(m_{\cal N}^{-2})$ multiplying the differential width.  
The contributions in \cref{sec:phasespace} result from multiplying the $\cO(m_{\cal N}^{-1})$ terms in this expression with the zero-th order correlations discussed in \cref{sec:template}.  These effects are in fact relevant to establish the matching with the results in Ref.~\cite{Holstein:1974zf}.

\newpage 
\subsection{$C_P^+$ Wilson coefficient}
\label{sec:subP}
\begin{equation}
    \cL^{(1)}\supset \frac{i C_P^+}{2m_N}    ( \psi_p^\dagger \sigma^k\psi_n) \nabla_k \big (\bar e_R \nu_L \big ).
\end{equation}
\begin{equation}
\begin{alignedat}{1}
\hat{\xi}  \xi_{b}(E_e) &= 
\frac{r^2}{3} \frac{1}{m_N}\lzs \lzm\frac{E_{e}^{\text{max}}}{E_e}-1\dzm m_e \re \lzm \capp \cppd \dzm \pm  \lzm E_{e}^{\text{max}}-2E_e+\frac{m_e^2}{E_e} \dzm \re \lzm \ctp\cppd  \dzm \dzs,\\
\hat{\xi} \Delta a(E_e) &= 
\frac{r^2}{3} \frac{1}{m_N} \lzs m_e \re\lzm \capp \cppd \dzm \pm (2E_e - \emax) \re \lzm \ctp\cppd  \dzm \dzs,\\
\hat{\xi} a'(E_e) &= 0,\\
\hat{\xi}\Delta A(E_e) &= 
 \frac{1}{3m_N} \bigg[- 3m_e r\sqrt{\frac{J}{J+1}}\re \lzm \cvp\cppd \dzm \pm r\sqrt{\frac{J}{J+1}}\lzm \emax - 4E_e \dzm \re \lzm \csp\cppd  \dzm  \\&+\lzm \emax-E_e\dzm \frac{r^2}{J+1} \re \lzm \ctp\cppd  \dzm\bigg],\\
\hat{\xi} A'(E_e) &= 
\frac{E_e}{2m_N}\lzs \pm 2r\sqrt{\frac{J}{J+1}}\re \lzm \csp\cppd  \dzm+\frac{r^2}{J+1}\re \lzm \ctp\cppd  \dzm  \dzs,\\
\hat{\xi}\Delta B(E_e) &= \frac{1}{3m_N}\bigg[
-3\lzm\frac{\emax}{E_e}-1\dzm m_e r\sqrt{\frac{J}{J+1}}\re \lzm \cvp\cppd  \dzm\\&
\mp r\sqrt{\frac{J}{J+1}}\lzm 3\emax - 4E_e + \frac{m_e^2}{E_e} \dzm \re \lzm \csp\cppd  \dzm 
-\frac{r^2}{J+1}\lzm E_e-\frac{m_e^2}{E_e}\dzm \re \lzm \ctp\cppd  \dzm \bigg],\\
\hat{\xi} B'(E_e) &= 
 \frac{\emax-E_e}{2m_N}\lzs 
\pm 2r\sqrt{\frac{J}{J+1}} \re \lzm \csp\cppd \dzm
-\frac{r^2}{J+1}\re \lzm \ctp\cppd  \dzm  \dzs,\\
\hat{\xi}\Delta D(E_e) &=
\frac{r^2}{J+1} \frac{1}{2m_N} \lzs 
\mp m_e \im \lzm \capp \cppd\dzm
-\emax \im \lzm \ctp\cppd  \dzm  \dzs,\\ 
\hat{\xi}D'(E_e)&=0,\\
\hat{\xi}\Delta \hat{c}(E_e) &= 
\frac{r^2}{2} \frac{1}{m_N} \lzs m_e \re \lzm \capp \cppd \dzm \pm (2E_e - \emax)\re \lzm \ctp\cppd \dzm \dzs,\\
\hat{\xi} c'(E_e)&=0,\\
\hat{\xi} c_1(E_e) &= 
\frac{r^2}{6} \frac{\emax-E_e}{m_N} \lzs \frac{m_e}{E_e} \re \lzm \capp \cppd \dzm \pm  \re \lzm \ctp\cppd  \dzm  \dzs,\\
\hat{\xi} c_2(E_e) &= 
\mp\frac{r^2}{6} \frac{E_e}{m_N}\re \lzm \ctp\cppd  \dzm, \\
\hat{\xi} c_3(E_e) &=
\mp \frac{ r^2}{2} \frac{\emax-E_e }{m_N}\im \lzm \ctp\cppd  \dzm,\\
\hat{\xi} c_4(E_e) &= 
    \mp \frac{ r^2}{2} \frac{\emax-E_e }{m_N}\im \lzm \ctp\cppd  \dzm.
\end{alignedat}
\end{equation}
\newpage
\subsection{$C_M^+$ Wilson coefficient}
\label{app:correlationsWM}
\begin{equation}
    \cL^{(1)}\supset -\frac{ C_M^+ }{2m_N}\epsilon^{ijk}   ( \psi_p^\dagger \sigma^j \psi_n)   
\nabla_i   \big  (\bar e_L \gamma^k  \nu_L \big ).
\end{equation}
\begin{equation}
\label{eq:SUB_CM}
\begin{alignedat}{1}
\hat{\xi} \xi_{b}(E_e) &= 
\frac{2r^2}{3}\frac{1}{m_N}\lzs \pm  \lzm E_e^{\text{max}} - 2E_e + \frac{m_e^2}{E_e} \dzm \re \lzm \capp \cmmd \dzm+\lzm\frac{E_e^{\text{max}}}{E_e}-1\dzm m_e  \re \lzm \ctp \cmmd \dzm  \dzs,\\
\hat{\xi} \Delta a(E_e) &=
\frac{2r^2}{3} \frac{1}{m_N}\lzs \pm \lzm 2E_e - \emax \dzm \re \lzm \capp \cmmd \dzm+m_e \re \lzm \ctp \cmmd \dzm  \dzs,\\
\hat{\xi} a'(E_e) &= 0,\\
\hat{\xi} \Delta A(E_e) &= 
 \frac{1}{m_N}\bigg[
 \mp {2 \over 3}\lzm\emax-E_e\dzm r\sqrt{\frac{J}{J+1}} \re \lzm \cvp \cmmd \dzm
 +\frac{r^2}{J+1} \lzm {5\over 3} E_e - {2 \over 3} \emax \dzm  \re \lzm \capp \cmmd \dzm  \\&
 \pm m_e \frac{r^2}{J+1} \re \lzm \ctp \cmmd \dzm  \bigg],\\
\hat{\xi} A'(E_e) &=  \frac{E_e}{m_N}\lzs 
\mp r \sqrt{J \over J+1} \re \lzm \cvp \cmmd \dzm
-\frac{r^2}{2(J+1)}\re \lzm \capp \cmmd \dzm  \dzs,\\
\hat{\xi} \Delta B(E_e) &= 
\frac{1}{m_N}\bigg[
\pm {2 \over 3} r\sqrt{\frac{J}{J+1}}\lzm E_e  -\frac{m_e^2}{E_e}\dzm \re\lzm \cvp \cmmd \dzm 
\pm \frac{r^2}{J+1}\lzm\frac{\emax}{E_e}-1\dzm m_e \re \lzm \ctp \cmmd \dzm
  \\
&+\frac{r^2}{J+1}\lzm \emax - {5 \over 3} E_e+\frac{2m_e^2}{3 E_e} \dzm \re \lzm \capp \cmmd \dzm \bigg],\\
\hat{\xi} B'(E_e) &= 
 \frac{\emax - E_e}{m_N}\lzs 
\pm r \sqrt{J \over J+1}   \re \lzm \cvp \cmmd\dzm
-\frac{r^2}{2(J+1)} \re \lzm \capp \cmmd \dzm  \dzs,\\
\hat{\xi} \Delta D(E_e) &= 
\frac{1}{2m_N}\bigg[\pm 2r\sqrt{\frac{J}{J+1}}\lzm 2E_e - \emax \dzm \im\lzm  \cvp \cmmd \dzm-\frac{r^2}{J+1}\emax \im\lzm \capp \cmmd \dzm \\&+2m_er \sqrt{\frac{J}{J+1}}\im \lzm  \csp\cmmd \dzm\mp m_e \frac{r^2}{J+1}  \im \lzm \ctp\cmmd  \dzm \bigg],\\
\hat{\xi} D'(E_e) &= 0,\\
\hat{\xi}\Delta \hat{c}(E_e) &= 
 \frac{r^2}{2}\frac{1}{m_N}\lzs 
 \mp \lzm 2E_e - \emax \dzm \re \lzm \capp \cmmd \dzm 
 - m_e \re \lzm \ctp \cmmd \dzm \dzs,\\
 \hat{\xi} c'(E_e)&=0,\\
\hat{\xi} c_1(E_e) &= 
{r^2 \over 6} \frac{\emax-E_e }{m_N}\lzs 
\mp \re \lzm \capp \cmmd \dzm 
- \frac{m_e}{E_e} \re\lzm \ctp \cmmd \dzm \dzs,\\
\hat{\xi} c_2(E_e) &=
\pm {r^2 \over 6} \frac{E_e }{m_N} \re \lzm \capp \cmmd \dzm,\\
\hat{\xi} c_3(E_e) &= 
\mp {r^2 \over 2} \frac{\emax-E_e }{m_N}\im \lzm \capp \cmmd \dzm,\\
\hat{\xi} c_4(E_e) &= 
\pm 
{r^2 \over 2} \frac{\emax-E_e }{m_N}\im \lzm \capp \cmmd \dzm.
\end{alignedat}
\end{equation}
The contribution of another Wilson coefficient $C_{FV}^+$ multiplied by the form factor $\beta_{FV}$ in \cref{eq:BETA1_dFGTmatrixelements} is exactly the same as that of $C_M^+$. The joint effect can be described by replacing above 
$C_M^+ \to C^+_{WM} \equiv C_M^+ + \beta_{FV} C^+_{FV}$. 

\newpage
\subsection{$C_E^+$ Wilson coefficient}
\label{app:correlationsIT}
\begin{equation}
    \cL^{(1)}\supset -\frac{i C_E^+}{2m_N}   ( \psi_p^\dagger \sigma^k\psi_n) \nabla_k \big (\bar e_L \gamma^{0} \nu_L \big ).
\end{equation}
\begin{equation}
\begin{alignedat}{1}
\hat{\xi} \xi_{b}(E_e) &= 
 \frac{r^2}{3}\frac{1}{m_N}\lzs 
 \mp \lzm E_e^{\text{max}}-\frac{m_e^2}{E_e} \dzm \re \lzm \capp \ceed \dzm 
 - m_e \lzm\frac{E_e^{\text{max}}}{E_e} - 1\dzm \re \lzm \ctp \ceed \dzm \dzs,\\
\hat{\xi}\Delta a(E_e) &= 
 \frac{r^2}{3}\frac{1}{m_N}\lzs 
 \mp\emax\re \lzm \capp \ceed \dzm 
 - m_e \re \lzm \ctp \ceed \dzm \dzs,\\
\hat{\xi} a'(E_e) &= 0,\\
\hat{\xi} \Delta A(E_e) &= 
\frac{1}{3m_N}\bigg[ 
\pm r\sqrt{\frac{J}{J+1}}\lzm \emax+2E_e \dzm \re\lzm \cvp \ceed \dzm
+\lzm\emax-E_e\dzm \frac{r^2}{J+1} \re \lzm \capp \ceed \dzm \\& 
+3m_er \sqrt{\frac{J}{J+1}}\re \lzm \csp \ceed \dzm \bigg],\\
\hat{\xi} A'(E_e) &= 
\frac{E_e}{2m_N}\lzs 
\pm 2r\sqrt{\frac{J}{J+1}} \re\lzm \cvp \ceed \dzm+\frac{r^2}{J+1}\re\lzm \capp \ceed \dzm  \dzs,\\
\hat{\xi} \Delta B(E_e) &=  
\frac{1}{3m_N}\bigg[\pm r\sqrt{\frac{J}{J+1}} \lzm 3\emax - 2E_e - \frac{m_e^2}{E_e} \dzm \re\lzm \cvp \ceed \dzm - \frac{r^2}{J+1}\lzm E_e - \frac{m_e^2}{E_e}\dzm \re \lzm \capp \ceed  \dzm  \\&+ 3r\sqrt{\frac{J}{J+1}} \lzm\frac{\emax}{E_e}-1\dzm m_e \re \lzm \csp \ceed \dzm \bigg],\\
\hat{\xi} B'(E_e) &= 
 \frac{\emax-E_e}{2m_N}\lzs 
\pm 2r\sqrt{\frac{J}{J+1}}\re \lzm \cvp \ceed\dzm
-\frac{r^2}{J+1}\re\lzm \capp \ceed \dzm  \dzs,\\
\hat{\xi} \Delta D(E_e) &= \frac{r^2}{J+1}\frac{1}{2m_N}\lzs \lzm 2E_e-\emax  \dzm \im\lzm \capp \ceed \dzm \pm m_e \im \lzm \ctp \ceed \dzm \dzs,\\
\hat{\xi} D'(E_e)&= 0,\\
\hat{\xi}\Delta \hat{c}(E_e) &= 
 \frac{r^2}{2}\frac{1}{m_N}\lzs 
 \mp\emax \re \lzm \capp \ceed \dzm 
 - m_e \re\lzm \ctp \ceed \dzm \dzs, \\
 \hat{\xi} c'(E_e)&=0,\\
\hat{\xi} c_1(E_e) &= \frac{r^2}{6} \frac{\emax-E_e}{m_N}\lzs 
\mp \re \lzm \capp \ceed \dzm 
- \frac{m_e}{E_e} \re \lzm \ctp \ceed \dzm \dzs,\\
\hat{\xi} c_2(E_e) &= \mp \frac{r^2}{6} \frac{E_e}{m_N}\re \lzm \capp \ceed \dzm,\\
\hat{\xi} c_3(E_e) &= \mp \frac{r^2}{2}\frac{\emax-E_e}{m_N} \im \lzm \capp \ceed \dzm,\\
\hat{\xi} c_4(E_e)&=\mp \frac{r^2}{2}\frac{\emax-E_e}{m_N} \im \lzm \capp \ceed \dzm.
\end{alignedat}
\end{equation}
\normalsize

\newpage 
\subsection{$C^+_{E'}$ Wilson coefficient}
\label{app:correlationsITprime}

\begin{equation}
{\cal L}^{(1)} \supset     
- {i C^+_{E'} \over 2 m_N}  ( \psi_p^\dagger \sigma^k\psi_n) \partial_t \big (\bar e_L \gamma^k \nu_L \big ) . 
\end{equation}
\begin{equation}
\begin{alignedat}{1}
\hat{\xi}  \xi_{b}(E_e) & =  
{ E_e^{\rm max}  \over m_N} r^2 \lzs \pm \re\lzm C_A^+ \bar{C}^+_{E'} \dzm+\frac{m_e}{E_e}\re\lzm  C_T^+ \bar{C}^+_{E'} \dzm   \dzs 
, \\  
\hat{\xi}  \Delta a(E_e) &=  
\mp  { E_e^{\rm max} \over  3 m_N} r^2 \re \lzm C_A^+ \bar{C}^+_{E'} \dzm  
, \\  
\hat{\xi}a'(E_e)  &=   0  
, \\  
\hat{\xi}  \Delta A(E_e) & = {E_e^{\rm max} \over m_N} \lzs 
\mp r \sqrt{J \over J+1} \re \lzm C_V^+ \bar{C}^+_{E'} \dzm  
- {r^2 \over J+1}\re \lzm C_A^+ \bar{C}^+_{E'} \dzm \dzs 
, \\
\hat{\xi}A'(E_e)  &=   0  
, \\ 
\hat{\xi}  \Delta B(E_e) & = {E_e^{\rm max} \over m_N} \bigg [ 
\mp r \sqrt{J \over J+1} \re \lzm C_V^+ \bar{C}^+_{E'} \dzm  
+ {r^2 \over J+1}\re \lzm C_A^+ \bar{C}^+_{E'} \dzm 
- r \sqrt{J \over J+1} {m_e \over E_e} \re \lzm C_S^+ \bar{C}^+_{E'} \dzm\\ & 
\pm {r^2 \over J+1}  {m_e \over E_e} \re \lzm C_T^+ \bar{C}^+_{E'} \dzm 
\bigg ] 
, \\ 
\hat{\xi}B'(E_e)  &=   0  
, \\ 
\hat{\xi}  \Delta D(E_e) & =  \mp {E_e^{\rm max} \over m_N} r \sqrt{J \over J+1}  \im \lzm C_V^+ \bar{C}^+_{E'} \dzm  
, \\ 
D'(E_e)  &=   0  
, \\ 
\hat{\xi}  \Delta \hat c(E_e) & =\pm  {E_e^{\rm max} \over m_N} r^2  \re \lzm C_A^+ \bar{C}^+_{E'} \dzm 
, \\ 
\hat{\xi}c'(E_e)  &=0,\\
\hat{\xi} c_1(E_e) &=0,\\
\hat{\xi} c_2(E_e) &=0,\\
\hat{\xi} c_3(E_e) &=0,\\
\hat{\xi} c_4(E_e) &=0.
\end{alignedat} 
\end{equation}

\newpage
\subsection{$C_{T1}^+$ Wilson coefficient}
\begin{equation}
    \cL^{(1)}\supset -\frac{i  C_{T1}^+}{2m_N}  ( \psi_p^\dagger \psi_n) 
\nabla_k \big  (\bar e_R \gamma^0 \gamma^k \nu_L  \big ).
\end{equation}
\begin{equation}
\label{eq:SUB_CT1}
\begin{alignedat}{1}
\hat{\xi} \xi_{b}(E_e) &= \frac{1}{m_N}\lzs \lzm\frac{E_e^{\text{max}}}{E_e} - 1\dzm m_e \re\lzm \cvp \Bar{C}^+_{T1}  \dzm \pm \lzm E_e^{\text{max}} - 2E_e + \frac{m_e^2}{E_e} \dzm \re\lzm \csp \Bar{C}^+_{T1} \dzm \dzs,\\
\hat{\xi}\Delta a(E_e) &= \frac{1}{m_N}\lzs m_e \re\lzm \cvp \Bar{C}^+_{T1} \dzm \pm \lzm 2E_e - \emax \dzm \re\lzm \csp \Bar{C}^+_{T1} \dzm \dzs,\\
\hat{\xi} a'(E_e)&=0,\\
\hat{\xi}\Delta A(E_e) &=  -
\frac{1}{m_N} r \sqrt{J \over J+1} \lzs m_e \re\lzm \capp \Bar{C}^+_{T1} \dzm \pm \lzm 2E_e - \emax \dzm \re\lzm \ctp \Bar{C}^+_{T1} \dzm \dzs,\\
\hat{\xi} A'(E_e) &= 0,\\
\hat{\xi}\Delta B(E_e) &=
-\frac{1}{m_N} r \sqrt{J \over J+1}  \lzs \lzm\frac{\emax}{E_e}-1\dzm m_e \re \lzm \capp \Bar{C}^+_{T1} \dzm  \pm \lzm \emax - 2E_e + \frac{m_e^2}{E_e} \dzm \re\lzm \ctp \Bar{C}^+_{T1} \dzm \dzs ,\\
\hat{\xi} B'(E_e) &= 0,\\
\hat{\xi}\Delta D(E_e) &=
\frac{1}{m_N} r \sqrt{J \over J+1} \lzs m_e \im\lzm \capp \Bar{C}^+_{T1} \dzm 
\pm  \lzm 2E_e - \emax \dzm\im\lzm \ctp\bar{C}^+_{T1} \dzm \dzs,\\
\hat{\xi} D'(E_e)&=0,\\
\hat{\xi}\Delta \hat{c}(E_e) &= 0,\\
\hat{\xi} c'(E_e)&=0,\\
\hat{\xi} c_1(E_e) &= 0,\\
\hat{\xi} c_2(E_e) &= 0,\\
\hat{\xi} c_3(E_e) &= 0,\\
\hat{\xi} c_4(E_e) &= 0.
\end{alignedat}
\end{equation}

The effects proportional to $\alpha_{FT} C_{FT}^+$ entering via the subleading tensor matrix element in \cref{eq:BETA1_dFGTmatrixelements} have the same functional form as in \cref{eq:SUB_CT1}. They can be obtained via the replacement 
$C_{T1}^+ \to C_{T1}^+ - \alpha_{FT} C_{FT}^+$ in \cref{eq:SUB_CT1}.

\newpage
\subsection{$C^+_{T2}$ Wilson coefficient}
\begin{equation}
    \cL^{(1)}\supset \frac{i C^+_{T2}}{2 m_N}   (\psi_p^\dagger \psi_n)  
(\bar e_R  \overleftrightarrow{\partial_t} \nu_L).
\end{equation}
\begin{equation}
\begin{alignedat}{1}
\hat{\xi} \xi_{b}(E_e) &= 
\frac{\emax - 2 E_e}{m_N} \bigg[ 
 \frac{m_e}{E_e}  \re \lzm \cvp\bar{C}^+_{T2}  \dzm
\pm \re \lzm \csp\bar{C}^+_{T2} \dzm    \bigg]
,\\
\hat{\xi}\Delta a(E_e) &= 
\pm \frac{2E_e - \emax}{m_N}\re \lzm \csp\bar{C}^+_{T2} \dzm
,\\
\hat{\xi} a'(E_e)&=0
,\\
\hat{\xi}\Delta A(E_e) &=
\pm r\sqrt{\frac{J}{J+1}}\frac{ \emax - 2 E_e}{m_N} \re \lzm \ctp\bar{C}^+_{T2}   \dzm
,\\
\hat{\xi} A'(E_e) &= 0
,\\
\hat{\xi}\Delta B(E_e) &= 
r\sqrt{\frac{J}{J+1}}\frac{2 E_e - \emax}{m_N} \lzs 
 \frac{m_e}{E_e}\re \lzm \capp \Bar{C}^+_{T2} \dzm
\pm \re \lzm \ctp\bar{C}^+_{T2}  \dzm  \dzs
,\\
            \hat{\xi} B'(E_e) &= 0,\\
\hat{\xi}\Delta D(E_e) &= 
\pm r\sqrt{\frac{J}{J+1}}\frac{2E_e - \emax}{m_N} \im \lzm  \ctp\bar{C}^+_{T2}  \dzm
,\\
            \hat{\xi} D'(E_e)&=0,\\
            \hat{\xi}\Delta \hat{c}(E_e) &= 0,\\
            \hat{\xi} c'(E_e)&=0,\\
            \hat{\xi} c_1(E_e) &=0 ,\\
            \hat{\xi} c_2(E_e) &=0 ,\\
            \hat{\xi} c_3(E_e) &=0 ,\\
            \hat{\xi} c_4(E_e) &=0 .\\
\end{alignedat}
\end{equation}
\newpage
\subsection{$C^+_{T3}$ Wilson coefficient}
\begin{equation}
    \cL^{(1)}\supset \frac{i C^+_{T3}}{m_N} (\psi_p^\dagger \sigma^k \psi_n)   
(\bar e_R  \overleftrightarrow{\nabla}_k \nu_L).
\end{equation}
\begin{equation}
\begin{alignedat}{1}
\hat{\xi} \xi_{b}(E_e) &= 
\frac{2r^2}{3}\frac{1}{m_N} \lzs 
m_e\lzm\frac{E_e^{\text{max}}}{E_e} - 1\dzm \re \lzm \capp \Bar{C}^+_{T3} \dzm
\pm \lzm E_e^{\text{max}} -\frac{m_e^2}{E_e}\dzm\re \lzm \ctp \Bar{C}^+_{T3} \dzm   \dzs
,\\
\hat{\xi}\Delta a(E_e) &= 
- \frac{2r^2}{3}\frac{1}{m_N} \lzs m_e \re\lzm \capp \Bar{C}^+_{T3} \dzm 
\mp \emax\re \lzm \ctp\bar{C}^+_{T3}  \dzm \dzs
,\\
\hat{\xi} a'(E_e)&=0
,\\
\hat{\xi}\Delta A(E_e) &= 
\frac{1}{m_N}\bigg[ 2r\sqrt{\frac{J}{J+1}}m_e\re \lzm \cvp\bar{C}^+_{T3}\dzm
\pm \frac{2r}{3}\sqrt{\frac{J}{J+1}}\lzm 2E_e+\emax \dzm\re \lzm \csp\bar{C}^+_{T3}  \dzm  \\&
+ \frac{2 r^2}{3(J+1)} \lzm \emax-E_e \dzm\re \lzm \ctp\bar{C}^+_{T3}  \dzm \bigg]
,\\
\hat{\xi} A'(E_e) &= \frac{1}{m_N}\bigg[
\mp 2r\sqrt{\frac{J}{J+1}}E_e \re \lzm \csp\bar{C}^+_{T3}  \dzm 
- \frac{r^2}{ J+1} E_e\re \lzm \ctp\bar{C}^+_{T3}  \dzm
\bigg]
,\\
\hat{\xi}\Delta B(E_e) &=\frac{1}{m_N}\bigg[
- 2r\sqrt{\frac{J}{J+1}} m_e\frac{\emax-E_e}{E_e}  \re \lzm \cvp\bar{C}^+_{T3}  \dzm  
\\&
\mp 2r\sqrt{\frac{J}{J+1}}  \frac{3E_e\emax - 2E_e^2 - m_e^2}{3E_e } \re \lzm \csp\bar{C}^+_{T3}  \dzm
+\frac{2 r^2}{3 (J+1)} \frac{E_e^2-m_e^2}{E_e } \re \lzm \ctp\bar{C}^+_{T3} \dzm
\bigg]
,\\
\hat{\xi} B'(E_e) &= \frac{1}{m_N}\bigg[
\pm 2r\sqrt{\frac{J}{J+1}}\lzm \emax-E_e \dzm\re \lzm \csp\bar{C}^+_{T3}  \dzm 
-\frac{ r^2}{J+1} \lzm \emax-E_e \dzm \re \lzm \ctp\bar{C}^+_{T3}  \dzm
\bigg]
,\\
\hat{\xi}\Delta D(E_e) &= 
\pm\frac{r^2}{J+1}\frac{1}{m_N}\lzs m_e \im \lzm \capp \Bar{C}^+_{T3} \dzm 
+ (2E_e - \emax)\im \lzm \ctp\bar{C}^+_{T3}  \dzm  \dzs
,\\
\hat{\xi} D'(E_e)&=0
,\\
\hat{\xi}\Delta \hat{c}(E_e) &= 
\frac{r^2}{m_N}\lzs 
- m_e \re \lzm \capp \Bar{C}^+_{T3} \dzm 
\mp \emax\re \lzm \ctp\bar{C}^+_{T3}  \dzm \dzs
,\\
\hat{\xi} c'(E_e)&=0
,\\
\hat{\xi} c_1(E_e) &= 
\frac{ r^2}{3} \frac{\emax-E_e}{m_N}\lzs 
\frac{m_e}{E_e} \re \lzm \capp \Bar{C}^+_{T3} \dzm
\pm\re \lzm \ctp\bar{C}^+_{T3}  \dzm  \dzs
,\\
\hat{\xi} c_2(E_e) &= 
\pm\frac{r^2}{3} \frac{E_e}{m_N} \re \lzm \ctp\bar{C}^+_{T3}  \dzm
,\\
\hat{\xi} c_3(E_e) &= 
\mp r^2 \frac{\emax-E_e}{m_N} \im \lzm \ctp\bar{C}^+_{T3}  \dzm
,\\
\hat{\xi} c_4(E_e) &= 
\pm r^2 \frac{\emax-E_e}{m_N} \im \lzm \ctp\bar{C}^+_{T3}  \dzm.
\end{alignedat}
\end{equation}

\newpage
\subsection{$C^+_{FV}$ Wilson coefficient}
\begin{equation}
    \cL^{(1)}\supset -\frac{i C^+_{FV}}{2m_N}  (\psi_p^\dagger  \overleftrightarrow{\nabla}_k \psi_n)   
 (\bar e_L \gamma^k \nu_L).
\end{equation}
We start with the contribution due to the first term in the 
$\langle \psi_p^\dagger  \overleftrightarrow{\nabla}_k \psi_n \rangle$ matrix element in \cref{eq:BETA1_dFGTmatrixelements}. 
\begin{equation}
\begin{alignedat}{1}
\hat{\xi} \xi_{b}(E_e) &= 
\frac{1}{m_\mathcal{N}}\lzs 
 \lzm E_e^{\text{max}} - \frac{m_e^2}{E_e}\dzm\re\lzm \cvp\cfvpd\dzm
\pm \lzm\frac{E_e^{\text{max}}}{E_e}-1\dzm m_e\re\lzm \csp\cfvpd\dzm  \dzs,\\
\hat{\xi}\Delta a(E_e) &= 
\frac{1}{m_\mathcal{N}}\lzs 
 E_e^{\text{max}} \re\lzm \cvp \cfvpd \dzm 
\pm m_e\re\lzm \csp\cfvpd\dzm\dzs,\\
\hat{\xi} a'(E_e)&=0,\\
\hat{\xi}\Delta A(E_e) &= 
 -r \sqrt{J \over J+1} \frac{1}{m_\mathcal{N}}\lzs 
E_e^{\text{max}}\re\lzm \capp\cfvpd\dzm 
\pm m_e\re\lzm \ctp\cfvpd\dzm\dzs,\\
            \hat{\xi} A'(E_e) &= 0,\\
\hat{\xi}\Delta B(E_e) &=
 -r \sqrt{J \over J+1} \frac{1}{m_\mathcal{N}}\lzs 
 \lzm E_e^{\text{max}} - \frac{m_e^2}{E_e}\dzm\re\lzm\capp\cfvpd\dzm 
\pm \lzm\frac{E_e^{\text{max}}}{E_e} - 1\dzm m_e\re\lzm\ctp\cfvpd\dzm\dzs,\\
            \hat{\xi} B'(E_e) &= 0,\\
\hat{\xi}\Delta D(E_e) &= 
 r \sqrt{J \over J+1}
\frac{1}{m_\mathcal{N}}\lzs 
 E_e^{\text{max}}\im\lzm \capp\cfvpd\dzm 
\pm m_e\im\lzm \ctp\cfvpd\dzm\dzs,\\
\hat{\xi} D'(E_e)&=0,\\
            \hat{\xi}\Delta \hat{c}(E_e) &= 0,\\
            \hat{\xi} c'(E_e)&=0,\\
            \hat{\xi} c_1(E_e) &=0 ,\\
            \hat{\xi} c_2(E_e) &=0 ,\\
           \hat{\xi} c_3(E_e) &=0 ,\\
            \hat{\xi} c_4(E_e) &=0 .
\end{alignedat}
\end{equation}
The contribution proportional to  $\beta_{FV} C_{FV}^+$ due to the second term in the $\langle \psi_p^\dagger  \overleftrightarrow{\nabla}_k \psi_n \rangle$ matrix element in \cref{eq:BETA1_dFGTmatrixelements} can be obtained from \cref{eq:SUB_CM} via the replacement $C_M^+ \to  C_M^+ + \beta_{FV} C^+_{FV}$. 

\newpage

\subsection{$C^+_{FA}$ Wilson coefficient}
\begin{equation}
    \cL^{(1)}\supset \frac{i C^+_{FA}}{2m_N}    (\psi_p^\dagger \sigma^k  \overleftrightarrow{\nabla}_k \psi_n)    (\bar e_L \gamma^{0} \nu_L).
\end{equation}
\begin{equation}
\begin{alignedat}{1}
\hat{\xi} \xi_{b}(E_e) &= 
\frac{r^2}{3}\frac{1}{m_\mathcal{N}}\lzs 
 \lzm E_e^{\text{max}} - \frac{m_e^2}{E_e}\dzm\re\lzm \capp\cfapd\dzm \pm m_e\lzm \frac{E_e^{\text{max}}}{E_e} - 1\dzm\re\lzm \ctp\cfapd\dzm\dzs,\\
\hat{\xi}\Delta a(E_e) &= 
\frac{r^2}{3}\frac{1}{m_\mathcal{N}}\lzs 
  E_e^{\text{max}}\re\lzm \capp\cfapd\dzm 
\pm  m_e\re\lzm \ctp\cfapd\dzm\dzs,\\
\hat{\xi}a'(E_e)&=0,\\
\hat{\xi}\Delta A(E_e) &= 
\frac{1}{m_\mathcal{N}}\bigg[ 
- r\sqrt{\frac{J}{J+1}} \frac{2E_e + \emax}{3}\re\lzm \cvp\cfapd\dzm
\mp \frac{r^2}{3(J+1)}(\emax-E_e)\re\lzm\capp\cfapd\dzm  \\&
\mp r\sqrt{\frac{J}{J+1}}m_e\re\lzm \csp\cfapd\dzm\bigg],\\
\hat{\xi} A'(E_e) &=  \frac{E_e}{m_\mathcal{N}} \lzs 
-  r \sqrt{J \over J+1}\re\lzm \cvp\cfapd\dzm
\mp  {r^2 \over 2(J+1)} \re\lzm \capp\cfapd\dzm \dzs ,\\
\hat{\xi}\Delta B(E_e) &= \frac{1}{m_\mathcal{N}}\bigg[
- r \sqrt{\frac{J}{J+1}}\frac{3E_e\emax - 2E_e^2-m_e^2}{3E_e}\re\lzm \cvp\cfapd\dzm
\pm \frac{r^2 }{3 (J+1)} \frac{E_e^2-m_e^2}{E_e} \re\lzm \capp\cfapd\dzm \\& 
\mp r\sqrt{\frac{J}{J+1}}\frac{m_e(\emax-E_e)}{E_e}\re\lzm \csp\cfapd\dzm \bigg],\\
\hat{\xi} B'(E_e) &=  \frac{\emax-E_e}{m_\mathcal{N}} \lzs 
- r \sqrt{J \over J+1} \re\lzm\cvp\cfapd\dzm 
\pm {r^2 \over 2(J+1)} \re\lzm\capp\cfapd\dzm  \dzs,\\
\hat{\xi}\Delta D(E_e) &=
\frac{r^2}{2(J+1)} \frac{1}{m_\mathcal{N}} \lzs
\pm \lzm \emax-2E_e  \dzm\im\lzm \capp \cfapd\dzm 
- m_e\im\lzm \ctp\cfapd\dzm\dzs,\\
\hat{\xi}D'(E_e)&=0,\\
\hat{\xi}\Delta \hat{c}(E_e) &= \frac{r^2}{2} \frac{1}{m_\mathcal{N}}\lzs E_e^{\text{max}}\re\lzm \capp\cfapd\dzm 
\pm m_e\re\lzm \ctp\cfapd\dzm\dzs,\\
\hat{\xi}c'(E_e)&=0,\\
\hat{\xi} c_1(E_e) &= 
  \frac{ \emax-E_e }{m_\mathcal{N}} {r^2 \over 6} \lzs 
  \re\lzm \capp\cfapd\dzm 
  \pm \frac{m_e}{E_e}\re\lzm \ctp\cfapd\dzm\dzs ,\\
\hat{\xi} c_2(E_e) &= 
 \frac{E_e}{m_\mathcal{N}} {r^2 \over 6} \re \lzm \capp \cfapd \dzm ,\\
\hat{\xi} c_3(E_e) &=
  \frac{\emax-E_e}{m_\mathcal{N}}  {r^2 \over 2} \im\lzm\capp \cfapd\dzm ,\\
\hat{\xi} c_4(E_e) &= 
  \frac{\emax-E_e}{m_\mathcal{N}} {r^2 \over 2} \im\lzm\capp \cfapd\dzm.
\end{alignedat}
\end{equation}

\newpage
\subsection{$C^+_{FT}$ Wilson coefficient}
\label{app:correlationsFT}
\begin{equation}
\cL^{(1)}\supset \frac{C^+_{FT}}{2m_N}  \epsilon^{ijk} (\psi_p^\dagger \sigma^i  \overleftrightarrow{\nabla}_j \psi_n)   (\bar e_R \gamma^{0} \gamma^k \nu_L).
\end{equation}
We start with the contribution due to the first term in the $\langle \psi_p^\dagger \sigma^{i}  \overleftrightarrow{\nabla}_j \psi_n \rangle$ matrix element in \cref{eq:BETA1_dFGTmatrixelements}. 
\begin{equation}
\begin{alignedat}{1}
\hat{\xi} \xi_{b}(E_e) &= 
\frac{2r^2}{3} \frac{1}{m_{\cal N} }\lzs
- \lzm\frac{E_e^{\text{max}}}{E_e}-1\dzm m_e \re\lzm\capp\cftpd\dzm 
\mp \lzm E_e^{\text{max}}- \frac{m_e^2}{E_e}\dzm \re\lzm\ctp\cftpd\dzm\dzs,
\\
\hat{\xi}\Delta a(E_e) &= 
\frac{2r^2}{3} \frac{1}{m_{\cal N} } \lzs 
- m_e\re\lzm\capp\cftpd\dzm 
\mp E_e^{\text{max}}\re\lzm \ctp\cftpd\dzm \dzs,
\\
\hat{\xi}a'(E_e)&=0,
\\
\hat{\xi}\Delta A(E_e) &= 
 \frac{1}{m_{\cal N} } \bigg[
\mp \frac{r^2}{J+1} m_e\re\lzm\capp\cftpd\dzm 
\mp \frac{2}{3}(\emax-E_e)r\sqrt{\frac{J}{J+1}}\re\lzm\csp\cftpd\dzm \\&
- \frac{r^2}{J+1}\frac{E_e + 2\emax}{3}\re\lzm\ctp\cftpd\dzm \bigg],
\\
\hat{\xi} A'(E_e) &= 
\frac{E_e}{m_{\cal N} } \lzs  
\mp r\sqrt{\frac{J}{J+1}}\re\lzm\csp\cftpd\dzm
- \frac{r^2}{2(J+1)}\re\lzm\ctp\cftpd\dzm \dzs,
\\
\hat{\xi}\Delta B(E_e) &=  
\frac{1}{m_{\cal N}} \bigg[
\mp \frac{r^2}{J+1} \frac{m_e}{E_e} (\emax-E_e) \re\lzm\capp\cftpd\dzm 
\pm \frac{2(E_e^2-m_e^2)}{3E_e} r\sqrt{\frac{J}{J+1}}\re\lzm\csp\cftpd\dzm
\\& - \frac{ 3E_e\emax  - E_e^2- 2m_e^2}{3E_e} \frac{r^2}{J+1}\re\lzm\ctp\cftpd\dzm\bigg],
\\
\hat{\xi} B'(E_e) &= 
\frac{1}{2} \frac{\emax-E_e}{m_N} \lzs
\pm 2r\sqrt{\frac{J}{J+1}}\re\lzm\csp\cftpd\dzm
-\frac{r^2}{J+1}\re\lzm\ctp\cftpd\dzm \dzs,
\\
\hat{\xi}\Delta D (E_e) &=  
 \frac{1}{m_{\cal N}} \bigg[
 - m_e r\sqrt{\frac{J}{J+1}}\im\lzm\cvp\cftpd\dzm 
 - { \emax - 2E_e \over 2 } \frac{r^2}{J+1} \im\lzm\ctp\cftpd \dzm 
 \\&
\pm {m_e \over 2} \frac{r^2}{J+1} \im\lzm\capp\cftpd\dzm 
\mp E_e^{\text{max}}r\sqrt{\frac{J}{J+1}}\im\lzm\csp\cftpd\dzm
             \bigg],
\\
\hat{\xi}D'(E_e)&=0,
\\
\hat{\xi}\Delta \hat{c}(E_e) &=
r^2 \frac{1}{2 m_{\cal N} } \lzs m_e\re\lzm\capp\cftpd\dzm \pm E_e^{\text{max}}\re\lzm\ctp\cftpd\dzm\dzs,
\\
\hat{\xi}c'(E_e)&=0,
\\
\hat{\xi} c_1(E_e) &=
 \frac{r^2}{6} \frac{\emax-E_e}{m_{\cal N} } \lzs \frac{m_e}{E_e}\re\lzm \capp\cftpd\dzm \pm  \re\lzm \ctp \cftpd\dzm \dzs ,\\
\hat{\xi} c_2(E_e) &=
\pm \frac{r^2}{6}\frac{E_e }{m_{\cal N} }\re\lzm \ctp \cftpd\dzm ,
\\
\hat{\xi} c_3(E_e) & =
\mp \frac{r^2}{2} \frac{\emax-E_e}{m_{\cal N} }\im\lzm \ctp\cftpd \dzm ,
\\
\hat{\xi} c_4(E_e) &= 
\pm \frac{r^2}{2} \frac{\emax-E_e}{m_{\cal N} } \im\lzm \ctp\cftpd \dzm.
\end{alignedat}
\end{equation}

The contributions proportional to $\alpha_{FT}$ due to the second term in the $\langle \psi_p^\dagger \sigma^{i}  \overleftrightarrow{\nabla}_j \psi_n \rangle$ matrix element in \cref{eq:BETA1_dFGTmatrixelements} can be obtained via the replacement 
$C_{T1}^+ \to C_{T1}^+ - \alpha_{FT} C_{FT}^+$ in \cref{eq:SUB_CT1}. 

We move to the contributions proportional to $\gamma_{FT}$ due to the last term in the $\langle \psi_p^\dagger \sigma^{i}  \overleftrightarrow{\nabla}_j \psi_n \rangle$ matrix element in \cref{eq:BETA1_dFGTmatrixelements}. 
These have to be treated separately because the resulting correlations do not fit into the template in \cref{eq:BETA_dGammaTemplate}. 
Instead,  they induce the following additional correlations in the differential  decay distribution: 
\begin{align}
\label{eq:BETA_FTdeltaGammaTemplate}
\Delta { d \Gamma \over  d E_e d \Omega_e   d \Omega_\nu  } = & 
 M_F^2 F(Z,E_e) (1 + \delta_R) {p_e E_e (E_e^{\rm max} - E_e)^2   \over 64 \pi^5  } 
 \hat \xi   \bigg \{ 
 \nnl & 
 \bigg  ( A''(E_e)  {\boldsymbol{J}  \cdot \boldsymbol{k}_e \over J E_e }  
 + B''(E_e) {\boldsymbol{J}  \cdot \boldsymbol{k}_\nu \over J E_\nu }\bigg ) 
{ (\boldsymbol{k}_e \cdot \boldsymbol{j}) (\boldsymbol{k}_\nu \cdot \boldsymbol{j}) \over  E_e E_\nu } 
- {B''(E_e) \over 3}   {\boldsymbol{J}  \cdot \boldsymbol{k}_e \over J E_e }   
- {A''(E_e) \over 3}  {p_e^2\over E_e^2} {\boldsymbol{J}  \cdot \boldsymbol{k}_\nu \over J E_\nu } 
\nnl +&  
{ (\boldsymbol{J}  \cdot \boldsymbol{j})^2\over J (J+1)} 
\bigg [  {\boldsymbol{J}  \cdot \boldsymbol{k}_e \over J E_e }   \bigg (
\hat A(E_e)  + \hat A'(E_e) \frac{\boldsymbol{k}_e \cdot \boldsymbol{k}_\nu}{E_e E_\nu}
+ \hat A''(E_e) { (\boldsymbol{k}_e \cdot \boldsymbol{j}) (\boldsymbol{k}_\nu \cdot \boldsymbol{j}) \over  E_e E_\nu } 
\bigg ) 
\nnl & +  {\boldsymbol{J}  \cdot \boldsymbol{k}_\nu \over J E_\nu }   \bigg (
\hat B(E_e)  + \hat B'(E_e) \frac{\boldsymbol{k}_e \cdot \boldsymbol{k}_\nu}{E_e E_\nu}
+ \hat B''(E_e) { (\boldsymbol{k}_e \cdot \boldsymbol{j}) (\boldsymbol{k}_\nu \cdot \boldsymbol{j}) \over  E_e E_\nu } 
\bigg ) 
\bigg ] \bigg \} , 
\end{align}
with 
\begin{align}
A''(E_e) = & 2 \gamma_{FT} r^2 {3 J^2 + 3J - 1 \over J(J+1)} {E_e \over  m_N} \re \lzm C_T^+ \bar C_{FT}^+\dzm,
\nnl 
B''(E_e) = & 2 \gamma_{FT} r^2 {3 J^2 + 3J - 1 \over J(J+1)} {E_e^{\rm max} - E_e \over  m_N} \re \lzm C_T^+ \bar C_{FT}^+\dzm,
\nnl 
\hat A(E_e)  = & 2 \gamma_{FT} r^2 {E_e^{\rm max} - E_e \over  m_N} \re \lzm C_T^+ \bar C_{FT}^+\dzm, 
\nnl
\hat A'(E_e)  = & 4 \gamma_{FT} r^2 {E_e \over  m_N} \re \lzm C_T^+ \bar C_{FT}^+\dzm,
\nnl
\hat A''(E_e)  = & - 10 \gamma_{FT} r^2 {E_e \over  m_N} \re \lzm C_T^+ \bar C_{FT}^+\dzm,
\nnl
\hat B(E_e)  = & 2 \gamma_{FT} r^2 {E_e^2 - m_e^2 \over  E_e m_N} \re \lzm C_T^+ \bar C_{FT}^+\dzm, 
\nnl
\hat B'(E_e)  = & 4 \gamma_{FT} r^2 {E_e^{\rm max} - E_e \over  m_N} \re \lzm C_T^+ \bar C_{FT}^+\dzm,
\nnl
\hat B''(E_e)  = &  -10 \gamma_{FT} r^2 {E_e^{\rm max} - E_e \over  m_N} \re \lzm C_T^+ \bar C_{FT}^+\dzm . 
\end{align}
On the other hand, the contributions proportional to $\gamma_{FT}$  to the usual correlations in  \cref{eq:BETA_dGammaTemplate}  are 
\begin{equation}
\begin{alignedat}{1}
\hat{\xi} \xi_{b}(E_e) &=  0 , 
\\
\hat{\xi}\Delta a(E_e) &=  0, 
\\
\hat{\xi}a'(E_e)&=0,
\\
\hat{\xi}\Delta A(E_e) &=  {\gamma_{FT} \over m_N } {r^2(2 J-1) (2 J+3) \over 3 J(J+1)}
\bigg [ \pm m_e \re \lzm C_A^+ \bar C_{FT}^+ \dzm 
+E_e \re \lzm C_T^+ \bar C_{FT}^+ \dzm  \bigg], 
\\
\hat{\xi} A'(E_e) &= 
- {\gamma_{FT} \over m_N } {r^2(4J^2 + 4J - 1) \over 2 J(J+1)}  
E_e \re \lzm C_T^+ \bar C_{FT}^+ \dzm  , 
\\
\hat{\xi}\Delta B(E_e) &=  
 {\gamma_{FT} (E_e^{\rm max} - E_e) \over m_N } {r^2(2 J-1) (2 J+3) \over 3 J(J+1)}
\lzs \pm {m_e \over E_e} \re \lzm C_A^+ \bar C_{FT}^+ \dzm 
+ \re \lzm C_T^+ \bar C_{FT}^+ \dzm  \dzs , 
\\
\hat{\xi} B'(E_e) &= 
- {\gamma_{FT}  (E_e^{\rm max} - E_e) \over m_N } {r^2(4J^2 + 4J - 1) \over 2 J(J+1)}   \re \lzm C_T^+ \bar C_{FT}^+ \dzm  , 
\\
\hat{\xi}\Delta D (E_e) &=  
 {\gamma_{FT} \over m_N } {r^2(2 J-1) (2 J+3) \over 6 J(J+1)}
\bigg[ \pm m_e \im \lzm C_A^+ \bar C_{FT}^+ \dzm 
+(E_e^{\rm max} - 2 E_e) \im \lzm C_T^+ \bar C_{FT}^+ \dzm  \bigg] , 
\\
\hat{\xi}D'(E_e)&=0,
\\
\hat{\xi}\Delta \hat{c}(E_e) &=
{\gamma_{FT} \over m_N }  \bigg [ 
\mp  m_e  r \sqrt{J+1 \over J} \re \lzm C_V^+ \bar C_{FT}^+ \dzm 
+ {3 m_e  \over 2}  {r^2 \over J}   \re \lzm C_A^+ \bar C_{FT}^+ \dzm 
\\ & 
+ (E_e^{\rm max} - 2 E_e)  r \sqrt{J+1 \over J}  \re \lzm C_S^+ \bar C_{FT}^+ \dzm  
\pm  {3 E_e^{\rm max} \over 2}  {r^2 \over J}  \re \lzm C_T^+ \bar C_{FT}^+ \dzm  \bigg ] , 
\\
\hat{\xi}c'(E_e)&=0,
\\
\hat{\xi} c_1(E_e) &=
{\gamma_{FT} (E_e^{\rm max} -  E_e) \over m_N }
\bigg [ 
\mp  {m_e \over 3 E_e} r \sqrt{J+1 \over J}  \re \lzm C_V^+ \bar C_{FT}^+ \dzm 
+ { m_e  \over 2 E_e}  {r^2 \over J}   \re \lzm C_A^+ \bar C_{FT}^+ \dzm 
\\ & 
- {1 \over 3} r \sqrt{J+1 \over J}  \re \lzm C_S^+ \bar C_{FT}^+ \dzm  
\pm {1 \over 2}  {r^2 \over J}  \re \lzm C_T^+ \bar C_{FT}^+ \dzm  \bigg ] , 
 \\
\hat{\xi} c_2(E_e) &=
{\gamma_{FT} E_e \over m_N }
\bigg [ 
 {1 \over 3}  r \sqrt{J+1 \over J}  \re \lzm C_S^+ \bar C_{FT}^+ \dzm  
\pm  {1 \over 2}  {r^2 \over J}  \re \lzm C_T^+ \bar C_{FT}^+ \dzm  \bigg ] , 
\\
\hat{\xi} c_3(E_e) & =
{\gamma_{FT}  (E_e^{\rm max} -  E_e) \over m_N }
\bigg [ 
-  r \sqrt{J+1 \over J}  \im \lzm C_S^+ \bar C_{FT}^+ \dzm  
\pm  {1 \over 2}  {r^2 \over J}  \im \lzm C_T^+ \bar C_{FT}^+ \dzm  \bigg ] , 
\\
\hat{\xi} c_4(E_e) &= 
{\gamma_{FT}  (E_e^{\rm max} -  E_e) \over m_N }
\bigg [ 
-  r \sqrt{J+1 \over J}  \im \lzm C_S^+ \bar C_{FT}^+ \dzm  
\mp   {1 \over 2}  {r^2 \over J}  \im \lzm C_T^+ \bar C_{FT}^+ \dzm  \bigg ] . 
\end{alignedat}
\end{equation}

\subsection{Phase space and normalization}
\label{sec:phasespace}

\begin{equation}
\begin{alignedat}{1}
\hat{\xi} \xi_{b}(E_e) &= 
{1 \over m_{\cal N}} \bigg [ 
\lzm 2 E_e - E_e^{\rm max} + {m_e^2 \over E_e} \dzm  |C_V^+|^2 
+ \lzm {10\over 3} E_e - E_e^{\rm max} - {m_e^2 \over 3 E_e} \dzm r^2  |C_A^+|^2
\\ & + \lzm 4 E_e - E_e^{\rm max} - {m_e^2 \over E_e} \dzm  |C_S^+|^2 
+ \lzm {8\over 3} E_e - E_e^{\rm max} + {m_e^2 \over 3 E_e} \dzm r^2  |C_T^+|^2
\\ & 
\pm 2 m_e \lzm 3 - {E_e^{\rm max} \over E_e} \dzm{\rm Re} \lzm C_V^+ \bar C_S^+ + r^2 C_A^+ \bar C_T^+ \dzm 
\bigg ] ,\\
\hat{\xi}\Delta a(E_e) &= 
{1 \over m_{\cal N}} \bigg [  
- E_e^{\rm max} |C_V^+|^2  
+ \lzm {1\over 3} E_e^{\rm max} - 4 E_e \dzm r^2 |C_A^+|^2 
\\ & 
+  \lzm E_e^{\rm max} - 6 E_e \dzm |C_S^+|^2 
- \lzm {1\over 3} E_e^{\rm max} + 2 E_e \dzm r^2 |C_T^+|^2 
 \mp 6 m_e {\rm Re} \lzm C_V^+ \bar C_S^+ + r^2 C_A^+ \bar C_T^+ \dzm 
\bigg ] ,\\
\hat{\xi} a'(E_e) &= 
\frac{E_e}{m_{ \mathcal{N}}} \lzs  -3\lzm |\cvp|^2-|\csp|^2 \dzm   + r^2 \lzm|\capp|^2 - |\ctp|^2\dzm   \dzs ,\\
\hat{\xi}\Delta A(E_e) &= 
2r\sqrt{\frac{J}{J+1}}  \lzs \frac{E_e^{\text{max}}-2E_e}{m_{ \mathcal{N}}}  \re\lzm\cvp\cappd \dzm    -  \frac{E_e^{\text{max}}-4 E_e}{m_{ \mathcal{N}}} \re\lzm\csp \ctpd\dzm \pm  \frac{m_e}{m_{ \mathcal{N}}} \re\lzm\cvp\ctpd + \capp\cspd\dzm   \dzs 
\\&
\pm\frac{r^2 }{J+1} \frac{1}{m_\mathcal{N}}  \lzs 
\big ( E_e^{\text{max}} - 4 E_e \big ) |C_A^+|^2 
- \big ( E_e^{\text{max}} - 2 E_e \big ) |C_T^+|^2 
\mp 2 m_e \re\lzm\cappd \ctp\dzm \dzs,\\
            \hat{\xi} A'(E_e) &= 6r\sqrt{\frac{J}{J+1}} \frac{E_e}{m_\mathcal{N}}  \re\lzm\cvp\cappd  - \csp \ctpd\dzm \pm\frac{3r^2 }{J+1} \frac{E_e}{m_\mathcal{N}}  \lzm |\capp|^2-|\ctp|^2  \dzm,\\
            \hat{\xi}\Delta B(E_e) &= 2r\sqrt{\frac{ J}{J+1}} \frac{1}{m_\mathcal{N}}  \Big[ (E_e^{\text{max}}-2E_e) \re\lzm\cvp\cappd \dzm   
            + (E_e^{\text{max}}-4 E_e)  \re\lzm\csp \ctpd\dzm 
            \\&\pm  m_e   \frac{E_e^{\text{max}}}{E_e} \re\lzm\cvp\ctpd  +  \capp\cspd\dzm   
            \mp 3 m_e \re\lzm\cvp\ctpd  +  \capp\cspd\dzm  -  \frac{m_e^2}{E_e}  \re\lzm\cvp\cappd  - \csp \ctpd\dzm 
            \Big]
            \\&\pm\frac{r^2 }{J+1} \frac{1}{m_\mathcal{N}}  \Big[-  E_e^{\text{max}}(|\capp|^2   +  |\ctp|^2)  
          + 2 E_e(2 |\capp|^2  +  |\ctp|^2)  \\&\mp 2  m_e \frac{E_e^{\text{max}}}{E_e}   
         \re\lzm\capp \ctpd\dzm   \pm 6 m_e \re\lzm\capp \ctpd\dzm   
           -  \frac{m_e^2}{E_e} \lzm |\capp|^2-|\ctp|^2\dzm  \Big] ,\\
            \hat{\xi} B'(E_e) &= 6r\sqrt{\frac{J}{J+1}} \frac{1}{m_\mathcal{N}}\lzs E_e \re\lzm\cvp\cappd  + \csp \ctpd\dzm  
          \pm  m_e \re\lzm\cvp\ctpd  + \capp\cspd \dzm
             \dzs
            \\&\mp\frac{3 r^2 }{J+1}  \frac{1}{m_\mathcal{N}}\lzs    E_e \lzm|\capp|^2  +  |\ctp|^2\dzm  
          \pm 2 m_e \re\lzm\capp \ctpd\dzm  \dzs,\\
        \hat{\xi}\Delta D(E_e) &=2r\sqrt{\frac{J}{J+1}}\frac{E_e^{\text{max}}-3E_e}{m_\mathcal{N}}\im\lzm\cvp\cappd  - \csp \ctpd\dzm,\\
\hat{\xi} D'(E_e) &=6r\sqrt{\frac{J}{J+1}} \frac{E_e}{m_\mathcal{N}} \im\lzm\cvp\cappd  - \csp \ctpd\dzm,\\
\hat{\xi}\Delta \hat{c}(E_e) &=  {3 E_e - E_e^{\rm max} \over m_{\cal N}} 
r^2 \lzm |C_A^+|^2 - |C_T^+|^2 \dzm  
,\\ 
\hat{\xi} c'(E_e) &= - {3 E_e \over  m_{\cal N}} r^2 \lzm |C_A^+|^2 - |C_T^+|^2 \dzm,\\
\hat{\xi}c_1(E_e) &=\hat{\xi}c_2(E_e)=\hat{\xi}c_3(E_e)=\hat{\xi}c_4(E_e)=0.
\end{alignedat}
\nonumber
\end{equation}

\section{Data used in the analysis}
\label{app:data}

In our numerical analysis we use the input from superallowed $0^+ \to 0^+$ beta transitions (\cref{tab:0plus}), neutron decay (\cref{tab:neutron}), mirror decays  (\cref{tab:mirror}), and  correlation measurements in pure Fermi decays (\cref{tab:pure}).

\begin{table}[h]
\caption{
\label{tab:0plus}
Input from superallowed $0^+ \to 0^+$ beta transitions~\cite{Hardy:2020qwl} used in our analysis.}
\begin{center}
\begin{tabular}{r@{\hspace{8mm}} r@{\hspace{8mm}}  c}
\hline\hline
Parent & ${\cal F}t$ [s] ~~~ & $\langle m_e/E_e \rangle$\\
\hline
$^{10}$C   & $3075.7 \pm 4.4$ &  0.619  \\
$^{14}$O   & $3070.2\pm 1.9$ &   0.438  \\
$^{22}$Mg  & $3076.2 \pm 7.0$ &  0.308  \\
$^{26m}$Al & $3072.4 \pm 1.1$ &  0.300  \\
$^{26}$Si & $3075.4 \pm 5.7$ &   0.264  \\
$^{34}$Cl  & $3071.6 \pm 1.8$ &  0.234  \\
$^{34}$Ar  & $3075.1\pm 3.1$ &   0.212  \\
$^{38m}$K  & $3072.9\pm 2.0$ &   0.213  \\
$^{38}$Ca  & $3077.8\pm 6.2$ &   0.195  \\
$^{42}$Sc  & $3071.7\pm 2.0$ &   0.201  \\
$^{46}$V   & $3074.3\pm 2.0$ &   0.183  \\
$^{50}$Mn  & $3071.1\pm 1.6$ &   0.169  \\
$^{54}$Co  & $3070.4\pm 2.5$ &   0.157  \\
$^{62}$Ga  & $3072.4\pm 6.7$ &   0.142  \\
$^{74}$Rb  & $3077\pm 11$ &      0.125  \\
\hline\hline
\end{tabular}
\end{center}
\end{table}

\begin{table}[h]
\caption{
\label{tab:neutron}
Input from neutron decay  used in our analysis.
}
\begin{center}
\begin{tabular}{
c@{\hspace{3mm}}  r@{\hspace{5mm}}  
c@{\hspace{3mm}}
c@{\hspace{3mm}}
c }
\hline\hline
Observable		& Value~~ 	& S factor		
& $\langle m_e/E_e \rangle$	& References \\
\hline
$\tau_n$~(s)	& 878.64(59) & 2.2
& 0.655					&\cite{Mampe:1993an,Byrne:1996zz,Serebrov:2004zf,Pichlmaier:2010zz,Steyerl:2012zz,Yue:2013qrc,Ezhov:2014tna,Arzumanov:2015tea,Pattie:2017vsj,Serebrov:2017bzo,UCNt:2021pcg} 
\\
$\tilde{A}_n$	&  $-0.11958(21)$ & 1.2 		
& 0.569					& \cite{Bopp:1986rt,Liaud:1997vu,Erozolimsky:1997wi,Mund:2012fq,Brown:2017mhw,Markisch:2018ndu,Zyla:2020zbs}\\
$\tilde{B}_n$	&  0.9805(30)	&			
& 0.591					& \cite{Kuznetsov:1995sk,Serebrov:1998aj,Kreuz:2005jz,Schumann:2007qe}\\
$\lambda_{AB}$&  $-1.2686(47)$ & 
& 0.581					& \cite{Mostovoi:2001ye}\\
$a_n$		& $-0.10426(82)$	& 		
& &\cite{Stratowa:1978gq,Byrne:2002tx,Beck:2019xye}  
\\
$\tilde{a}_n$	& $-0.1078(18)$	&  		
& 0.695					& \cite{Hassan:2020hrj}
\\
\hline\hline
\end{tabular}
\end{center}
\end{table}

\begin{table}[h]
\bc
\setlength{\tabcolsep}{3pt}
\setlength{\extrarowheight}{7pt}
\begin{tabular}{ccrcrrr}
\hline\hline
Parent & $J=J'$
& $\Delta$~[MeV]~~~
& $\langle m_e/E_e \rangle$
& $f_A/f_V$~~
& ${\cal F}t$ [s]~~~~~ 
& Correlation~~~~~~  
 \\  \hline
${}^{17}$F  & 5/2 
&  2.24947(25) 
& 0.447 & 1.0007(1)  & 2292.4(2.7)~\cite{Brodeur:2016spm}
&  $\tilde A=0.960(82)$~\cite{PhysRevLett.63.1050,Severijns:2006dr} 
 \\
${}^{19}$Ne  & 1/2 
& 2.72849(16)
& 0.386 & 1.0012(2) & 1721.44(92)~\cite{Rebeiro:2018lwo}
  &  $\tilde{A}_{0} =-0.0391(14)$~\cite{Calaprice:1975zz}  
  \\
  & & & & & &$\tilde A_0 = -0.03871(91)$~\cite{Combs:2020ttz}
   \\
${}^{21}$Na & 3/2 
&  3.035920(18) 
& 0.355 & 1.0019(4)   &  4071(4)~\cite{Karthein:2019bss} 
&  $\tilde a=0.5502(60)$~\cite{Vetter:2008zz}   
       \\
${}^{29}$P & 1/2 
& 4.4312(4) 
& 0.258 & 0.9992(1) & 4764.6(7.9)~\cite{Long:2020lby}
  &  $\tilde A=0.681(86)$~\cite{Masson:1990zz}
   \\
${}^{35}$Ar & 3/2 
& 5.4552(7)
&  0.215  & 0.9930(14)  & 5688.6(7.2)~\cite{Severijns:2008ep}     
& $\tilde A=0.430(22)$~\cite{Garnett:1987gw,Converse:1993ba,NaviliatCuncic:2008xt} 
\\
${}^{37}$K & 3/2 
&  5.63647(23)
&  0.209 & 0.9957(9)    & 4605.4(8.2) \cite{Shidling:2014ura}

& $\tilde A =-0.5707(19)$~\cite{Fenker:2017rcx}  \\
&&&&&&     
$\tilde B =  -0.755(24)$~\cite{Melconian:2007zz}
     \\  
\hline\hline
\end{tabular} 
\ec 
\caption{\label{tab:mirror}
Input from mirror beta decays used in our analysis.}
\end{table}
\begin{table}[h]
\caption{
\label{tab:pure}
Input from correlation measurements in pure Fermi  decays used in our analysis. 
}
\begin{center}
\begin{tabular}{
l@{\hspace{5mm}} c @{\hspace{3mm}} 
c@{\hspace{3mm}}  c@{\hspace{3mm}}
 r@{\hspace{5mm}}  
c@{\hspace{3mm}}  c 
}
\hline\hline
Parent		& $J =J'$	& Type	& Observable	& Value~~
& $\langle m_e/E_e \rangle$	& Ref.
\\ \hline
$^{32}$Ar		& 0 		&  F/$\beta^+$		& $\tilde{a}$	& 0.9989(65)
&  0.210	& \cite{Adelberger:1999ud} 
\\
$^{38m}$K	& 0	 		&  F/$\beta^+$		& $\tilde{a}$	& 0.9981(48)	
&  0.161  & \cite{Gorelov:2004hv}
\\
\hline\hline
\end{tabular}
\end{center}
\end{table}
%

\section{On the many-body currents}
\label{app:twobody}

In this appendix we briefly discuss how the many-body currents affect our analysis.
At the leading order in recoil, we can calculate the amplitudes for nuclear beta transitions in two different ways, corresponding to distinct effective theory expansions. 
One way involves matrix elements of {\em quark} operators in \cref{eq:TH_Lweft} between the initial and final nuclear states. 
The other involves matrix elements of {\em nucleon} operators in \cref{eq:TH_NRleeyang0}.  
These two ways will in general yield different results, because the latter misses the contribution of operators quartic and higher in the nucleon field (and still bi-linear in the lepton fields), which are neglected in our pionless EFT Lagrangian. 
The contributions of these operators are referred to as {\em many-body currents} in the nuclear literature~\cite{Butler:2001jj,Cirigliano:2012pq,Hoferichter:2015ipa,Savage:2016kon,De-Leon:2016wyu,Korber:2017ery,Parreno:2021ovq,Detmold:2021oro};  two-body currents corresponding to quartic nucleon operators, etc. 
The difference between the two ways of calculating matrix elements therefore gives us an insight into the structure and magnitude of many-body effects. 

For simplicity, let us restrict to $\beta_-$ transitions with parent and daughter nucleus of spin $J=1/2$; the general case is qualitatively similar and will be discussed in a separate future publication. 
In analogy to \cref{eq:TH_matrixelements}, 
Lorentz symmetry and parity determines the matrix elements of the quark bi-linears occurring in \cref{eq:TH_Lweft} to be 
\begin{align}
\bra{{\cal N}'}  \bar{u}   \gamma_\mu  d  \ket{{\cal N}}  =& 
   g_V^i   \bar  u_{\cal N'}  \gamma_\mu  u_{\cal N}  ~\left( 1 \right.+ \cO(q/m_{\cal N}) \left. \right) , 
\nnl 
\bra{ {\cal N}' }  \bar{u}   \gamma_\mu  \gamma_5 d  \ket{ {\cal N}}  =& 
  g_A^i \bar  u_{\cal N'}  \gamma_\mu \gamma_5   u_{\cal N}  ~\left( 1 \right.+ \cO(q/m_{\cal N}) \left. \right), 
\nnl 
\bra{ {\cal N}'  }  \bar{u}   d  \ket{ {\cal N} }  =& 
g_S^i  \bar u_{\cal N'}  u_{\cal N}~\left( 1 \right.+ \cO(q/m_{\cal N})  \left. \right)  , 
\nnl 
\bra{ {\cal N}' } \bar u \gamma_5  d  \ket{ {\cal N} }  =& 
g_P^i \bar  u_{\cal N'} \gamma_5 u_{\cal N} ~\left( 1 \right.+ \cO(q/m_{\cal N} \left. \right),
\nnl 
\bra{ {\cal N}' }  \bar{u}   \sigma_{\mu\nu}  d  \ket{ {\cal N} }  =& 
 g_T^i  \bar  u_{\cal N'}  \sigma_{\mu\nu}  u_{\cal N} ~\left( 1 \right.+ \cO(q/m_{\cal N}) \left. \right). 
\end{align}
Here, $u_{\cal N} = u(p,J_z,m_{\cal N})$  ($u_{\cal N'} = u(k',J_z',m_{\cal N'})$)  is the  spinor wave function of the mother (daughter) nucleus,
$p \approx k'$  is the momentum of the mother and daughter nucleus,  
$J_z$, $J_z'$ are the projection of their polarizations on the $z$ axis, 
and $q = p - k'$ corrections are neglected. 
The nuclear charges $g_X^i$ are numerical parameters depending on the strong nuclear dynamics, 
and a priori they are transition dependent, which is indicated by the index $i$.  
However, for some of the charges one can determine the transition dependence from symmetry arguments. 
In particular, if ${\cal N}$ and ${\cal N}'$ reside within the same isospin multiplet then 
$g_V^i = M_F g_V$ 
in the limit of unbroken isospin symmetry.  
For the scalar charges, we can relate them to the vector ones using the equations of motion in the limit $e \to 0$ where QED is decoupled, 
$i \partial^\mu [\bar u \gamma^\mu d ] =  (m_d - m_u )\bar u d$. 
Then, along the lines of Ref.~\cite{Gonzalez-Alonso:2013ura},  
\begin{align}
\bra{ {\cal N}'  }  \bar{u}   d  \ket{ {\cal N} } = & 
{ \bra{{\cal N}'} i \partial_\mu [\bar{u}   \gamma_\mu  d]  \ket{{\cal N}}   \over m_d - m_u} 
={[m_{\cal N} - m_{\cal N'} ]_{\rm QCD }  \over m_d - m_u}  g_V^i   \bar  u_{\cal N'}    u_{\cal N} 
\nnl  = & 
{[m_n - m_p ]_{\rm QCD }  \over m_d - m_u}  g_V^i   \bar  u_{\cal N'}    u_{\cal N}  , 
\end{align}
where the QCD subscript denotes the mass difference in the limit  $e \to 0$. 
The last step relies on the fact that in this limit the isospin breaking interactions are negligible. 
Thus, $g_S^i ={[m_n - m_p ]_{\rm QCD }  \over m_d - m_u}  M_F g_V =M_F\,g_S$, 
where we used $g_S=  {[m_n - m_p ]_{\rm QCD}  \over (m_d - m_u) }$~\cite{Gonzalez-Alonso:2013ura}. 
Similar discussion applies to the pseudoscalar charges. 
The equation of motion 
$i \partial^\mu [\bar u \gamma^\mu \gamma_5 d ] = -  (m_d + m_u )\bar u\gamma_5 d$ implies 
\begin{align}
\bra{ {\cal N}'  }  \bar{u}   \gamma_5 d  \ket{ {\cal N} } =  & 
- { \bra{{\cal N}'} i \partial_\mu [\bar{u}   \gamma_\mu \gamma_5  d]  \ket{{\cal N}}   \over m_d + m_u} 
={[m_{\cal N} + m_{\cal N'} ]_{\rm QCD }  \over m_d + m_u}  g_A^i   \bar  u_{\cal N'}   \gamma_5 u_{\cal N} 
\nnl  \approx & 
{2 A g_A^i m_N \over m_d + m_u}    \bar  u_{\cal N'}   \gamma_5 u_{\cal N} ~,
\end{align}
that is, $g_P^i = A\,g_P\,g_A^i/g_A$.
All in all, for the vector and scalar charges the transition dependence  is fixed by the isospin quantum numbers, 
while for the pseudoscalar charges it can be related to the transition dependence of the axial charges. 
On the other hand, for the tensor charges the transition dependence cannot be established from simple arguments. 
The beta decay amplitude in the quark level EFT,  at leading order in recoil for each coupling, takes the form 
\begin{align}
\cM_{\rm QEFT} = & - {V_{ud} \over v^2} \bigg \{ 
M_F g_V (1 + \epsilon_L + \epsilon_R)   L^\mu (\bar  u_{\cal N'}  \gamma_\mu  u_{\cal N})
- g_A^i  (1 + \epsilon_L - \epsilon_R)  L^\mu  (\bar  u_{\cal N'}  \gamma_\mu  \gamma_5 u_{\cal N})
\nnl &
~~~~~~~+  M_F g_S \epsilon_S  L  (\bar  u_{\cal N'}u_{\cal N})
- g_P \epsilon_P  A\,\frac{g_A^i}{g_A}   L  (\bar  u_{\cal N'} \gamma_5 u_{\cal N}) 
+  g_T^i {\epsilon_T \over 2  } i L^{\mu\nu}  (\bar  u_{\cal N'}\sigma_{\mu\nu}u_{\cal N})  \bigg \} 
\nnl & ~~~~~~~\times 
~\left\{ 1 \right.+ \cO(q/m_{\cal N}) \left. \right \}
,  \end{align}
where the leptonic currents $L^\mu$, $L$, and $L^{\mu\nu}$ are defined under \cref{eq:BETA_M1}. 
Taking the non-relativistic limit 
\begin{align}
\cM_{\rm QEFT}=  & - {2  \sqrt{m_{\cal N} m_{\cal N'}}  V_{ud} \over v^2} \bigg \{ 
M_F g_V (1 + \epsilon_L + \epsilon_R)   L^0 [\boldsymbol{1}]_{J_z'}{}^{J_z}
+ g_A^i (1 + \epsilon_L - \epsilon_R)   L^k  [\sigma^k]_{J_z'}{}^{J_z}
\nnl &
+ M_F g_S L [\boldsymbol{1}]_{J_z'}{}^{J_z}
-  g_T^i   \epsilon_T   L^{0k}  [\sigma^k]_{J_z'}{}^{J_z} \bigg \}
~\left( 1 + \cO(q/m_{\cal N}) \right)
. \end{align}
Using the $\cO(m_{\cal N}^0)$ part of the matching equations in \cref{eq:TH_matching0}, 
 \begin{align}
\cM_{\rm QEFT}=  & - 2    \sqrt{m_{\cal N} m_{\cal N'}}   \bigg \{ 
M_F    [\boldsymbol{1}]_{J_z'}{}^{J_z} \bigg [ 
  C_V^+  L^0 
  + C_S^+   L  \bigg ] 
\nnl  &
-   [\sigma^k]_{J_z'}{}^{J_z} \bigg  [
{g_A^i \over g_A} C_A^+  L^k 
+ C_T^+ { g_T^i  \over g_T}  L^{0k}   \bigg ]  \bigg \}
~\left( 1 + \cO(q/m_{\cal N}) \right)
. \end{align}
Comparing this result with the one obtained in the non-relativistic nucleon-level EFT, 
cf. \cref{eq:BETA_M1,eq:BETA1_FGTmatrixelements}, which we refer to as  $\cM_{\rm NEFT}$, 
 we can determine the leading contribution of the many-body currents to the beta decay amplitude. 
 Defining $\cM_{\rm MB} \equiv \cM_{\rm QEFT} - \cM_{\rm NEFT}$ we find 
 \begin{align}
\cM_{\rm MB}=  &  2    \sqrt{m_{\cal N} m_{\cal N'}} 
[\sigma^k]_{J_z'}{}^{J_z} \bigg [
 \bigg ( {g_A^i \over g_A}  - {M_F   r_i \over \sqrt 3} \bigg )   C_A^+  L^k  
+  \bigg ( { g_T^i  \over g_T} - {M_F  r_i \over \sqrt 3}  \bigg ) C_T^+  L^{0k}   \bigg ]
~\left( 1 + \cO(q/m_{\cal N}) \right)
. \end{align}
The first observation is that many-body currents do not affect the vector (as is well known) and scalar contributions, 
up to effects suppressed by the ratio of typical beta decay momenta and the nuclear masses.
On the other hand, many-body currents do affect the axial and tensor contributions.  
For the axial ones, however, this does not have any practical consequences for our analysis. 
The reason is that the parameter $r_i$ in the nucleon EFT, defined by \cref{eq:BETA1_FGTmatrixelements} is not known accurately from first principles (except for neutron decay), and is thus fixed by beta decay data. 
From the phenomenological point of view it does not matter whether $r_i$ or $g_A^i$ is fitted - the constraints on the parameters of interest remain the same. 
From the theoretical perspective one concludes that the global fit effectively constrains $r_i^{eff}\equiv \frac{\sqrt{3}}{g_A M_F} g_A^i$ (which include many-body effects) rather than $r_i$ (which is the parameter defined in the nucleon EFT where many-body effects are neglected). 
Finally, many-body currents also affect tensor contributions, and the effect is physical except for neutron decay (where $g_T^i = g_T$ and many-body effects are absent by definition) and for  pure Fermi transitions (which are only sensitive to vector and scalar contributions and thus are free of many-body effects at the discussed order).  
In order to precisely determine the sensitivity of mirror decays  to  $C_T^+$ one would have to know the parameters $g_T^i$ for each given transition. 
Currently, no such calculations exist. 
In our analysis we effectively used 
$g_T^i  =  {g_T g_A^i  \over g_A}$,  
which assumes that tensor and axial many body corrections are the same. 
The difference between $g_T^i $ and $g_T^{i,\rm 1B}$  might not be a small effect.\footnote{
For the axial case, the many body corrections are estimated to be $\cO(10\%)$ for light nuclei~\cite{Butler:2001jj}, 
and are expected to grow as $\sim A$ for heavier nuclei~\cite{Cirigliano:2012pq}. }
This introduces an uncertainty to our fit. 
Its practical consequence are however not important at this point,  because the constraints on the tensor Wilson coefficient $C_T^+$ are dominated by neutron data, where many-body effects are absent by definition. 
On the other hand, the many-body contributions entering through the tensor currents are estimated to be larger than the recoil effects. 
Therefore, any future fit including the subleading tensor parameters will require precise determinations of the parameters $g_T^i$, perhaps from ab-initio calculations.

\newpage

\bibliographystyle{JHEP}
\bibliography{pseudoV3}

\end{document}